\documentclass[aps,a4paper,amssymb,amsmath,reprint,showpacs,superscriptaddress]{revtex4-1}
\usepackage{amsmath,amssymb}
\usepackage{bm,braket,latexsym,multirow}
\usepackage{bbding,pifont,wasysym}
\usepackage{colortbl,array}
\usepackage[hiresbb,pdftex]{graphicx} 
\usepackage[pdftex]{xcolor}

\usepackage[pdftex,
draft=false,
bookmarks=true,
bookmarksnumbered=false,
bookmarksopen=false,
bookmarksopenlevel=4,
colorlinks=true,
anchorcolor=black,
citecolor=blue,
filecolor=magenta,
linkcolor=blue,
linkbordercolor={1 0 0},
urlcolor=violet,
pdfborder={0 0 1},
pdftitle={},
pdfauthor={},
pdfsubject={},
pdfkeywords={},
pdfpagemode=UseThumbs,
]{hyperref}

\newcommand{\Pa}[0]{$\mathcal{P}$}
\newcommand{\T}[0]{$\mathcal{T}$}
\newcommand{\PT}[0]{$\mathcal{PT}$}

\newcommand{\mr}[2]{#1 \, {\mathrm{ \,  #2 \,}}}
\newcommand{\trace}[1]{\mathrm{Tr}[{#1}]}
\newcommand{\dmat}[2]{d_{#1}^{\,#2}}
\newcommand{\bhline}[1]{\noalign{\hrule height #1}}
\newcommand{\bk}{\bm{k}}
\newcommand{\cm}{$\checkmark$}
\newcommand{\cmd}{($\checkmark$)}
\newcommand{\ud}{\mathfrak{D}}
\newcommand{\bce}{\mathcal{A}}

\begin{document}

\title{Chiral photocurrent in parity-violating magnet \\
			and enhanced response in topological antiferromagnet}
\author{Hikaru Watanabe}
\email[]{watanabe.hikaru.43n@st.kyoto-u.ac.jp}
\affiliation{Department of Physics, Graduate School of Science, Kyoto University, Kyoto 606-8502, Japan}
\author{Youichi Yanase}
\affiliation{Department of Physics, Graduate School of Science, Kyoto University, Kyoto 606-8502, Japan}
\affiliation{Institute for Molecular Science, Okazaki,444-8585, Japan}
\date{\today}

\begin{abstract}
Rectified electric current induced by irradiating light, so-called photocurrent, is an established phenomenon in optoelectronic physics. In this paper, we present a comprehensive classification of the photocurrent response arising from the parity violation in bulk systems. We clarify the contrasting role of \T{}- and \PT{}-symmetries and consequently find a new type of photocurrent phenomena characteristic of parity-violating magnets, \textit{intrinsic Fermi surface effect} and \textit{gyration current}. Especially, the gyration current is induced by the circularly-polarized light and it is the counterpart of the shift current caused by the linearly-polarized light. This photocurrent adds a new functionality of materials studied in various fields of condensed matter physics such as multiferroics and spintronics. A list of materials is provided. Furthermore, we show that the gyration current is strongly enhanced by topologically nontrivial band dispersion. On the basis of the microscopic analysis of Dirac models, we demonstrate the divergent photocurrent response and elucidate the importance of tilting of Dirac cones.
\end{abstract}
\maketitle


\section{Introduction}
Optical responses have been providing a lot of interests in condensed matter physics. The optical probes are extensively implemented in the spectroscopy such as the angle-resolved photo-emission spectroscopy and real-space imaging of material phases. Recent studies have clarified exotic phenomena where light and electron are strongly coupled to each other; for instance, photo-induced phase transitions and higher harmonic generations in solids~\cite{Buzzi2019,Ghimire2010,Yoshikawa2019}. Among the nonlinear optical responses, the photocurrent response is constantly offering renewed interests.

The photocurrent phenomenon was historically attributed to the internal field and surface effects of ferroelectric materials~\cite{Chynoweth1956SurfacePyroelectricity,Chen1969LNOinternalfield,Glass1974BPVEinLiNbO3,Choi2009BiFeO3PVE} or to the heterostructure whose prototypical example is the p-n junction device~\cite{Sturman1992Book,Fridkin2001Review}. On the other hand, the photocurrent response originating from the bulk electronic structure has also been clarified. The discovery of the bulk photocurrent can be traced back to the study of a well-known ferroelectric system, BaTiO$_3$~\cite{Koch1976BTO}. The bulk photocurrent has been theoretically investigated by perturbative calculations~\cite{Kraut1981Photovoltaiceffect,Sipe1993,Aversa1995,Sipe2000secondorder}. Subsequently, a first-principles calculation has successfully explained the photocurrent response in ferroelectric materials~\cite{YoungRappe2012_FirstPrincipleBTO,YoungRappe2012BiFeO3}.

Whereas the basic formalism~\cite{Sipe2000secondorder,Ventura2017,Passos2018,Parker2019} and first-principles calculations~\cite{YoungRappe2012_FirstPrincipleBTO,YoungRappe2012BiFeO3,Cook2017design,AzpirozSouza2018} have been established, recent developments in topological science have provided us with new insights into the photocurrent response. The system hosting a topologically nontrivial electronic structure shows enhanced photoelectronic responses due to diverging geometric quantities~\cite{DeJuan2016,Ishizuka2016WeylPGE,Raguchi2016PCMinWeyl,TanRappe2016BPEinTI,ChanPALee2017PGEinWeyl,Yang2017DivergentBPVEinWeyl,Chang2019Fermiarc_photocurrent_theory}. Importantly, robustness of the nontrivial band dispersion may be ensured by its topological property, and it is beneficial for invulnerable and high-performance optoelectronic devices~\cite{Liu2020SemimetalReview}. Recent experiments have actually supported the enhanced photoelectronic responses in various topological materials~\cite{McIver2011,Kastl2015UltrafastPGE_Bi2Se3,Wu2016SHG_in_TaAs,Ma2019_TaIrTe4_photocurrent,Rees2020_fermiarc_photocurrent_experiment}. 

In general, the photocurrent response is allowed when the parity symmetry is violated. This symmetry requirement was satisfied by noncentrosymmetric crystal structures in the previous studies. On the other hand, we have overlooked the other type of parity violation, that is, the magnetic parity violation~\cite{Spaldin2008a,Watanabe2018grouptheoretical,Hayami2018Classification}. In the case of the magnetic parity violation, the magnetic order breaks not only the parity symmetry(\Pa{}-symmetry) but also the time-reversal symmetry (\T{}-symmetry). In a class of such parity-violating magnet the combined symmetry, namely, \PT{}-symmetry is preserved~\cite{Spaldin2008a,Watanabe2018grouptheoretical,Hayami2018Classification}. This symmetry is a striking property of the parity-violating magnets distinct from conventional noncentrosymmetric systems where the \T{}-symmetry is preserved. According to the group-theoretical classification combined with model studies~\cite{Yanase2014zigzag,Zelezny2014NeelorbitTorque,Hayami2014h,Sumita2016,hikaruwatanabe2017}, the \T{} and \PT{} are fundamental symmetries characterizing quantum phases, and essentially distinguish the electronic structure and physical responses unique to the parity violation~\cite{Watanabe2018grouptheoretical,Hayami2018Classification}. The magnetic parity violation has already been discussed in the contexts of multiferroics and spintronics. The candidate materials actually exist in a broad range of magnetic compounds~\cite{Gallego2016,Watanabe2018grouptheoretical,Watanabe2018}. In spite of these findings, there is few studies focusing on the photocurrent in magnetic systems except for a few recent theoretical works~\cite{Zhang2019switchable,Holder2020consequences,fei2020giant}. Thus, it is highly desirable for promoting the functionality of matter to understand a role of the magnetic parity violation in the photoelectronic phenomena. 

This work mainly consists of two parts. Firstly, we present a systematic classification of the photocurrent responses from the viewpoint of \T{}- and \PT{}-symmetries. Following the established perturbative treatment based on the spinless free fermions, we clarify the contrasting roles of these fundamental symmetries and complete all the photocurrent responses. It is shown that the photocurrent is clearly classified on the basis of these symmetries. Furthermore, the classification result leads us to discovery of new linearly- and circularly-polarized photo-induced currents which we name \textit{intrinsic Fermi surface effect} and \textit{gyration current}, respectively. These photocurrents are unique to the magnetically-parity-violating systems and show different properties from the known photocurrent arising from the magnetic parity violation~\cite{Zhang2019switchable,Holder2020consequences}. We also generalize our classification scheme to spinful systems. Especially, owing to the Kramers degeneracy, careful treatment is required to obtain gauge-invariant formulas for the \PT{}-symmetric systems.

Secondly, we clarify basic properties of the gyration current. The gyration current is the counterpart of the shift current and closely related to quantum geometry of the electronic structure. Using the spinful Hamiltonian having the magnetic parity violation, we present microscopic calculations of the gyration current, and compare it with the attenuation coefficient and joint density of states which contribute to the optoelectronic phenomena. Moreover, we show that the gyration current is strongly enhanced by topologically nontrivial electronic structures. We introduce a model Hamiltonian mimicking a real topological antiferromagnet CuMnAs, and show analytical expressions for the gyration current coefficient. A divergent behavior in the low-frequency regime results from the nontrivial quantum geometry. We also show numerical calculations indicating that slightly-massive Dirac electrons also realize an enhanced gyration current. Note that CuMnAs is a promising material for antiferromagnetic spintronics~\cite{Wadley2016}. Thus, our results may motivate interdisciplinary investigations between topological science, optoelectronics, and antiferromagnetic spintronics.

The outline of the paper is as follows. In Sec.~\ref{Sec_formalism_NLO}, we introduce the formalism based on the perturbative calculation in terms of the electric field. Sec.~\ref{Sec_derivation_photocurrent} presents the classification of photocurrent responses in spinless systems by making use of the \T{}- and \PT{}-symmetries. In Secs.~\ref{Sec_photocurrent_Drude} and \ref{Sec_photocurrent_Berry_curvature_dipole}, we describe the photocurrent unique to metals, and Secs.~\ref{Sec_injection_current} and \ref{Sec_shift_current} are devoted to the formulation of the photocurrent allowed in both metals and insulators. Table~\ref{Table_photocurrent_circular_linear_classification} summarizes the classification result of Sec.~\ref{Sec_derivation_photocurrent}. The fomulation is generalized to the spinful case in Sec.~\ref{Sec_classification_spinful}. 
In Sec.~\ref{Sec_gyration_current}, we study the gyration current in details. We first discuss basic properties [Sec.~\ref{Sec_basic_of_gyration_current}], and  next study a simple model [Sec.~\ref{Sec_Rashba_Dressel_model}]. Furthermore, divergent enhancement of the gyration current response in topological antiferromagnet is proposed in Sec.~\ref{Sec_gyration_in_topological_materials}. Finally, we summarize this work in Sec.~\ref{Sec_discussion_summary}.

		\begin{table*}[htbp]
		\caption{Classification of photocurrent responses in terms of \T{}- and \PT{}-symmetries and of linearly-polarized  ($\updownarrow$) and circularly-polarized ($\circlearrowleft$) lights. Note that the responses with the superscript `$\ast$' are allowed in metals. The bold-faced class is clarified by this work.}
		\label{Table_photocurrent_circular_linear_classification}
		\centering
		\begin{tabular}{ccc}\bhline{0.5pt}\vspace{3pt}
		&\T{}			&\PT{}			 \\ \bhline{0.5pt}
		\multirow{3}{*}{($\updownarrow$)}
			&\multirow{3}{*}{Shift current}	&Drude term$^\ast$	 			\\ 
			&&Magnetic injection current	 			\\ 
			&&\textbf{Intrinsic Fermi surface effect}$^\ast$	 			\\ 
		\multirow{3}{*}{($\circlearrowleft$)}
			&Berry curvature dipole effect$^\ast$&\multirow{3}{*}{\textbf{Gyration current}}\\	 
			&Electric injection current&\\
			&Intrinsic Fermi surface effect$^\ast$&\vspace{3pt}\\
		\bhline{0.5pt}		
		\end{tabular}
		\end{table*}

\section{formulation} \label{Sec_formalism_NLO}
This section shows the formalism of perturbative calculations of nonlinear optical responses within the free particle approximation. Although the calculation has been done in previous theoretical studies~\cite{Sipe1993,Aversa1995,Sipe2000secondorder,Ventura2017,Passos2018,Parker2019}, the derivation is shown below for completeness. The noninteracting Hamiltonian is given by
		\begin{equation}
 		H_0 = \int \frac{d\bm{k}}{\left( 2\pi \right)^d} \sum_a \epsilon_{\bm{k}a} c^\dagger_{\bm{k}a}c_{\bm{k}a},\label{noninteracting_Hamiltonian}
 		\end{equation}
where we define the annihilation and creation operators $c_{\bm{k}a},c^\dagger_{\bm{k}a}$ of the Bloch state $\ket{\psi_{\bm{k}a}} = \exp{(i\bm{k}\cdot \hat{\bm{r}})} \ket{u_a (\bm{k})}$ labeled by the crystal momentum $\bm{k}$ and band index $a$. The periodic part of the Bloch state satisfies a Bloch equation,
		\begin{equation}
		H_0 (\bm{k}) \ket{u_{a} (\bm{k})} = \epsilon_{\bm{k}a} \ket{u_{a} (\bm{k})}. \label{Bloch_equation}
		\end{equation}
Next, we consider interaction between electrons and electromagnetic fields. Since the illuminating light is spatially uniform in the length scale of a lattice constant and photo-electric field is much more strongly coupled to electrons than photo-magnetic field, the effect of electromagnetic field is approximated by an uniform electric field, that is written as $\bm{E} (t)$. This is the so-called electric-dipole approximation~\cite{Sturman1992Book}. The applied electric field can be introduced to the Hamiltonian by two approaches; length gauge and velocity gauge approaches~\cite{Ventura2017,Passos2018,Parker2019}.

In the velocity gauge approach~\cite{Kraut1981Photovoltaiceffect,Passos2018,Parker2019}, the electric field modifies the kinetic part of the noninteracting Hamiltonian. The canonical momentum $\bm{p}$ is replaced as
		\begin{equation}
 		\bm{p} \rightarrow \bm{p} -q \bm{A}(t),\label{velocity_gauge_minimal_coupling}
 		\end{equation}
where $\bm{E} (t) = - \partial_t \bm{A} (t)$ and $q$ is the charge of carriers. In this framework, the electric field gives rise to a shift of the momentum. Thus, we can make use of well-established diagrammatic techniques to calculate the nonlinear optical responses~\cite{Parker2019,Joao2019SpectralMethod,Holder2020consequences}. On the other hand, with the length gauge approach, the electric field is taken into account by the dipole Hamiltonian written as
		\begin{equation}
 		H_\text{E} =  - q \bm{r}\cdot \bm{E} (t).\label{dipole_Hamiltonian}
 		\end{equation} 	
In a general sense, the position operator breaks the translation symmetry of solids and may make the Bloch representation less convenient to describe the Hamiltonian under the electric field. In the infinite volume limit, however, the position operator is written in the Bloch representation as~\cite{Adams1959,Blount1962}
		\begin{equation}
		\left[ \bm{r}_{\bm{k}} \right]_{ab} =  i \partial_{\bm{k}} \delta_{ab} + \bm{\xi}_{ab} .  \label{position_operator}
		\end{equation}
The position operator consists of the derivative of crystal momentum $\partial_\mu = \partial/ \partial k_\mu$ and the Berry connection $\bm{\xi}_{ab} = i \Braket{u_a (\bm{k}) | \partial_{\bm{k}}u_b (\bm{k})}$ defined in the manifold of the Brillouin zone. Especially, the Berry connection is a characteristic term of crystalline systems. Although the position operator obtained in Eq.~\eqref{position_operator} is not diagonal in the band index, we can proceed to the perturbative calculations without discarding the Bloch basis. These two gauge choices should be identical to respect the gauge invariance. The equivalence has been confirmed in noninteracting systems by explicitly carrying out the time-dependent gauge transformation~\cite{Aversa1995,Ventura2017}. In the following, we adopt the length gauge. In fact, by using the length gauge approach, various contributions to the nonlinear optical responses are clearly divided in terms of intraband and interband transitions.

To obtain the expectation value of the nonlinear electric current, we derive the current density operator $q \bm{v}$ where $\bm{v}$ is the velocity operator. In the framework of the first quantization with the Heisenberg picture, the velocity operator in the length gauge is given by
		\begin{equation}
 		\left[  v^{(\text{E})} (t)\right]^\mu   = \left[ \dot{r}^{(\text{E})}(t)  \right]^\mu = \frac{1}{i\hbar } \left[ r^\mu (t), H (t) \right],
 		\end{equation} 
where the Hamiltonian $H(t)$ consists of Eqs.~\eqref{noninteracting_Hamiltonian} and \eqref{dipole_Hamiltonian} in the length gauge. Because of the commutative property between the dipole Hamiltonian and the position operator, the electric field does not make any correction to the velocity operator of the unperturbed Hamiltonian [Eq.~\eqref{noninteracting_Hamiltonian}]. Thus, the velocity operator in the Bloch representation is obtained as  
		\begin{equation}
 		\bm{v}_{ab}  \left[ =  \bm{v}^{(\text{E})}_{ab} \right]
		=  \hbar^{-1} \nabla_{\bm{k}} \epsilon_a \delta_{ab} + i \hbar^{-1}  \epsilon_{ab} \bm{\xi}_{ab}.\label{velocity_operator}
 		\end{equation} 
We note that the velocity operator in the velocity gauge is expressed in a modified form since the perturbative part arising from Eq.~\eqref{velocity_gauge_minimal_coupling} does not commute with the position operator~\cite{Ventura2017,Passos2018,Parker2019}.

The perturbative calculations are straightforwardly conducted in the same way as the linear response theory~\cite{Kubo1957}. Here, we derive the nonlinear optical conductivity by following the density matrix approach~\cite{Sipe1993,Aversa1995,Sipe2000secondorder,Ventura2017}. Introducing the density matrix operator $P = \sum_n e^{-H(t)/(k_\text{B}T) } \ket{n}\bra{n}$, we obtain the time-evolution as
		\begin{equation}
		i\hbar \partial_t P \left( t \right) = [ H\left( t \right), P \left( t \right) ].\label{von-Neumann_eq}
		\end{equation}
Note that we adopt the Schr\"{o}dinger picture in the following calculations. When the perturbative calculations are conducted in the Bloch representation, it is convenient to use the reduced density matrix defined by
		\begin{equation}
		\rho_{\bm{k},a b} (t) =  \mathrm{Tr}[ c^\dagger_{\bm{k}b} c_{\bm{k}a} P (t)].
		\end{equation}
In the following, the momentum dependence of the reduced density matrix $\rho_{\bm{k}}$ is implicit unless otherwise mentioned. Equation~\eqref{von-Neumann_eq} in the frequency domain is obtained as
		\begin{equation}
		\left( \hbar \omega -\epsilon_{ab }  \right) \rho_{ab} \left( \omega \right) = -q  \int  \frac{d\Omega}{2\pi} E^\mu \left( \Omega \right)  [ r^\mu_{\bm{k}} , \rho \left( \omega -\Omega \right) ]_{ab},
		\end{equation}
where $\epsilon_{ab} = \epsilon_{\bm{k}a} - \epsilon_{\bm{k}b}$ and we adopt a convention for the Fourier transformation given by
		\begin{equation}
		\rho_{ab}\left( t \right) = \int \frac{d\omega}{2\pi} e^{-i\omega t} \rho_{ab} \left( \omega \right).\label{density_matrix_time_fourier_convention}
 		\end{equation}
Regarding the magnitude of the electric field $|\bm{E}|$ as the perturbation parameter, the reduced density matrix is expanded by powers of the electric field, $\rho= \sum_n \rho^{(n)}$ with $\rho^{(n)} = O(|\bm{E}|^n)$. Thus, we obtain the recursive equation,
		\begin{equation}
		\left( \hbar \omega -\epsilon_{ab }  \right) \rho^{(n+1)}_{ab} \left( \omega \right) = -q  \int  \frac{d\Omega}{2\pi} E^\mu \left( \Omega \right)  [ r^\mu_{\bm{k}}, \rho^{(n)} \left( \omega -\Omega \right) ]_{ab},\label{sequencial_equation_density_matrix}
		\end{equation}
where the zeroth component is given by $\rho^{(0)}_{ab} (\omega)= 2\pi \delta (\omega) f (\epsilon_{\bm{k}a}) \delta_{ab}$ with the Fermi distribution function $f (\epsilon) = \left[ 1+ \exp{(\epsilon-\mu)/(k_\text{B}T)}\right]^{-1}$ and the chemical potential $\mu$. Following Refs.~\cite{Ventura2017,Passos2018}, we introduce the matrix $\hat{d}^{\,\omega}$ defined by
		\begin{equation}
		d_{ab}^{\,\omega} = \frac{1}{\hbar \omega + i 0 - \epsilon_{ab}},
		\end{equation}
where $+i0$ is the infinitesimal and positive scalar derived from the adiabatic application of the external field~\cite{Kubo1957}. Then, Eq.~\eqref{sequencial_equation_density_matrix} is recast as
		\begin{equation}
		\rho^{(n+1)}_{ab} \left( \omega \right) = -q \int \frac{d\Omega}{2\pi} d_{ab }^{\,\omega}  E^\mu \left( \Omega \right) [ r^\mu_{\bm{k}} , \rho^{(n)} \left( \omega -\Omega \right) ]_{ab}.\label{sequencial_equation_density_matrix_with_dmat}
		\end{equation}

For classification of contributions to nonlinear optical conductivity, we make use of the intraband-interband decomposition of the position operator~\cite{Aversa1995,Sipe2000secondorder}. The position operator in the Bloch representation $r^\mu_{\bm{k}}$ [Eq.~\eqref{position_operator}] is divided into the diagonal and off-diagonal components in the band index as $\bm{r}_i$ and $\bm{r}_e$. The perturbation by the electric field is classified into the intraband effect $-q \bm{r}_i \cdot \bm{E}$ and interband effect $-q \bm{r}_e \cdot \bm{E}$. Sequentially calculating the corrections to the reduced density matrix $\rho^{(n)}$ ($n>0$), we obtain the second-order correction $\rho^{(2)}$ as
		\begin{equation}
		\rho^{(2)}_{ab} (\omega) = \rho^\text{(ii)}_{ab} (\omega) + \rho^\text{(ei)}_{ab} (\omega)+ \rho^\text{(ie)}_{ab} (\omega) + \rho^\text{(ee)}_{ab} (\omega),\label{density_matrix_2nd_contribution}
		\end{equation}
where we classify the components by intraband (i) and interband (e) effects. Each term is explicitly given by
		\begin{widetext}
		\begin{align}
		&\rho^\text{(ii)}_{ab} (\omega) 
		= (-iq)^2 \int \frac{d\Omega d\Omega'}{(2\pi)^2}  E^\mu (\Omega) E^\nu (\Omega') d_{ab }^{\,\omega} d_{ab}^{\,\omega-\Omega}  \partial_\mu \partial_\nu f(\epsilon_{\bm{k}a}) \times 2\pi \delta_{ab} \delta (\omega - \Omega -\Omega'), \label{density_matrix_2nd_intra_intra_contribution} \\
		&\rho^\text{(ei)}_{ab} (\omega) 
		= -iq^2 \int \frac{d\Omega d\Omega'}{(2\pi)^2}  E^\mu (\Omega) E^\nu (\Omega') d_{ab }^{\,\omega}   d_{aa }^{\,\omega-\Omega}  \xi^\mu_{ab} \partial_\nu f_{ab }  \times 2\pi  \delta (\omega - \Omega -\Omega'),\\ 
		&\rho^\text{(ie)}_{ab} (\omega) 
		= -iq^2 \int \frac{d\Omega d\Omega'}{(2\pi)^2}  E^\mu (\Omega) E^\nu (\Omega') d_{ab }^{\,\omega}  \left[  \partial_\mu \left(  d_{ab }^{\,\omega-\Omega}  f_{ab } \xi^\nu_{ab }  \right) - i \left( \xi^\mu_{aa} - \xi^\mu_{bb}  \right) d_{ab }^{\,\omega-\Omega}  f_{ab } \xi^\nu_{ab }  \right] \times 2\pi  \delta (\omega - \Omega -\Omega'), \\
		&\rho^\text{(ee)}_{ab} (\omega) 
		= q^2 \sum_{c} \int \frac{d\Omega d\Omega'}{(2\pi)^2} E^\mu (\Omega) E^\nu (\Omega') d_{ab }^{\,\omega}  \left[  d_{cb }^{\,\omega-\Omega}  \xi^\mu_{ac} \xi^\nu_{cb}  f_{ bc} - d_{ac}^{\,\omega-\Omega}   \xi^\mu_{cb} \xi^\nu_{ac} f_{ ca}   \right] \times 2\pi  \delta (\omega - \Omega -\Omega').
		\end{align}
		\end{widetext}
Summation over the repeated Greek indices such as $\mu = x,y,z$ is implicit, and $f_{ab } = f(\epsilon_{\bm{k}a}) - f(\epsilon_{\bm{k}b})$. Note that the components $\rho^\text{(ii)}$ and $\rho^\text{(ei)}$ are finite only when the low-energy carriers are present owing to the Fermi surface or thermal excitations as implied by the Fermi surface factor $\partial_\mu f$~\cite{Ideue2017,deyo2009semiclassical,Moore2010,Sodemann2015}. On the other hand, the other terms ($\rho^\text{(ie)}$ and $\rho^\text{(ee)}$) contribute to the nonlinear optical conductivity even in insulating systems at the absolute zero temperature~\cite{Aversa1995}. In the perturbative calculation of the nonlinear response, the result should not be affected by an arbitrary permutation of applied external fields~\cite{Parker2019}. Thus, we symmetrize the indices and frequencies of electric fields. Exemplified by Eq.~\eqref{density_matrix_2nd_intra_intra_contribution}, the expression is modified as

		\begin{align}
		&\rho^\text{(ii)}_{ab} (\omega) \notag \\
		&= \frac{(-iq)^2}{2!} \int \frac{d\Omega d\Omega'}{(2\pi)^2}  E^\mu (\Omega) E^\nu (\Omega') d_{ab }^{\,\omega} d_{ab }^{\omega-\Omega}  \partial_\mu \partial_\nu f(\epsilon_{\bm{k}a})\notag \\
		&~~~~~\times 2\pi \delta_{ab} \delta (\omega - \Omega -\Omega') + \left[ \left( \mu, \Omega \right) \leftrightarrow \left( \nu, \Omega' \right) \right].
		\end{align}

Finally, we obtain the full expression 
		\begin{align}
		J^{\mu}_{(2)} (\omega) 
			&= \int \frac{d\bm{k}}{\left( 2\pi \right)^d} \sum_{a,b} q v^\mu_{ab} \rho^{(2)}_{ba} (\omega),  \\
			&\equiv  \int \frac{d\omega_1d\omega_2}{(2\pi)^2}  \tilde{\sigma}^{\mu;\nu\lambda} (\omega; \omega_1,\omega_2 ) E^\nu \left( \omega_1  \right)E^\lambda \left( \omega_2  \right),\label{second_order_electric current}
		\end{align}
for the second-order nonlinear electric current density. Considering the common factor, we take a convention for the second-order optical conductivity $\sigma^{\mu;\nu \lambda} \left( \omega; \omega_1,\omega_2 \right)$ given by
		\begin{equation}
 		\tilde{\sigma}^{\mu;\nu \lambda} \left( \omega; \omega_1,\omega_2 \right)= 2\pi \delta (\omega -\omega_1 -\omega_2) ~\sigma^{\mu;\nu \lambda} \left( \omega; \omega_1,\omega_2 \right).
 		\end{equation} 
Classifying the components by following the decomposition in Eq.~\eqref{density_matrix_2nd_contribution}, the conductivity tensor is divided as
		\begin{align}
		\sigma^{\mu;\nu \lambda} 
				&= \sigma^{\mu;\nu \lambda}_\text{ii} + \sigma^{\mu;\nu \lambda}_\text{ei} + \sigma^{\mu;\nu \lambda}_\text{ei}+ \sigma^{\mu;\nu \lambda}_\text{ee}, \label{second_order_nonlinear_conductivity}
		\end{align}
where each component is obtained as
		\begin{widetext}
		\begin{align}
		\sigma^{\mu;\nu \lambda}_\text{ii} \left( \omega; \omega_1,\omega_2 \right)
			&= \frac{q^3}{2}\int \frac{d\bm{k}}{\left( 2\pi \right)^d} \sum_a -v^\mu_{aa} d_{aa}^{\,\omega} d_{aa}^{\,\omega_2}  \partial_\nu \partial_\lambda f(\epsilon_{\bm{k}a}) + \left[ \left( \nu, \omega_1 \right) \leftrightarrow \left( \lambda, \omega_2 \right) \right], \label{second_order_conductivity_ii} \\
		\sigma^{\mu;\nu \lambda}_\text{ei} \left( \omega; \omega_1,\omega_2 \right)
			&=\frac{q^3}{2}\int \frac{d\bm{k}}{\left( 2\pi \right)^d} \sum_{a,b}   -i v^\mu_{ab} d_{ba }^{\,\omega}   d_{aa}^{\,\omega_2}   \xi^\nu_{ba } \partial_\lambda f_{ba} + \left[ \left( \nu, \omega_1 \right) \leftrightarrow \left( \lambda, \omega_2 \right) \right], \label{second_order_conductivity_ei}\\
		\sigma^{\mu;\nu \lambda}_\text{ie} \left( \omega; \omega_1,\omega_2 \right)
			&= \frac{q^3}{2}\int \frac{d\bm{k}}{\left( 2\pi \right)^d} \sum_{ab} -i  v^\mu_{ab}  d_{ ba}^{\,\omega}  \left[\partial_\nu \left( d_{ba}^{\,\omega_2}  f_{ba} \xi^\lambda_{ba } \right) - i \left( \xi^\nu_{bb}- \xi^\nu_{aa} \right)  d^{\omega_2}_{ba} f_{ba} \xi^\lambda_{ba}  \right] + \left[ \left( \nu, \omega_1 \right) \leftrightarrow \left( \lambda, \omega_2 \right) \right],\label{second_order_conductivity_ie} \\
 		\sigma^{\mu;\nu \lambda}_\text{ee} \left( \omega; \omega_1,\omega_2 \right)
			&= \frac{q^3}{2}\int \frac{d\bm{k}}{\left( 2\pi \right)^d} \sum_{a,b,c} v^\mu_{ab}  d_{ ba}^{\,\omega}  \left(  d_{ca }^{\,\omega_2}  \xi^\nu_{bc} \xi^\lambda_{ca}  f_{ac} - d_{bc}^{\,\omega_2}   \xi^\nu_{ca} \xi^\lambda_{bc} f_{cb}  \right) +
			 \left[ \left( \nu, \omega_1 \right) \leftrightarrow \left( \lambda, \omega_2 \right) \right].\label{second_order_conductivity_ee}
		\end{align}
		\end{widetext}
The expression is consistent with the previous results~\cite{Aversa1995,Sipe2000secondorder,MatsyshynSodemann2019}. Although the above formula is generally applicable to second-order optical responses such as second harmonic generation~\cite{Fiebig2005SHG_review} and parametric generation process~\cite{Sturman1992Book}, we only focus on the photocurrent response in the following sections.

\section{Photocurrent formula}\label{Sec_derivation_photocurrent}
In this section, we derive the photocurrent formulas in \T{}-/\PT{}-symmetric systems. For the photocurrent response, the frequencies are taken as 
		\begin{equation}
		\omega = 0,~\omega_1 = -\Omega,~\omega_2 = \Omega,\label{photocurrent_frequency_condition}
		\end{equation}
where we assume $\Omega >0$ without loss of generality. In this section, we consider spinless systems to clarify the contrasting role of \T{} and \PT{}-symmetries. Note that the formulas are extended to the spinful systems later [Sec.~\ref{Sec_classification_spinful}]. 

Firstly, we present a basic symmetry consideration of the photocurrent. The photocurrent response is classified into the linearly-polarized and circularly-polarized light-induced currents which we call LP-photocurrent and CP-photocurrent, respectively. Owing to the fact that the time-domain electric field is real, the electric field in the frequency domain satisfies the relation,
		\begin{equation}
  		\bm{E} (\omega) = \bm{E}^\ast (-\omega).
  		\end{equation}  
The electric current in Eq.~\eqref{second_order_electric current} is transformed as
		\begin{align}
		&J^{\mu}_{(2)} (\omega = 0)\notag \\
			&= \int \frac{d \omega_2}{2\pi}  \sigma^{\mu;\nu\lambda} (0;-\Omega,\Omega) E^{\nu} (-\Omega) E^{\lambda} (\Omega),\\
			&= \int \frac{d \Omega}{2\pi}  \sigma^{\mu;\nu\lambda} (0;-\Omega,\Omega) (E^{\nu} (\Omega))^\ast E^{\lambda} (\Omega),\\ 
			&= \int \frac{d \Omega}{2\pi}  \sigma^{\mu;\nu\lambda} (0;-\Omega,\Omega) \left[ L^{\nu\lambda} \left( \Omega \right)+ i \epsilon_{\nu\lambda\tau}F^{\tau} \left( \Omega \right) \right]. 
		\end{align}
Here we decomposed the product of electric fields into real and imaginary components defined by
		\begin{align}
			&L^{\nu\lambda} (\Omega)= \text{Re} \left[  E^{\nu} (\Omega) (E^{\lambda} (\Omega) )^\ast \right],\\
			&\bm{F} (\Omega) = \frac{i}{2} \bm{E}(\Omega) \times \bm{E}^\ast (\Omega),	
		\end{align}
which are related to the Stokes parameters~\cite{Wolf2007Book_coherence}. Thus, by taking the linearly-polarized light corresponding to the equator of the Poincar\'{e} sphere, $L^{\nu\lambda} \neq 0$ and $\bm{F} = \bm{0}$ are satisfied. To the contrary, in the case of the circularly-polarized light described by the north and south poles of the Poincar\'{e} sphere, $L^{\nu\lambda} = 0$ and $\bm{F} \neq \bm{0}$ are satisfied. The sign of the vector $\bm{F}$ represents handedness of the circularly-polarized light; for the left-handed circularly-polarized light along the $z$-direction, $\bm{E} =E_0 (1, i,0)$ leads to $\bm{F} = |E_0|^2 \hat{z} $.

In the case of the LP-photocurrent, the indices of irradiating electric fields are symmetric. Thus, the LP-photocurrent response is rewritten as
		\begin{equation}
		J^{\mu}_{\text{LP}}= \int \frac{d \Omega}{2\pi} \eta^{\mu;\nu\lambda} (\Omega)  L^{\nu\lambda}(\Omega),
		\end{equation}
where we introduced the symmetrized photocurrent conductivity 
		\begin{equation}
		\eta^{\mu;\nu\lambda} (\Omega) =\frac{1}{2}\left[  \sigma^{\mu;\nu\lambda}(0;-\Omega,\Omega) + \sigma^{\mu;\lambda\nu} (0;-\Omega,\Omega) \right].
		\end{equation}
The symmetry of the LP-photocurrent tensor $\eta^{\mu;\nu\lambda}$ is the same as that of the piezoelectric tensor. Hence, the LP-photocurrent is allowed in noncentrosymmetric systems belonging to the piezoelectric class~\cite{halasyamani1998noncentrosymmetric}.

On the other hand, the indices of irradiating electric fields are anti-symmetric for the CP-photocurrent tensor. The response formula is obtained as
\begin{equation}
	J^{\mu}_{\text{CP}} 
		= \int \frac{d \Omega}{2\pi}  \kappa^{\mu\tau} (\Omega)  F^{\tau}(\Omega),
	\end{equation}
where we introduced an axial tensor
		\begin{equation}
		\kappa^{\mu\tau} (\Omega) = i \epsilon_{\nu\lambda\tau } \sigma^{\mu;\nu\lambda} (0;-\Omega,\Omega).\label{CP_photocurrent_coefficient}
		\end{equation}
The noncentrosymmetric crystallographic point groups with the non-zero $\hat{\kappa}$ are called gyrotropic (optically-active) point groups~\cite{halasyamani1998noncentrosymmetric}. Therefore, the piezoelectric and gyrotropic point groups having the \T{} or \PT{}-symmetry are shown in Appendix~\ref{App_Sec_point_group_list} with a list of materials. With the LP/CP-photocurrent decomposition, we finally obtain the photocurrent response by
		\begin{align}
			&J^{\mu}_{(2)} (\omega = 0) = J^{\mu}_{\text{LP}} + J^{\mu}_{\text{CP}},\\
			&= \int \frac{d \Omega}{2\pi} \left[  \eta^{\mu;\nu\lambda} (\Omega)  L^{\nu\lambda}(\Omega)  + \kappa^{\mu\nu} (\Omega)  F^{\nu}(\Omega) \right].
		\end{align}

Now, we proceed to the derivation of photocurrent responses. As shown in seminal works, the photocurrent responses in the \T{}-symmetric systems have already been clarified in both insulators~\cite{Sipe1993,Aversa1995,Sipe2000secondorder} and metals~\cite{Moore2010}. On the other hand, the photocurrent phenomenon arising from the magnetic order remains unexplored except for a few recent theoretical studies~\cite{Zhang2019switchable,Holder2020consequences,fei2020giant}. Although we reproduce some of the known results in the following subsections, our calculation is distinct from the previous theoretical studies because of the following reasons; we systematically investigate all the photocurrent responses from the viewpoint of the \T{}- and \PT{}-symmetries, unify the reported works, and importantly clarify new photocurrents, named \textit{intrinsic Fermi surface effect} and \textit{gyration current}. In the following, we analyze Eqs.~\eqref{second_order_conductivity_ii}-\eqref{second_order_conductivity_ee} one by one. Frequency dependence of the conductivity tensor is implicit unless otherwise explicitly denoted. 
Table~\ref{Table_photocurrent_classification} shows the classification result of the photocurrent responses in the \T{}- and \PT{}-symmetric systems.

		\begin{table}[htbp]
		\caption{Classification of the photocurrent conductivity in the \T{}-/\PT{}-symmetric systems. The symbols $\updownarrow$ and $\circlearrowleft$ denote photocurrents induced by linearly-polarized and circularly-polarized lights, respectively. The photocurrent denoted by `(this work)' is clarified in this work. The symbols `d' and `o' in the term $\sigma_\text{ee}$ represent the diagonal and off-diagonal components of the velocity matrix $v^{\mu}_{ab}$ in Eq.~\protect\eqref{second_order_conductivity_ee}, while (P) and ($\delta$) denote the terms consisting of the principal integration (reactive part) and delta function (absorptive part), respectively.}
		\label{Table_photocurrent_classification}
		\centering
		\begin{tabular}{lll}
				\bhline{0.5pt}
				&\multicolumn{1}{l}{$\mathcal{T}$}&\multicolumn{1}{l}{$\mathcal{PT}$}	 \\ \hline
				$\sigma_\text{ii}					$&$\times$&$\updownarrow$~\cite{Holder2020consequences}	\\
				$\sigma_\text{ei} 					$&$\circlearrowleft$~\cite{Moore2010}&$\times$		\\
				$\sigma_\text{ee;d}	(\delta) $&$\circlearrowleft$~\cite{Kraut1981Photovoltaiceffect,Sipe2000secondorder}	&$\updownarrow$~\cite{Zhang2019switchable} \\	
				$\sigma_\text{ie}+\sigma_\text{ee;o} (\delta) $&$\updownarrow$~\cite{Sipe2000secondorder}	&$\circlearrowleft \text{(this work)}\vspace{3pt}$\\
				$\sigma_\text{ie}+\sigma_\text{ee} (\text{P})$&$\circlearrowleft$~\cite{Fernando2020Difference}	&$\updownarrow \text{(this work)}$\vspace{3pt}\\
				\bhline{0.5pt}
		\end{tabular}
		\end{table}

\subsection{Fermi surface effect I : Drude term} \label{Sec_photocurrent_Drude}

We first consider the intraband-only contribution [Eq.~\eqref{second_order_conductivity_ii}] which we call Drude term~\cite{Holder2020consequences,Watanabe2020NLC}. The Drude term does not essentially require the multi-band structures and can be captured by the conventional Boltzmann's transport theory where the single band is treated~\cite{Ideue2017}. The photocurrent response is evaluated as
		\begin{align}
		&\sigma^{\mu;\nu \lambda}_\text{ii} \left( \omega; \omega-\omega_2,\omega_2 \right)\notag \\
			&=  \frac{q^3}{2 }\int \frac{d\bm{k}}{\left( 2\pi \right)^d} \sum_a -v^\mu_{aa} d_{aa}^{\,\omega} d_{aa}^{\,\omega_2}  \partial_\nu \partial_\lambda f(\epsilon_{\bm{k}a}) \notag \\
			&~~~+ \left[ \left( \nu, \omega-\omega_2 \right) \leftrightarrow \left( \lambda, \omega_2 \right) \right], \\
			&=  -\frac{q^3}{2\hbar^2 \omega} \left(  \frac{1}{\omega_2 } + \frac{1}{ \omega - \omega_2 } \right)\int \frac{d\bm{k}}{\left( 2\pi \right)^d} \sum_a v^\mu_{aa}  \partial_\nu \partial_\lambda f(\epsilon_{\bm{k}a}),\\
			&\xrightarrow{\omega \rightarrow 0,\omega_2 \rightarrow \Omega} \frac{q^3}{2\hbar^2 \Omega^2}  \int \frac{d\bm{k}}{\left( 2\pi \right)^d} \sum_a v^\mu_{aa}  \partial_\nu \partial_\lambda f(\epsilon_{\bm{k}a}).\label{photocurrent_ii}
		\end{align}
$\sigma^{\mu;\nu \lambda}_\text{ii}$ is therefore classified as the LP-photocurrent response since we can interchange the order of partial derivatives $\partial_\nu \partial_\lambda$. We hence rewrite
		\begin{align}
		\eta^{\mu;\nu \lambda}_\text{D} 
		 	&= \frac{q^3}{2\hbar^2 \Omega^2}  \int \frac{d\bm{k}}{\left( 2\pi \right)^d} \sum_a v^\mu_{aa}  \partial_\nu \partial_\lambda f(\epsilon_{\bm{k}a}). \label{photocurrent_Drude}
		\end{align}
The subscript `D' denotes `Drude' term. It is noteworthy that the magnitude diverges as $\sim \Omega^{-2}$ in the low-frequency regime $\Omega \ll 1$. Owing to Eq.~\eqref{velocity_operator}, the momentum integral in Eq.~\eqref{photocurrent_ii} is recast as 
		\begin{equation}
		\int \frac{d\bm{k}}{\left( 2\pi \right)^d} v^\mu_{aa}  \partial_\nu \partial_\lambda f(\epsilon_{\bm{k}a}) = \frac{1}{\hbar} \int \frac{d\bm{k}}{\left( 2\pi \right)^d} \left( \partial_\mu  \partial_\nu \partial_\lambda \epsilon_{\bm{k}a}  \right) f(\epsilon_{\bm{k}a}),
		\end{equation}
which is finite if and only if both of the \Pa{}- and \T{}-symmetries are broken~\cite{Watanabe2020NLC}. In fact, the \T{}-symmetry ensures the degeneracy between $\pm \bm{k}$ points in the Brillouin zone. Thus, third derivative of the energy spectrum, $\partial_\mu  \partial_\nu \partial_\lambda \epsilon_{\bm{k}a}$, is canceled out by the integration over $\bm{k}$. On the other hand, the \PT{}-symmetry does not forbid the anti-symmetric band dispersion and allows the Drude term [see Table~\ref{Table_photocurrent_classification}].

\subsection{Fermi surface effect II: Berry curvature dipole term}\label{Sec_photocurrent_Berry_curvature_dipole}
In this subsection, we consider the photocurrent derived from the $\sigma_\text{ei}$ term [Eq.~\eqref{second_order_conductivity_ei}]. Although this component is characteristic to metals as the Drude term is, the response needs the multi-band effect. A derivation has successfully been obtained by the semiclassical theory~\cite{Moore2010,Morimoto2016}. Supposing Eq.~\eqref{photocurrent_frequency_condition}, the expression is rewritten by
		\begin{align}
		&\sigma^{\mu;\nu \lambda}_\text{ei} \notag\\
			&=\frac{q^3}{2\hbar^2 \Omega}\int \frac{d\bm{k}}{\left( 2\pi \right)^d} \sum_{a\neq b}  \xi^\mu_{ab}  \xi^\nu_{ba } \partial_\lambda f_{ba} + \left[ \left( \nu, -\Omega \right) \leftrightarrow \left( \lambda, \Omega \right) \right],\\
			&=\frac{q^3}{2\hbar^2 \Omega}\int \frac{d\bm{k}}{\left( 2\pi \right)^d} \sum_{a\neq b}  \left(  \xi^\mu_{ba}  \xi^\nu_{ab} - \xi^\mu_{ab}  \xi^\nu_{ba }  \right) \partial_\lambda f(\epsilon_{\bm{k}a}) \notag \\
			&~~~+ \left[ \left( \nu, -\Omega \right) \leftrightarrow \left( \lambda, \Omega \right) \right],\\
			&=\frac{q^3}{2\hbar^2 \Omega}\int \frac{d\bm{k}}{\left( 2\pi \right)^d} \sum_a i\epsilon_{\mu\nu\tau} \Omega^\tau_a \, \partial_\lambda f(\epsilon_{\bm{k}a}) \notag \\
			&~~~+ \left[ \left( \nu, -\Omega \right) \leftrightarrow \left( \lambda, \Omega \right) \right],
		\end{align}
where we introduced the Berry curvature for the $a$-th band as
		\begin{equation}
		\Omega^\mu_{a} = \epsilon_{\mu\nu\lambda} \partial_\nu \xi^\lambda_{aa} = \frac{i}{2} \sum_{b\neq a} \epsilon_{\mu\nu\lambda} \left(  \xi^\nu_{ab}  \xi^\lambda_{ba }- \xi^\lambda_{ab}  \xi^\nu_{ba} \right).
		\end{equation}
Conducting a partial derivative in the last line, the formula is transformed to the well-known form
		\begin{align}
		&\sigma^{\mu;\nu \lambda}_\text{ei}\notag \\
			&=- \frac{i q^3}{2\hbar^2 \Omega}\int \frac{d\bm{k}}{\left( 2\pi \right)^d} \sum_a \left( \epsilon_{\mu\nu\tau} \partial_\lambda \Omega^\tau_a - \epsilon_{\mu\lambda\tau} \partial_\nu \Omega^\tau_a  \right)  f(\epsilon_{\bm{k}a}),\label{photocurrent_BCD_bare}
		\end{align}
which is called Berry curvature dipole term~\cite{Moore2010,Sodemann2015}. Here we introduce the Berry curvature dipole defined by 
		\begin{equation}
		\mathcal{D}^{\,\mu\nu} = \int \frac{d\bm{k}}{\left( 2\pi \right)^d} \sum_{a} f(\epsilon_{\bm{k}a}) \partial_\mu \Omega^\nu_a.
		\end{equation}
The Berry curvature dipole is allowed when the \Pa{}-symmetry is broken and the Berry curvature in the momentum space shows a dipolar distribution in the Brillouin zone~\cite{Sodemann2015,Ma2019BCD_experiment_WTe2}. The photocurrent arising from the Berry curvature dipole is anti-symmetric under $\nu \leftrightarrow \lambda$, and it is therefore a CP-photocurrent. Thus, we describe the formula of the Berry curvature dipole effect [Eq.~\eqref{photocurrent_BCD_bare}] as
		\begin{align}
		\kappa^{\mu\nu}_\text{BCD} 
			&= i \epsilon_{\nu\lambda\tau} \sigma^{\mu;\lambda \tau}_\text{ei},\\
			&= -\frac{q^3}{ \hbar^2 \Omega} \left( \mathcal{D}^{\,\mu\nu} - \delta_{\mu\nu} \trace{\mathcal{D}} \right),\label{photocurrent_BCD}
		\end{align}
which depends on the frequency of irradiating lights as $O(\Omega^{-1})$. 

The symmetry of the Berry curvature dipole is the same as that of the CP-photocurrent tensor, and hence it is allowed in the \T{}-preserved gyrotropic crystals~\cite{Moore2010,Sodemann2015}. In contrast, in the \PT{}-symmetric systems, the Berry curvature $\Omega^\mu_a$ vanishes at each $\bk$ point since it is odd-parity under the \PT{}-operation. Thus, the photocurrent response derived from $\sigma_\text{ei}$ is regarded as the Berry curvature dipole effect which is unique to the \T{}-symmetric and metallic systems, whereas it is forbidden in the \PT{}-symmetric or insulating systems.

\subsection{Interband effect I : injection current}\label{Sec_injection_current}

We next consider the $\sigma_\text{ee}$ term [Eq.~\eqref{second_order_conductivity_ee}]. Especially, in this subsection we focus on the diagonal component of the velocity operator $v^\mu_{ab}$ ($a=b$) and denote the corresponding conductivity tensor as $\sigma_\text{ee;d}$. The expression is given by
		\begin{align}
		&\sigma^{\mu;\nu \lambda}_\text{ee;d} \left( \omega; \omega_1,\omega_2 \right) \notag\\
			&= \frac{q^3}{2 \hbar  \omega}\int \frac{d\bm{k}}{\left( 2\pi \right)^d} \sum_{a\neq c} \Delta^\mu_{ac}  \xi^\nu_{ac} \xi^\lambda_{ca}  f_{ac} d_{ca }^{\,\omega_2} + \left[ \left( \nu, \omega_1 \right) \leftrightarrow \left( \lambda, \omega_2 \right) \right],\\
			&= \frac{q^3}{2 \hbar  \omega}\int \frac{d\bm{k}}{\left( 2\pi \right)^d} \sum_{a\neq c} \Delta^\mu_{ac}  \xi^\nu_{ac} \xi^\lambda_{ca}  f_{ac} \left(  d_{ca }^{\,\omega_2} + d_{ac}^{\,\omega_1}  \right),\label{second_conductivity_ee;d}
		\end{align}
where $\Delta^\mu_{ac} = v^\mu_{aa} - v^\mu_{cc} = \partial_\mu \epsilon_{ac}/\hbar$ represents the group velocity difference between the $a$-th and $c$-th band electrons at momentum $\bm{k}$~\cite{Zhang2019switchable}. Supposing the condition Eq.~\eqref{photocurrent_frequency_condition}, the resulting expression diverges due to the pre-factor $\omega^{-1}$. Thus, $O(\omega^0)$ and $O(\omega)$ terms in the integrand of Eq.~\eqref{second_conductivity_ee;d} will survive in the limit of $\omega \rightarrow 0$~\cite{NastosSipe2010,Fernando2020Difference}. Accordingly, we perform Taylor expansion
		\begin{equation}
		d_{ac}^{\,\omega_1} = d_{ac}^{\, -\omega_2} + \frac{-\hbar}{(\hbar\omega_1 - \epsilon_{ca})^2}_{| \omega_1  = -\omega_2} (\omega_1 +\omega_2) + O((\omega_1 +\omega_2)^2), 
		\end{equation}
and we rewrite Eq.~\eqref{second_conductivity_ee;d} as
\begin{align}
	&\sigma^{\mu;\nu \lambda}_\text{ee;d} \left( \omega; \omega_1,\omega_2 \right)\notag \\
	&= \frac{q^3}{2 \hbar  \omega}\int \frac{d\bm{k}}{\left( 2\pi \right)^d} \sum_{a\neq c} \Bigl[ \Delta^\mu_{ac}  \xi^\nu_{ac} \xi^\lambda_{ca}  f_{ac} \left(  d_{ca }^{\,\omega_2} + d_{ac}^{\,-\omega_2} \right) \notag \\
	&+\Delta^\mu_{ac} \xi^\nu_{ac} \xi^\lambda_{ca}  f_{ac}  \frac{-\hbar}{(-\hbar\omega_2 - \epsilon_{ca})^2} (\omega_1+\omega_2) \Bigr] + O((\omega_1+\omega_2)^2),\\
	&= \frac{q^3}{2 \hbar  \omega}\int \frac{d\bm{k}}{\left( 2\pi \right)^d} \sum_{a\neq c} \Bigl[ \Delta^\mu_{ac}  \xi^\nu_{ac} \xi^\lambda_{ca}  f_{ac} \left(  d_{ca }^{\,\omega_2} + d_{ac}^{\,-\omega_2} \right) \notag \\
	&+\xi^\nu_{ac} \xi^\lambda_{ca}  f_{ac}  \left(  \partial_\mu d_{ca }^{\,\omega_1} \right)_{| \omega_1  = -\omega_2}(\omega_1+\omega_2) \Bigr] + O((\omega_1+\omega_2)^2),\label{second_conductivity_ee;d_expanded}\\
	&= \sigma^{\mu;\nu \lambda}_\text{inj} \left( \omega; \omega_1,\omega_2 \right) + \sigma^{\mu;\nu \lambda}_\text{intI} \left( \omega; \omega_1,\omega_2 \right) + O((\omega_1+\omega_2)^2),
\end{align}
where we denote the $O(\omega^{-1})$ and $O(\omega^0)$ components by $\sigma^{\mu;\nu \lambda}_\text{inj}$ and $\sigma^{\mu;\nu \lambda}_\text{intI}$, respectively.

With the condition Eq.~\eqref{photocurrent_frequency_condition}, we take the first line in Eq.~\eqref{second_conductivity_ee;d_expanded}
		\begin{equation}
		\sigma^{\mu;\nu \lambda}_\text{inj}= \lim_{\omega \rightarrow 0} \frac{q^3}{2 \hbar  \omega}\int \frac{d\bm{k}}{\left( 2\pi \right)^d} \sum_{a\neq b} \Delta^\mu_{ab}  \xi^\nu_{ab} \xi^\lambda_{ba}  f_{ab} \left( d_{ba }^{\,\Omega}  + d_{ab}^{\,-\Omega}  \right).\label{injection_current_bare}	
		\end{equation}
The optical response is strongly enhanced under the resonant condition that $\hbar \Omega = \pm \epsilon_{ab}$. Thus, we decompose the matrix $\dmat{ab}{\Omega}$ as
		\begin{equation}
		\dmat{ab}{\Omega} = \frac{1}{\hbar \Omega- \epsilon_{ab}} = \text{P} \frac{1}{\hbar \Omega - \epsilon_{ab}} - i\pi \delta (\hbar \Omega  - \epsilon_{ab} ),	\label{principal_integral}
		\end{equation}    
where $\text{P}$ symbolically represents the principal integral for $\Omega$. Note that the infinitesimal parameter $+i0$ is implicitly assumed in the form of $\hbar \Omega + i 0$. 

Eq.~\eqref{injection_current_bare} is rewritten as
		\begin{align}
		&\sigma^{\mu;\nu \lambda}_\text{inj}\notag \\
			&= \lim_{\omega \rightarrow 0} \frac{- i\pi  q^3}{\hbar  \omega}\int \frac{d\bm{k}}{\left( 2\pi \right)^d} \sum_{a\neq b} \Delta^\mu_{ab}  \xi^\nu_{ab} \xi^\lambda_{ba}  f_{ab} \delta (\hbar \Omega  - \epsilon_{ba} ),\\
			&= \lim_{\omega \rightarrow 0} \frac{- i\pi  q^3}{\hbar  \omega}\int \frac{d\bm{k}}{\left( 2\pi \right)^d} \sum_{a\neq b} \Delta^\mu_{ab} \left(  g^{\nu\lambda}_{ab} -\frac{i}{2} \Omega^{\nu\lambda}_{ab}  \right)  f_{ab} \delta (\hbar \Omega  - \epsilon_{ba} ),\label{injection_current_bare2}
		\end{align}
where we introduce the band-resolved quantum metric and Berry curvature which are respectively given by
		\begin{align}
		&g^{\mu\nu}_{ab} = \frac{1}{2}\left( \xi^\mu_{ab} \xi^\nu_{ba} + \xi^\mu_{ab} \xi^\nu_{ba} \right),\label{band_resolved_quantum_metric}\\
		&\Omega^{\mu\nu}_{ab} = i \left(  \xi^\mu_{ab}  \xi^\nu_{ba }- \xi^\nu_{ab}  \xi^\mu_{ba} \right). \label{band_resolved_Berry_curvature}
		\end{align}
These geometric quantities are related to the U(1) quantum metric and Berry curvature as $ g^{\nu\lambda}_{a} = \sum_{b\neq a}  g^{\nu\lambda}_{ab}$ and $\Omega^{\mu}_{a} = \sum_{b\neq a} \epsilon_{\mu\nu\lambda} \Omega^{\nu\lambda}_{ab}/2$~\cite{Gao2020TunablePGE}. The band-resolved quantum metric (Berry curvature) is symmetric (anti-symmetric) under $\nu \leftrightarrow \lambda$ and contributes to the LP-photocurrent (CP-photocurrent). 

Eq.~\eqref{injection_current_bare2} is the general formula for the photocurrent arising from the component $\sigma_\text{inj}$. Then, we proceed to the classification by the \T{}- and \PT{}-symmetries below. Beforehand, we investigate the transformation property of geometric quantities under those symmetry operations. As shown in Appendix~\ref{AppSec_Symmetry_requirement}, the Berry connection is transformed as $\xi^{\nu}_{ab} (\bm{k})  = \xi^{\nu}_{ba} (-\bm{k}) $ for the \T{}-symmetry while $\xi^{\nu}_{ab} (\bm{k})  = -\xi^{\nu}_{ba} (\bm{k}) $ for the \PT{}-symmetry. Accordingly, the band-resolved geometric quantities are transformed as 
		\begin{equation}
		g^{\mu\nu}_{ab} (\bk) = g^{\mu\nu}_{ab} (-\bk),~\Omega^{\mu\nu}_{ab} (\bk) = -\Omega^{\mu\nu}_{ab} (-\bk),\label{quantum_geometric_tensor_T_sym_spinless}
		\end{equation}
for the \T{}-symmetry while
		\begin{equation}
		g^{\mu\nu}_{ab} (\bk) = g^{\mu\nu}_{ab} (\bk),~\Omega^{\mu\nu}_{ab} (\bk) = -\Omega^{\mu\nu}_{ab} (\bk),\label{quantum_geometric_tensor_PT_sym_spinless}
		\end{equation}
for the \PT{}-symmetry. Making use of the fact that the group velocity difference $\Delta^\mu_{ab}$ is odd/even under \T{}/\PT{}-symmetry, we can show that either of the band-resolved quantum metric or Berry curvature contributes to the photocurrent response~\cite{Zhang2019switchable}.

In the \T{}-symmetric systems, the corresponding photocurrent is obtained as 
		\begin{align}
		&\sigma^{\mu;\nu \lambda}_\text{inj} (\mathcal{T}) \notag \\
			&= \lim_{\omega \rightarrow 0} \frac{- \pi  q^3}{2\hbar  \omega}\int \frac{d\bm{k}}{\left( 2\pi \right)^d} \sum_{a\neq b} \Delta^\mu_{ab}  \Omega^{\nu\lambda}_{ab}  f_{ab} \delta (\hbar \Omega  - \epsilon_{ba} ),\label{electric_injection_current_bare}
		\end{align}
which satisfies the anti-symmetric condition under the permutation $\nu\leftrightarrow \lambda$, and hence represents the CP-photocurrent. This is called ``injection current"~\cite{Sipe2000secondorder}. Following the definition in Eq.~\eqref{CP_photocurrent_coefficient}, we obtain the CP-photocurrent tensor
		\begin{align}
		&\kappa^{\mu\nu}_\text{inj} \notag = i \epsilon_{\nu\lambda\tau }\sigma^{\mu;\lambda \tau}_\text{inj} (\mathcal{T}) \\ 
			&=\lim_{\omega \rightarrow 0}  \frac{-i\pi q^3}{2\hbar \omega} \int \frac{d\bm{k}}{\left( 2\pi \right)^d} \sum_{a\neq b} \epsilon_{\nu\lambda\tau} \Delta^\mu_{ab} \Omega^{\lambda\tau}_{ab}  f_{ab}  \delta (\hbar \Omega  - \epsilon_{ba}  ).\label{electric_injection_current}
		\end{align}
The band-resolved Berry curvature is further simplified by the circular representation of the Berry connection given by~\cite{Souze2008Dichroic}
		\begin{equation}
		\xi^\pm_{ab} = \frac{1}{\sqrt{2}} \left( \xi^x_{ab} \pm i \xi^y_{ab} \right).\label{BC_circular_representation}
		\end{equation}
On the basis of this representation, Eq.~\eqref{band_resolved_Berry_curvature} is recast as
		\begin{equation}
		\Omega^{xy}_{ab}  = |\xi^-_{ab}|^2 - |\xi^+_{ab}|^2,
		\end{equation}
which indicates the difference of the dipole-transition amplitude between left- and right-handed circularly-polarized lights~\cite{Souze2008Dichroic}. Accordingly, Eq.~\eqref{electric_injection_current} with $\nu = z$ is rewritten as
		\begin{align}
		&\kappa^{\mu z}_\text{inj} \notag \\ 
			&=\lim_{\omega \rightarrow 0}  \frac{i \pi q^3}{\hbar \omega} \int \frac{d\bm{k}}{\left( 2\pi \right)^d} \sum_{a\neq b}  \left( |\xi^+_{ab}|^2 - |\xi^-_{ab}|^2 \right) \Delta^\mu_{ab} f_{ab}  \delta (\hbar \Omega  - \epsilon_{ba}  ).\label{electric_injection_current_with_BCV}
		\end{align}
The injection current in the \T{}-symmetric systems arises from the band-resolved Berry curvature. Therefore, nonmagnetic Weyl semimetals hosting the divergent Berry curvature are potential candidates which show a giant injection current response in the low-frequency regime. For instance, a well-known Weyl semimetal TaAs exerts a large photocurrent response under mid-infrared lights which may be attributed to the large Berry curvature near Weyl nodes~\cite{Ma2017CPGE_TaAs}, while the enhanced response has also been observed in the higher frequency regime (near-infrared regime) where the group velocity difference may be responsible for the enhanced photocurrent~\cite{Gao2020ChiralTerahertz}. Such topological effect may appear more prominently in the presence of the chiral Weyl fermions~\cite{DeJuan2016,Bradlyn2016chiralWeyl,Chang2018Chiralweylfermions,Cano2019}. Only recently, a related experimental work has been done with a chiral Weyl system RhSi~\cite{Rees2020_fermiarc_photocurrent_experiment}.

On the other hand, the \PT{}-symmetry requires that the Berry curvature vanishes at each $\bk$. Hence, the injection current in the \PT{}-symmetric systems originates from the band-resolved quantum metric.  The formula is written by
		\begin{align}
		&\sigma^{\mu;\nu \lambda}_\text{inj} (\mathcal{PT}) \notag \\
			&= \lim_{\omega \rightarrow 0} \frac{- i\pi  q^3}{\hbar  \omega}\int \frac{d\bm{k}}{\left( 2\pi \right)^d} \sum_{a\neq b} \Delta^\mu_{ab} g^{\nu\lambda}_{ab} f_{ab} \delta (\hbar \Omega  - \epsilon_{ba} ).\label{magnetic_injection_current_bare}
		\end{align}
This expression satisfies the symmetric property for the permutation $\nu\leftrightarrow \lambda$. Thus, the photocurrent is classified as a LP-photocurrent. This result is consistent with Refs.~\cite{Zhang2019switchable,Holder2020consequences,fei2020giant}. The response tensor is given by $\eta^{\mu;\nu \lambda}_\text{inj}   = \sigma^{\mu;\nu \lambda}_\text{inj} (\mathcal{PT})/2 + \sigma^{\mu;\lambda\nu }_\text{inj} (\mathcal{PT})/2$ with Eq.~\eqref{magnetic_injection_current_bare}.
In contrast to the band-resolved Berry curvature, the band-resolved quantum metric represents the dipole-transition amplitude under the linearly-polarized light.

As shown above, the geometric property related to the injection current is different between the \T{}-symmetric and \PT{}-symmetric systems. Whereas the CP-photocurrent in the former is owing to the band-resolved Berry curvature, the LP-photocurrent in the latter arises from the band-resolved quantum metric. Thus, we distinguish the injection currents allowed in the \T{}- and \PT{}-symmetric systems as ``electric injection current" and ``magnetic injection current", respectively [see Table~\ref{Table_photocurrent_circular_linear_classification}]. 

The general formula in Eq.~\eqref{injection_current_bare2} is decomposed as
		\begin{equation}
		\sigma^{\mu;\nu \lambda}_\text{inj} = \eta^{\mu;\nu \lambda}_\text{inj} -\frac{i}{2} \epsilon_{\nu\lambda\tau }\kappa^{\mu\tau}_\text{inj},\label{ee_diagonal_general_decomposition}
		\end{equation}
and both of the electric and magnetic injection currents are allowed in the absence of the \T{} and \PT{}-symmetry. We will see a parallel discussion for the intrinsic Fermi surface effect and shift current in Sec.~\ref{Sec_shift_current}.

In addition to the quantum geometric quantities, two factors are responsible for these injection currents; joint density of states and group-velocity difference $\Delta^\mu_{ab}$. The joint density of states is defined as
		\begin{equation}
		J \left( \Omega \right)= \sum_{a\neq b}\int \frac{d\bm{k}}{\left( 2\pi \right)^d} \delta (\hbar \Omega - \epsilon_{ab}).\label{joint_density_of_states}
		\end{equation}
It measures the number of electrons excited by illuminating light having the frequency $\Omega$ and also plays a crucial role in linear optical responses~\cite{Grosso2013Book}. $J (\Omega)$ is strongly enhanced in the presence of the generalized van Hove singularity where the following condition is satisfied
		\begin{equation}
		\partial_{\bk} \epsilon_{ab}\equiv 0.\label{generalized_van_Hove_singularity}
		\end{equation}
The generalized van Hove singularity originates not only from a pair of usual van Hove singularities given by $\partial_{\bk} \epsilon_{\bm{k}a} = \partial_{\bk} \epsilon_{\bm{k}b} \equiv 0$ but also from the subspace in the Brillouin zone satisfying $\partial_{\bk} \epsilon_{\bm{k}a} = \partial_{\bk} \epsilon_{\bm{k}b} \neq 0$. The factor $\Delta^\mu_{ab}$, however, weakens the contribution from the latter singularity points. Thus, it may be important for a sizable injection current to make use of the normal van Hove singularities satisfying 
		\begin{equation}
		\partial_{\bk} \epsilon_{\bm{k}a} \equiv 0,~\partial_{\bk} \epsilon_{\bm{k}b} \equiv 0,~ \partial_{\mu}^2 \epsilon_{\bm{k}b} \cdot \partial_{\mu}^2 \epsilon_{\bm{k}a} < 0,  
		\end{equation}
where the coordinate $k_\mu$ denotes the direction of the injection current. Such dispersion can be found in prototypical direct-gap semiconductors.

Peculiarly, response coefficients of the injection currents diverge in the limit of $\omega \rightarrow 0$. This seemingly unphysical behavior can be bounded by the scattering rate $\gamma$~\cite{Passos2018}, while our calculation assumes the optical regime, $\hbar \omega \gg \gamma $, for simplicity. Since the induced photocurrent suffers from scatterings before it diverges, the resulting current converges to a finite value~\cite{Rees2020_fermiarc_photocurrent_experiment}. By introducing the scattering rate $\gamma$, the matrix $\dmat{ab}{\omega}$ is modified as
		\begin{equation}
   		\dmat{ab}{\omega}  = \frac{1}{\hbar \omega + i0 -\epsilon_{ab}} \rightarrow  \frac{1}{\hbar \omega + i \gamma -\epsilon_{ab}}.
		\end{equation}
Accordingly, for instance, the formula of the electric injection current in Eq.~\eqref{electric_injection_current} is replaced with
		\begin{align}
		&\kappa^{\mu\nu}_\text{inj} 
		\rightarrow  -\frac{\pi q^3}{2} \int \frac{d\bm{k}}{\left( 2\pi \right)^d} \sum_{a\neq b}  \epsilon_{\nu\lambda\tau}\Delta^\mu_{ab} \Omega^{\lambda\tau}_{ab}  f_{ab} \frac{1}{ \left( \hbar \Omega  - \epsilon_{ba} \right)^2 + \gamma^2 }.\label{electric_injection_with_scattering}
		\end{align}
The expression converges in the limit $\omega \rightarrow 0$. This phenomenological treatment is known to be justified in the calculation based on Floquet formalism~\cite{DeJuan2016}.

Next, the remaining term $\sigma_\text{intI}$ in Eq.~\eqref{second_conductivity_ee;d_expanded} is decomposed into the LP and CP-photocurrents. It is given by
		\begin{align}
		&\sigma^{\mu;\nu \lambda}_\text{intI}\notag \\
			&=  \frac{q^3}{2 \hbar}\int \frac{d\bm{k}}{\left( 2\pi \right)^d} \sum_{a\neq b} \xi^\nu_{ab} \xi^\lambda_{ba}  f_{ab}  \left(  \partial_\mu d_{ba }^{\,\omega'} \right)_{| \omega'  = -\Omega},\\	
			&=  \frac{q^3}{2 \hbar}\int \frac{d\bm{k}}{\left( 2\pi \right)^d} \sum_{a\neq b} \left( g^{\nu\lambda}_{ab} - \frac{i}{2}\Omega^{\nu\lambda}_{ab} \right)  f_{ab} \frac{\hbar \Delta_{ab}}{\left( \hbar \Omega -\epsilon_{ab} \right)^2}.\label{intrinsic_term_ee;d_current_bare}
		\end{align}
aking use of Eqs.~\eqref{quantum_geometric_tensor_T_sym_spinless},~\eqref{quantum_geometric_tensor_PT_sym_spinless} and following the parallel discussion of the injection current, we identify that the CP-photocurrent (LP-photocurrent) is allowed in the \T{}-symmetric (\PT{}-symmetric) systems as
		\begin{align}
		&\sigma^{\mu;\nu \lambda}_\text{intI} \left( \mathcal{T} \right) 
			= \frac{-i q^3}{4 \hbar}\int \frac{d\bm{k}}{\left( 2\pi \right)^d} \sum_{a\neq b} \Omega^{\nu\lambda}_{ab}  f_{ab}  \partial_\mu \frac{1}{\hbar \Omega -\epsilon_{ab} },\label{intrinsic_term_ee;d_current_T_symmetric}\\
		&\sigma^{\mu;\nu \lambda}_\text{intI} \left( \mathcal{PT} \right) 
			= \frac{q^3}{2 \hbar}\int \frac{d\bm{k}}{\left( 2\pi \right)^d} \sum_{a\neq b} g^{\nu\lambda}_{ab}  f_{ab}  \partial_\mu \frac{1}{\hbar \Omega -\epsilon_{ab} },\label{intrinsic_term_ee;d_current_PT_symmetric}
		\end{align}
which will be discussed in the next subsection.

\subsection{Interband effect II :\\  shift current and intrinsic Fermi-surface effect}\label{Sec_shift_current}
Finally, we analyze the remaining terms, that is, the $\sigma_\text{ee}$ term with the off-diagonal component of $v^\mu_{ab}$ in Eq.~\eqref{second_order_conductivity_ee} and the $\sigma_\text{ie}$ term in Eq.~\eqref{second_order_conductivity_ie}. We denote the former contribution by $\sigma_\text{ee;o}$.

When we consider the photocurrent response by adopting Eq.~\eqref{photocurrent_frequency_condition}, the formula for $\sigma_\text{ie}$ is recast with use of  Eq.~\eqref{velocity_operator} as
		\begin{align}
		&\sigma^{\mu;\nu \lambda}_\text{ie}\notag \\
			&= \frac{q^3}{2\hbar }\int \frac{d\bm{k}}{\left( 2\pi \right)^d} \sum_{a\neq b} \left[ - \partial_\nu  \xi^\mu_{ab}  + i \left( \xi^\nu_{aa}- \xi^\nu_{bb} \right) \xi^\mu_{ab} \right]  \xi^\lambda_{ba} f_{ba} d_{ba}^{\,\Omega} \notag \\
			&~~~ + \left[ \left( \nu, -\Omega \right)\leftrightarrow \left( \lambda, \Omega \right) \right]. \label{cond_ie_bare}
		\end{align}
It is convenient to introduce the U(1)-covariant derivative $\bm{D}$ which acts on the physical quantity in the Bloch representation as~\cite{Aversa1995,Sipe2000secondorder,Ventura2017}
		\begin{equation}
		\left[ D_\mu O \right]_{ab} = \partial_{\mu} O_{ab} - i \left( \xi^\nu_{aa}- \xi^\nu_{bb} \right) O_{ab}. \label{covariant_derivative}
		\end{equation}
Then, we rewrite Eq.~\eqref{cond_ie_bare} as 
		\begin{align}
		\sigma^{\mu;\nu \lambda}_\text{ie}
		&= \frac{q^3}{2\hbar}\int \frac{d\bm{k}}{\left( 2\pi \right)^d} \sum_{a\neq b} - \left[ D_\nu  \xi^\mu\right]_{ab}  \xi^\lambda_{ba} f_{ba} d_{ba}^{\,\Omega} \notag \\
			&~~~ + \left[ \left( \nu, -\Omega \right)\leftrightarrow \left( \lambda, \Omega \right) \right].\label{cond_ie_bare2}
		\end{align}
Similar expression can be found in the term $\sigma_\text{ee;o}$ which is given by
		\begin{align}
 		&\sigma^{\mu;\nu \lambda}_\text{ee;o} \left( \omega; \omega_1,\omega_2 \right)\notag \\
			&= \frac{q^3}{2}\int \frac{d\bm{k}}{\left( 2\pi \right)^d} \sum_{a\neq b \neq c} v^\mu_{ab}  d_{ ba}^{\,\omega}  \left(  d_{ca }^{\,\omega_2}  \xi^\nu_{bc} \xi^\lambda_{ca}  f_{ac} - d_{bc}^{\,\omega_2}   \xi^\nu_{ca} \xi^\lambda_{bc} f_{cb}  \right) \notag \\
			& ~~~+ \left[ \left( \nu, \omega_1 \right) \leftrightarrow \left( \lambda, \omega_2 \right) \right].
		\end{align}
In the condition Eq.~\eqref{photocurrent_frequency_condition}, the formula is recast as
		\begin{align}
 		&\sigma^{\mu;\nu \lambda}_\text{ee;o} \notag \\
			&= \frac{q^3}{2\hbar}\int \frac{d\bm{k}}{\left( 2\pi \right)^d} \sum_{a\neq b \neq c}  i \xi^\mu_{ab}  \left(  d_{ca }^{\,\Omega}  \xi^\nu_{bc} \xi^\lambda_{ca}  f_{ac} - d_{bc}^{\,\Omega}   \xi^\nu_{ca} \xi^\lambda_{bc} f_{cb}  \right) \notag \\
			&~~~+ \left[ \left( \nu, -\Omega \right) \leftrightarrow \left( \lambda, \Omega \right) \right],\\
			&= \frac{q^3}{2\hbar}\int \frac{d\bm{k}}{\left( 2\pi \right)^d} \sum_{a\neq b \neq c}  i \left( \xi^\mu_{ab} \xi^\nu_{bc} - \xi^\nu_{ab} \xi^\mu_{bc} \right) \xi^\lambda_{ca}  f_{ac} d_{ca}^{\,\Omega} \notag \\
			&~~~ + \left[ \left( \nu, -\Omega \right) \leftrightarrow \left( \lambda, \Omega \right) \right],\label{cond_ee_o_bare}
		\end{align}
where we use Eq.~\eqref{velocity_operator} in the first line. As for the summation over the band index $b$, we can use the following formula~\cite{Aversa1995}
		\begin{equation}
		\left[ D_\mu  \xi^\nu \right]_{ac} - \left[ D_\nu  \xi^\mu \right]_{ac}  = \sum_{b \neq a,c} i \left( \xi^\mu_{ab} \xi^\nu_{bc}  - \xi^\nu_{ab} \xi^\mu_{bc} \right).
		\end{equation}
The $\sigma_\text{ee;o}$ term is therefore rewritten by
		\begin{align}
		\sigma^{\mu;\nu \lambda}_\text{ee;o} 
		&= \frac{q^3}{2\hbar}\int \frac{d\bm{k}}{\left( 2\pi \right)^d} \sum_{a\neq c}  \left( \left[ D_\mu  \xi^\nu \right]_{ac} - \left[ D_\nu  \xi^\mu \right]_{ac}  \right) \xi^\lambda_{ca}  f_{ac} d_{ca}^{\,\Omega} \notag \\
		&~~~+ \left[ \left( \nu, -\Omega \right) \leftrightarrow \left( \lambda, \Omega \right) \right].\label{cond_ee_o_bare2}
		\end{align}
Summing up Eqs.~\eqref{cond_ie_bare2} and \eqref{cond_ee_o_bare2}, we obtain a simplified expression as
		\begin{align}
		&\sigma^{\mu;\nu \lambda}_\text{ee+ie} = \sigma^{\mu;\nu \lambda}_\text{ee;o} + \sigma^{\mu;\nu \lambda}_\text{ie} \notag \\
		&= \frac{q^3}{2\hbar}\int \frac{d\bm{k}}{\left( 2\pi \right)^d} \sum_{a\neq b}   \left[ D_\mu  \xi^\nu \right]_{ab} \xi^\lambda_{ba}  f_{ab} d_{ba}^{\,\Omega}   + \left[ \left( \nu, -\Omega \right) \leftrightarrow \left( \lambda, \Omega \right) \right].
		\end{align}
Using Eq.~\eqref{principal_integral}, the formula is decomposed into 
		\begin{align}
		&\sigma^{\mu;\nu \lambda}_\text{ee+ie}
			= \frac{q^3}{2\hbar}\int \frac{d\bm{k}}{\left( 2\pi \right)^d} \sum_{a\neq b} f_{ab} \notag \\
			& ~~~\times \left[  S^{\mu;\nu\lambda}_{ab}   \text{P} \frac{1}{\hbar \Omega - \epsilon_{ba}} - i\pi A^{\mu;\nu\lambda}_{ab} \delta (\hbar \Omega - \epsilon_{ba}) \right].\label{shift_current_bare}
		\end{align}
Here, we introduced 
		\begin{align}
		S^{\mu;\nu\lambda}_{ab}= \left[ D_\mu  \xi^\nu \right]_{ab} \xi^\lambda_{ba}  + \left[ D_\mu  \xi^\lambda \right]_{ba} \xi^\nu_{ab},\\
		A^{\mu;\nu\lambda}_{ab} = \left[ D_\mu  \xi^\nu \right]_{ab} \xi^\lambda_{ba}  - \left[ D_\mu  \xi^\lambda \right]_{ba} \xi^\nu_{ab}.
		\end{align}
Owing to the Hermitian property of the Berry connection, general formulas for the LP and CP-photocurrent coefficients are obtained as
		\begin{align}
		&\eta^{\mu;\nu\lambda}_\text{ee+ie} 
			=\frac{q^3}{2\hbar}\int \frac{d\bm{k}}{\left( 2\pi \right)^d} \sum_{a\neq b} f_{ab} \notag \\
			&\times \left[ \text{Re}\, S^{\mu;\nu\lambda}_{ab}    \text{P} \frac{1}{\hbar \Omega - \epsilon_{ba}} + \pi \text{Im} \,  A^{\mu;\nu\lambda}_{ab}  \delta (\hbar \Omega - \epsilon_{ba})  \right],\label{ie_ee_LP_photocurrent} 
		\end{align}
and
		\begin{align}
		&\kappa^{\mu\tau}_\text{ee+ie} 
			=\epsilon_{\tau\nu\lambda} \frac{q^3}{2\hbar}\int \frac{d\bm{k}}{\left( 2\pi \right)^d} \sum_{a\neq b} f_{ab} \notag \\
			&\times \left[  \text{Im} \, S^{\mu;\nu\lambda}_{ab}   \text{P} \frac{1}{\hbar \Omega - \epsilon_{ba}} - \pi \text{Re}\,  A^{\mu;\nu\lambda}_{ab}   \delta (\hbar \Omega - \epsilon_{ba}) \right] ,\label{ie_ee_CP_photocurrent}
		\end{align}
which do not include any imaginary component. 

Now, we present a symmetry classification of the general expressions, Eqs.~\eqref{ie_ee_LP_photocurrent} and \eqref{ie_ee_CP_photocurrent}, as we did for the injection current. The \T{}-symmetry leads to the relation~\footnote{
        $\left[ D_\mu (\bm{k}) \xi^\nu (\bm{k}) \right]_{ab}$ is explicitly written as $\left[ D_\mu (\bm{k}) \xi^\nu (\bm{k}) \right]_{ab} = \partial\xi^\nu_{ab} (\bm{k})/\partial k^\mu  - i \left[ \xi^\mu_{aa} (\bm{k}) - \xi^\mu_{bb} (\bm{k}) \right] \xi^\nu_{ab} (\bm{k})$. 
		}
		\begin{equation}
		\left[ D_\mu (\bm{k}) \xi^\nu (\bm{k}) \right]_{ab} \xi^\lambda_{ba} (\bm{k}) = -\left[ D_\mu (-\bm{k}) \xi^\nu (-\bm{k}) \right]_{ba} \xi^\lambda_{ab} (-\bm{k}).\label{BCderiv_BC_prod_T_sym}
		\end{equation}
Combining this with the relation $\epsilon_{\bm{k}a}  = \epsilon_{-\bm{k}a} $ ensured by the \T{}-symmetry, Eq.~\eqref{shift_current_bare} is transformed as
		\begin{align}
		&\sigma^{\mu;\nu \lambda}_\text{ee+ie} (\mathcal{T})= \frac{q^3}{2\hbar}\int \frac{d\bm{k}}{\left( 2\pi \right)^d} \sum_{a\neq b} f_{ab}\notag \\
				&\times\Bigl[ i \text{Im}\, S^{\mu;\nu\lambda}_{ab} \, \text{P} \frac{1}{\hbar \Omega - \epsilon_{ba}} + \pi \text{Im}\, A^{\mu;\nu\lambda}_{ab}   \delta (\hbar \Omega - \epsilon_{ba}) \Bigr].\label{electric_shift_current_and_NL_permittivity}
		\end{align}
This is the photoconductivity formula in the \T{}-symmetric systems. The integrand including the principal value and that with delta function are anti-symmetric and symmetric under the permutation $\nu\leftrightarrow \lambda$, respectively. Thus, the former corresponds to the CP-photocurrent given by
		\begin{align}
			&\kappa_\text{intII}^{\mu\tau} = i\epsilon_{\nu\lambda\tau} \sigma^{\mu;\nu \lambda}_\text{ee+ie} (\mathcal{T}), \\
			&=\frac{-q^3}{\hbar}\int \frac{d\bm{k}}{\left( 2\pi \right)^d} \sum_{a\neq b} \epsilon_{\nu\lambda\tau} \mathrm{Im}\,\left( \left[ D_\mu  \xi^\nu \right]_{ab} \xi^\lambda_{ba} \right) f_{ab} \text{P} \frac{1}{\hbar \Omega - \epsilon_{ba}}. \label{intrinsic_effect_ee;o}
		\end{align}
By using the band-resolved Berry curvature, the formula is rewritten as 
		\begin{equation}
			\kappa_\text{intII}^{\mu\nu}= \frac{q^3}{4\hbar}\int \frac{d\bm{k}}{\left( 2\pi \right)^d} \sum_{a\neq b} \epsilon_{\nu\lambda\tau}\partial_\mu \Omega^{\lambda\tau}_{ab} f_{ab} \text{P} \frac{1}{\hbar \Omega - \epsilon_{ba}}.\label{intrinsic_effect_ee;o_z}
		\end{equation}
On the other hand, the latter is the LP-photocurrent called shift current~\cite{Kraut1981Photovoltaiceffect,Sipe2000secondorder},
		\begin{align}
		&\eta^{\mu;\nu \lambda}_\text{shift} =\frac{\pi q^3}{2\hbar}\int \frac{d\bm{k}}{\left( 2\pi \right)^d} \sum_{a\neq b} f_{ab} \delta (\hbar \Omega - \epsilon_{ba}) \notag \\
			&~~~~~\times \text{Im}\,\left(  \left[ D_\mu  \xi^\nu \right]_{ab} \xi^\lambda_{ba}  + \left[ D_\mu  \xi^\lambda \right]_{ab} \xi^\nu_{ba}  \right).
			\label{shift_current}
		\end{align}
Taking both compoments into account, we denote the total photoconductivity as follows
		\begin{equation}
		\sigma^{\mu;\nu \lambda}_\text{ee+ie} (\mathcal{T}) = \eta^{\mu;\nu \lambda}_\text{shift} -\frac{i}{2} \epsilon_{\nu\lambda\tau} \kappa_\text{intII}^{\mu\tau}.\label{electric_ie_ee_decomposition}
		\end{equation}
The CP-photocurrent $\kappa_\text{intII}^{\mu\tau}$ is simplified by combining it with Eq.~\eqref{intrinsic_term_ee;d_current_T_symmetric}. The expression is obtained as
		\begin{align}
		&\kappa_\text{IFS}^{\mu\nu}= i \epsilon_{\nu\lambda\tau} \sigma_\text{intI}^{\mu;\lambda\tau} \left( \mathcal{T} \right) +\kappa_\text{intII}^{\mu\nu},\\
		&=-\frac{q^3}{4\hbar}\int \frac{d\bm{k}}{\left( 2\pi \right)^d} \sum_{a\neq b} \epsilon_{\nu\lambda\tau} \Omega^{\lambda\tau}_{ab}  \text{P} \frac{1}{\hbar \Omega - \epsilon_{ba}} \partial_\mu f_{ab},\\
		&=-\frac{q^3}{2\hbar}\int \frac{d\bm{k}}{\left( 2\pi \right)^d} \sum_{a\neq b} \epsilon_{\nu\lambda\tau} \Omega^{\lambda\tau}_{ab} \frac{\hbar \Omega}{\hbar^2 \Omega^2 - \epsilon_{ab}^2} \partial_\mu f \left( \epsilon_{\bk a} \right),\label{intrinsic_fermi_surface_effect_T_symmetric}
		\end{align}
which we denote intrinsic Fermi surface effect in Table~\ref{Table_photocurrent_classification}. The formula represents a Fermi surface effect while it is not sensitive to the relaxation time in contrast to usual Fermi surface effects such as the Drude conductivity. The resulting formula is consistent with Ref.~\cite{Fernando2020Difference} where the nearly-static photocurrent in the \T{}-symmetric systems has been elucidated.

Here, we discuss the shift current term in details. Following the prescription presented in Ref.~\cite{Sipe2000secondorder}, we decompose the Berry connection into the magnitude and phase
		\begin{equation}
		\xi^\nu_{ab} = |\xi^\nu_{ab}| \exp{(-i \phi_{ab}^\nu)}.
		\end{equation}
$|\xi^\nu_{ab}| = |\xi^\nu_{ba}|$ and $\phi_{ab}^\nu = -\phi_{ba}^\nu $ are satisfied by the Hermitian property. The shift current formula Eq.~\eqref{shift_current} is recast as
		\begin{widetext}
		\begin{align}
        \eta^{\mu;\nu \lambda}_\text{shift}
				&= - \frac{\pi q^3}{2\hbar} \int \frac{d\bm{k}}{\left( 2\pi \right)^d} \sum_{a\neq b} \left( R^{\mu}_{ab;\nu} + R^{\mu}_{ab;\lambda} \right)g^{\nu\lambda}_{ab}   f_{ab} \delta (\hbar \Omega - \epsilon_{ba}) \notag \\
				&~~-\frac{\pi q^3}{2\hbar} \int \frac{d\bm{k}}{\left( 2\pi \right)^d} \sum_{a\neq b} \left[   \left( \partial_\mu |\xi^\nu_{ab}| \right) |\xi^\lambda_{ba}| - |\xi^\nu_{ab}| \left(  \partial_\mu |\xi^\lambda_{ba}| \right) \right] \sin{(\phi_{ab}^\nu +\phi_{ba}^\lambda)} f_{ab} \delta (\hbar \Omega - \epsilon_{ba}), \label{electric_shift_current_bare}
		\end{align}
		\end{widetext}
where we introduced so-called shift vector defined by
		\begin{equation}
		R^{\mu}_{ab;\nu} =  \partial_\mu \phi_{ab}^\nu +\xi^\mu_{aa} - \xi^\mu_{bb}.
		\end{equation}
This vector implies the wave-packet shift of the excited electron along the $\mu$-direction through the interband transition $a \leftrightarrow b$~\cite{Morimoto2016topological,FregosoMorimotoMoore2017Quantitative}. We can take coordinate axes so that the polarization of the linearly-polarized light is parallel to one of the axes. Thus, taking $\nu = \lambda$ without loss of generality, we obtain the well-known formula for the shift current~\cite{Kraut1981Photovoltaiceffect,Sipe2000secondorder}
		\begin{equation}
		\eta^{\mu;\nu \nu}_\text{shift} = - \frac{\pi q^3}{\hbar} \int \frac{d\bm{k}}{\left( 2\pi \right)^d} \sum_{a\neq b} R^{\mu}_{ab;\nu} g^{\nu\nu}_{ab} f_{ab} \delta (\hbar \Omega - \epsilon_{ba}).\label{electric_shift_current}
		\end{equation}
Note that the shift vector and band-resolved quantum metric are individually invariant under the U(1)-gauge transformation. The shift current [Eq.~\eqref{electric_shift_current}] is in sharp contrast to the magnetic injection current [Eq.~\eqref{magnetic_injection_current_bare}], another LP-photocurrent allowed in insulators. The shift current is described by the shift vector in the real-space picture,  whereas the magnetic injection current arises from the group-velocity difference $\Delta^\mu_{ab}$ which is a characteristic property in the momentum-space (See also Table~\ref{Table_photocurrent_ingredients}). The joint density of states and band-resolved quantum metric play important roles in both LP-photocurrents.

Now we move on to the photocurrent in the \PT{}-symmetric systems, a main topic of this paper. We can simplify Eq.~\eqref{shift_current_bare} by making use of the \PT{}-symmetry. After the parallel discussion, we obtain
		\begin{align}
		&\sigma^{\mu;\nu \lambda}_\text{ee+ie} (\mathcal{PT})= \frac{q^3}{2\hbar}\int \frac{d\bm{k}}{\left( 2\pi \right)^d} \sum_{a\neq b} f_{ab}\notag \\
				&\times\Bigl[ \text{Re}\, S^{\mu;\nu\lambda}_{ab} \, \text{P} \frac{1}{\hbar \Omega - \epsilon_{ba}} - i\pi \text{Re}\, A^{\mu;\nu\lambda}_{ab}   \delta (\hbar \Omega - \epsilon_{ba}) \Bigr],\label{magnetic_shift_current_andddd_NL_permittivity}
		\end{align}
for the photoconductivity $\sigma_\text{ie+ee}$ in the \PT{}-symmetric systems. We notice the \T{}-/\PT{}-correspondence of the $\sigma^{\mu;\nu \lambda}_\text{ee+ie}$ term. In the \PT{}-symmetric system, the reactive term including the principal integrand represents the response to the linearly-polarized light, while the absorptive term containing the delta function represents the circularly-polarized light-induced photocurrent.

The formula for the LP-photocurrent is obtained as
		\begin{align}
		&\eta^{\mu;\nu \lambda}_\text{intII}\notag \\ 
			&= \frac{q^3}{2\hbar}\int \frac{d\bm{k}}{\left( 2\pi \right)^d} \sum_{a\neq b} f_{ab}\text{P} \frac{1}{\hbar \Omega - \epsilon_{ba}} \notag \\
			&~~~~~\times \text{Re}\,\left(  \left[ D_\mu  \xi^\nu \right]_{ab} \xi^\lambda_{ba}  + \left[ D_\mu  \xi^\lambda \right]_{ab} \xi^\nu_{ba}  \right),\\
			&= \frac{q^3}{2\hbar}\int \frac{d\bm{k}}{\left( 2\pi \right)^d} \sum_{a\neq b} \partial_\mu g^{\nu\lambda}_{ab} f_{ab}\text{P} \frac{1}{\hbar \Omega - \epsilon_{ba}}.
		\end{align}
Combining this equation with Eq.~\eqref{intrinsic_term_ee;d_current_PT_symmetric}, we finally obtain the formula for an intrinsic Fermi surface effect
	\begin{align}
	&\eta_\text{IFS}^{\mu;\nu\lambda} = \sigma_\text{intI}^{\mu;\nu\lambda} \left( \mathcal{PT} \right) +\eta_\text{intII}^{\mu;\nu\lambda},\\
	&=\frac{q^3}{\hbar}\int \frac{d\bm{k}}{\left( 2\pi \right)^d} \sum_{a\neq b} g^{\nu\lambda}_{ab} \frac{\epsilon_{ab}}{\hbar^2 \Omega^2 - \epsilon_{ab}^2} \partial_\mu f \left( \epsilon_{\bk a} \right).\label{intrinsic_fermi_surface_effect_PT_symmetric}
	\end{align}
This term comprises the Fermi surface term and quantum metric, and it is therefore the counterpart of Eq.~\eqref{intrinsic_fermi_surface_effect_T_symmetric} which is characterized by the Berry curvature instead of the quantum metric. In the static limit ($\Omega \rightarrow 0$), the formula for the LP-photocurrent is recast as
	\begin{align}
	\eta_\text{IFS}^{\mu;\nu\lambda} \rightarrow - \frac{q^3}{\hbar}\int \frac{d\bm{k}}{\left( 2\pi \right)^d} \sum_{a\neq b}  \frac{g^{\nu\lambda}_{ab}}{\epsilon_{ab}} \partial_\mu f \left( \epsilon_{\bk a} \right).
	\end{align}
The expression is similar to the semiclassically-derived (static) nonlinear conductivity~\cite{Gao2014}, which is interpreted as a correction to the quantum geometry by the electric field. However, we note that the nonlinear conductivity in Ref.~\cite{Gao2014} shows only the Hall response. Contrary to that, Eq.~\eqref{intrinsic_fermi_surface_effect_PT_symmetric} indicates that the induced photocurrent can be parallel as well as perpendicular to the incident direction of lights.

Here, we show the CP-photocurrent, which is the counterpart of the shift current. This photocurrent has properties distinguished from the shift current: it is induced by the circularly-polarized photon instead of the linearly-polarized photon, and unique to the magnetically-parity-violating system. We therefore call the response \textit{gyration current}. The gyration current formula is given by
		\begin{align}
		&\kappa^{\mu\nu}_\text{gyro} = i \epsilon_{\nu\lambda\tau} \sigma^{\mu;\lambda\tau}_\text{ee+ie} (\mathcal{PT}) \\
			 &= \frac{\pi q^3}{\hbar} \int \frac{d\bm{k}}{\left( 2\pi \right)^d} \sum_{a\neq b} \epsilon_{\nu\lambda\tau } \text{Re}\, \left( \left[ D_\mu  \xi^\lambda \right]_{ab} \xi^\tau_{ba} \right) f_{ab} \delta (\hbar \Omega - \epsilon_{ba}).\label{gyration_current_second_tensor_form}
		\end{align}
We will discuss the gyration current in Sec.~\ref{Sec_gyration_current} in details.
It is noteworthy that only the gyration current is induced by the circularly-polarized light in the \PT{}-symmetric systems. Therefore, we can unambiguously detect the gyration current by measuring the CP-photocurrent. This is not the case of the CP-photocurrent of \T{}-symmetric systems because of the admixture of various CP-photocurrents such as the Berry curvature dipole term and electric injection current~\cite{DeJuan2016} [see Table~\ref{Table_photocurrent_circular_linear_classification}]. Furthermore, the photocurrent measurements may be useful to identify the symmetry of a parity-violating order parameter in magnetic materials because the response tensor is sensitive to the symmetry.

Combining the gyration current with a part of the intrinsic Fermi surface term, we obtain the photoconductivity in the \PT{}-symmetric systems,
		\begin{equation}
		\sigma^{\mu;\nu \lambda}_\text{ee+ie} (\mathcal{PT}) = \eta^{\mu;\nu \lambda}_\text{intII} -\frac{i}{2} \epsilon_{\nu\lambda\tau} \kappa_\text{gyro}^{\mu\tau}.\label{magnetic_ie_ee_decomposition}
		\end{equation}
Collecting Eqs.~\eqref{electric_ie_ee_decomposition} and \eqref{magnetic_ie_ee_decomposition}, we rewrite the general formula for the $\sigma_\text{ie+ee}$ term and decompose it into the LP-photocurrent and CP-photocurrent,
		\begin{align}
		&\eta^{\mu;\nu \lambda}_\text{ie+ee} = \eta^{\mu;\nu \lambda}_\text{shift} + \eta^{\mu;\nu \lambda}_\text{intII}, \label{ee_ie_diagonal_general_LP_decomposition}\\
		&\kappa^{\mu\nu}_\text{ie+ee} = \kappa^{\mu\nu}_\text{intII} + \kappa^{\mu\nu}_\text{gyro}. \label{ee_ie_diagonal_general_CP_decomposition}
		\end{align}
Thus, the \T{}/\PT{} correspondence also holds in the case of the intrinsic Fermi surface effect and the shift current mechanism.
When both \T{}- and \PT{}-symmetries are broken, the photocurrent allowed by each symmetry is admixed with each other. Similar discussion can be found in the second-order nonlinear conductivity~\cite{Watanabe2020NLC}. 

Summarizing this section, we reproduced the formulas for several known photocurrent responses, and uncovered new photocurrents, the intrinsic Fermi surface effect and the gyration current. Although the contrasting role of the \T{}- and \PT{}-symmetries has been implied for several photocurrent responses studied very recently~\cite{Zhang2019switchable,Holder2020consequences}, it remained unclear whether the \T{}-/\PT{}-correspondence is generally applicable to the photocurrent classification. Our classification, however, systematically classifies the photocurrent responses and verifies the \T{}/\PT{} correspondence in a rigorous way. The obtained classification completes all the photocurrent responses within the independent-particle approximation and provides clear decomposition of the general photoconductivity coefficients [See Eqs.~\eqref{ee_diagonal_general_decomposition},~\eqref{ee_ie_diagonal_general_LP_decomposition}, and \eqref{ee_ie_diagonal_general_CP_decomposition}]. The decomposition has naturally led to the finding of the intrinsic Fermi surface effect and the gyration current.

Furthermore, it is also shown in Ref.~\cite{Watanabe2020NLC} that the \T{}- and \PT{}-symmetries play important roles in classifying the extrinsic contributions~\cite{Du2019Disorder,Isobe2020Rectification} to the photocurrent response. Interestingly, the extrinsic contributions arising from the impurity scattering are strongly suppressed by the \PT{}-symmetry~\cite{Watanabe2020NLC}, while they can be main terms in the \T-symmetric systems. Therefore, the \PT-symmetric systems focused on in this paper are more favorable to investigate the intrinsic photocurrent.

\section{Generalization to spinful systems}\label{Sec_classification_spinful}

The formulation is straightforwardly generalized to the spinful system. Classification of the LP/CP-photocurrent in Table~\ref{Table_photocurrent_classification} does not depend on whether the system is spinless or spinful. The photoconductivity formula in the \PT{}-symmetric systems, however, is slightly modified due to the Kramers degeneracy appearing at each $\bk$.

Owing to the double degeneracy ensured by the \PT{}-symmetry, the Bloch states have U(2)-gauge degree of freedom at least. Note that the gyration current formula in spinless systems [Eq.~\eqref{gyration_current_second_tensor_form}] is not invariant under the U(2)-gauge transformation. Thus, we modify the decomposition of the nonlinear conductivity tensor in Eq.~\eqref{second_order_nonlinear_conductivity} to be U(2)-gauge invariant. Firstly, the Berry connection is divided as
		\begin{equation}
		\xi^\mu_{ab} = \alpha_{ab}^\mu + \bce^\mu_{ab},\label{U2_BC_decomposition}
		\end{equation}
where the intraband Berry connection $\alpha_{ab}^\mu$ is introduced for the degenerate bands satisfying $\epsilon_{\bk a} = \epsilon_{\bk b}$. With the decomposition of the Berry connection, the intraband position operator $r^\mu_i$ is modified as
		\begin{equation}
		\left( r_i^\mu \right)_{ab} = i\partial_\mu \delta_{ab} + \alpha^\mu_{ab},\label{U2_intra_position_operator}
		\end{equation}
and the interband position operator is given by $\left( r_e^\mu \right)_{ab} = \bce_{ab}^\mu $. Accordingly, we define the band-resolved quantum metric and Berry curvature by 
		\begin{align}
		&g^{\mu\nu}_{ab} = \frac{1}{2} \left(  \mathcal{A}^\mu_{ab} \mathcal{A}^\nu_{ba} + \mathcal{A}^\nu_{ab} \mathcal{A}^\mu_{ba}  \right),\label{u1_quantum_metric_u2ver}\\
		&\Omega^{\mu\nu}_{ab} = i \left(  \mathcal{A}^\mu_{ab} \mathcal{A}^\nu_{ba} - \mathcal{A}^\nu_{ab} \mathcal{A}^\mu_{ba}  \right).
		\end{align}
Based on the U(2)-type position operators, we divide the nonlinear optical conductivity into four terms. The calculation can be done as in the spinless systems, and hence we give the derivation in Appendix~\ref{Sec_app_U2gauge}. 

In the following, we consider formulas for the photocurrent in the \PT{}-symmetric and spinful systems, that is, the Drude term, magnetic injection current, intrinsic Fermi surface effect, and gyration current.
The Drude term is the same as Eq.~\eqref{photocurrent_Drude} except for the Kramers degree of freedom included in the summation over the band indices. In the case of spinful systems, the anti-symmetrically distorted band structure causing a finite Drude term is realized by the coupling between the parity-violating magnetic order and the sublattice-dependent spin-orbit coupling~\cite{Yanase2014zigzag,Zelezny2014NeelorbitTorque}. This will be exemplified by the model study in Secs.~\ref{Sec_Rashba_Dressel_model} and~\ref{Sec_gyration_in_topological_materials}.

Similarly, the photoconductivity formulas for the magnetic injection current and intrinsic Fermi surface effect are respectively obtained by replacing the band-resolved quantum metric in Eqs.~\eqref{magnetic_injection_current_bare} and \eqref{intrinsic_fermi_surface_effect_PT_symmetric} with Eq.~\eqref{u1_quantum_metric_u2ver}, whereas the formula for the gyration current is obtained as
		\begin{align}
		&\kappa^{\mu\nu}_\text{gyro} = \frac{\pi q^3}{\hbar} \int \frac{d\bm{k}}{\left( 2\pi \right)^d} \sum_{a\neq b} f_{ab} \delta (\hbar \Omega - \epsilon_{ba}) \notag \\
		&~~~~~\times \epsilon_{\nu\lambda\tau } \text{Re} \left( \left[ \ud_\mu  \mathcal{A}^\lambda \right]_{ab} \mathcal{A}^\tau_{ba} \right),\label{gyration_current_second_tensor_form_U2}
		\end{align}
where $\ud_\mu$ is the U(2)-covariant derivative. We can straightforwardly show that all the obtained expressions are U(2)-gauge invariant.

In conclusion, although the photocurrent formulas for the spinful system are mostly the same as those for the spinless system, the gyration current is modified due to the different gauge symmetry. Note that the formulation can be easily generalized to the system having $n$-fold degenerate bands. In particular, in a high-symmetric subspace of the Brillouin zone manifold, a high degeneracy with $n = 4,6$ may exist in a symmetry-enforced way. Hence, our formulation gives insights into the photocurrent responses arising from such multi-fold degenerate fermions~\cite{Flicker2018Multifoldfermions,YangNagaosa2014,Cano2019}. 

We comment that the U(2)-gauge invariant formulation becomes unnecessary when the \PT{}-symmetry is absent and the Kramers degeneracy is lifted. Then, the U(2)-covariant derivative is replaced by that for the U(1)-gauge [Eq.~\eqref{covariant_derivative}]. In particular, calculations of \T{}-symmetric spinful systems can be conducted as in the spinless case. Thus, the formulas for the photocurrent are the same as those for spinless systems.

\section{Analysis of Gyration current}\label{Sec_gyration_current}
In this section, we investigate the gyration current response [Eq.~\eqref{gyration_current_second_tensor_form} for spinless systems and Eq.~\eqref{gyration_current_second_tensor_form_U2} for spinful systems] in details. After revealing basic properties in Sec.~\ref{Sec_basic_of_gyration_current}, we present a microscopic study based on a spinful model in Sec.~\ref{Sec_Rashba_Dressel_model}. Furthermore, we show a giant gyration current arising from divergent geometric quantities in a topological antiferromagnet  [Sec.~\ref{Sec_gyration_in_topological_materials}].

\subsection{Basic property}\label{Sec_basic_of_gyration_current}
Firstly, we consider the spinless system for simplicity. Since the gyration current is induced by the circularly-polarized light, it is convenient to adopt the circular representation as in the electric injection current. With the circularly-polarized light along the $z$-direction, the response formula is rewritten as
		\begin{align}
		&\kappa^{\mu z}_\text{gyro}=  \frac{\pi q^3}{\hbar} \int \frac{d\bm{k}}{\left( 2\pi \right)^d} \sum_{a\neq b} f_{ab} \delta (\hbar \Omega - \epsilon_{ba}) \notag \\
		&~~~\times  \text{Re}\, \left( i \left[ D_\mu  \xi^+ \right]_{ab} \xi^-_{ba}  - i \left[ D_\mu  \xi^- \right]_{ab} \xi^+_{ba}  \right).
		\end{align}
Note that this formula can be applied to the system without \PT{}-symmetry. We write the left/right-handed Berry connections $\xi^\pm$ by
		\begin{equation}
		\xi^\pm_{ab} =  |\xi^\pm_{ab}| \exp{(-i\phi^\pm_{ab})},
		\end{equation}
which satisfy the relation $\phi^+_{ab} =-\phi^-_{ba} $ due to the definition of $\xi^\mu_{ab}$ ($\mu=x,y$ and $a\neq b$). Then, the gyration current formula is recast as
		\begin{align}
		\kappa^{\mu z}_\text{gyro}
		&=  \frac{\pi q^3}{\hbar} \int \frac{d\bm{k}}{\left( 2\pi \right)^d} \sum_{a\neq b} f_{ab} \delta (\hbar \Omega - \epsilon_{ba})\notag \\
		&~~~ \times \left(  R_{ab;+}^\mu | \xi^+_{ab}|^2 - R_{ab;-}^\mu | \xi^-_{ab}|^2  \right).\label{gyration_current_CP_rep}
		\end{align}
Here, we introduced \textit{chiral shift vector} given by
		\begin{equation}
		R_{ab;\pm}^\mu = \partial_\mu \phi^\pm_{ab} + \xi^\mu_{aa} - \xi^\mu_{bb},
		\end{equation}
which is invariant under the U(1)-gauge transformation.

The meaning of Eq.~\eqref{gyration_current_CP_rep} is clear. Corresponding to the handedness of the dipole-transition amplitude denoted by $| \xi^\pm_{ab}|^2$, the circularly-polarized light excites the electrons. Through the interband transition $a \leftrightarrow b$, the excited electron makes positional shift determined by the chiral shift vector. The resulting electrons' flow gives rise to the gyration current. Interestingly, a similar expression has been obtained in a recent study of a circular-photo-induced nonlinear polarization in a layered system~\cite{Gao2020TunablePGE}. 

The transition amplitudes, $| \xi^\pm_{ab}|^2$, are further decomposed into 
		\begin{align}
  		| \xi^\pm_{ab}|^2 
  			&= (g^{xx}_{ab} + g^{yy}_{ab}) \mp \Omega^{xy}_{ab},\label{handed_BC_decomposition}
  		\end{align}  
which consist of the band-resolved quantum metric and Berry curvature. Although other photocurrents allowed in insulators are related to either of the band-resolved quantum metric or Berry curvature, the gyration current is derived from both geometric quantities. Using the decomposition in Eq.~\eqref{handed_BC_decomposition}, Eq.~\eqref{gyration_current_CP_rep} is transformed as
		\begin{align}
		&\kappa^{\mu z}_\text{gyro} 
			=  \frac{\pi q^3}{\hbar} \int \frac{d\bm{k}}{\left( 2\pi \right)^d} \sum_{a\neq b} f_{ab} \delta (\hbar \Omega - \epsilon_{ba}) \notag \\
			&\times \left[ \left( R_{ab;+}^\mu -R_{ab;-}^\mu  \right) \left(  g^{xx}_{ab} + g^{yy}_{ab}\right) -  \left( R_{ab;+}^\mu + R_{ab;-}^\mu  \right) \Omega^{xy}_{ab}  \right]. \label{gyration_current_decomposition}
		\end{align}
When we impose the \PT{}-symmetry, we have
		\begin{align}
		&\kappa^{\mu z}_\text{gyro} \notag \\
			&=\frac{2 \pi q^3}{\hbar} \int \frac{d\bm{k}}{\left( 2\pi \right)^d} \sum_{a\neq b} f_{ab} \delta (\hbar \Omega - \epsilon_{ba})  R_{ab;+}^\mu \left(  g^{xx}_{ab} + g^{yy}_{ab}\right) .\label{gyration_current_decomposition_no_BCV}
		\end{align}
where we used the relations, $\Omega^{\mu\nu}_{ab} = 0$ and $\phi_{ab}^- = -\phi^+_{ab} + \pi$. 

Next, we consider the gyration current in the spinful system [Eq.~\eqref{gyration_current_second_tensor_form_U2}]. For the U(2)-gauge description, we assume the \PT{}-symmetric system below. The formula is recast as
		\begin{align}
		&\kappa^{\mu z}_\text{gyro}=  \frac{\pi q^3}{\hbar} \int \frac{d\bm{k}}{\left( 2\pi \right)^d} \sum_{a\neq b} f_{ab} \delta (\hbar \Omega - \epsilon_{ba})  \notag \\
			&~~~\times \text{Re}\, \left( i  \left[ \ud_\mu  \mathcal{A}^+ \right]_{ab} \mathcal{A}^-_{ba}  - i\left[ \ud_\mu  \mathcal{A}^- \right]_{ab} \mathcal{A}^+_{ba}  \right),\label{gyration_current_decomposition_U2_z_CP}\\
			&=  \frac{\pi q^3}{\hbar} \int \frac{d\bm{k}}{\left( 2\pi \right)^d} \sum_{a\neq b} f_{ab} \delta (\hbar \Omega - \epsilon_{ba}) \notag \\
			&~~~\times \text{Re}\, \Bigl[ \left(  R_{ab;+}^\mu | \mathcal{A}^+_{ab}|^2 - R_{ab;-}^\mu | \mathcal{A}^-_{ab}|^2  \right) \notag \\
			&~~~+ \alpha^\mu_{a\bar{a}} \left( \mathcal{A}^+_{\bar{a}b} \mathcal{A}^-_{ba} - \mathcal{A}^-_{\bar{a}b} \mathcal{A}^+_{ba} \right) - \alpha^\mu_{\bar{b}b} \left( \mathcal{A}^+_{a\bar{b}} \mathcal{A}^-_{ba} - \mathcal{A}^-_{a\bar{b}} \mathcal{A}^+_{ba} \right) \Bigr],
		\end{align}
where $(a, \bar{a})$ denotes the Kramers pair ensured by the \PT{}-symmetry and we introduced the circular representation of the Berry connection $\mathcal{A}^\pm_{ab}$ as in Eq.~\eqref{BC_circular_representation}. Taking the gauge where $\alpha^\mu_{a\bar{a}}=0$ is satisfied, the formula is recast as
		\begin{align}
		&\kappa^{\mu z}_\text{gyro}\notag \\
		&=  \frac{\pi q^3}{\hbar} \int \frac{d\bm{k}}{\left( 2\pi \right)^d} \sum_{a\neq b} f_{ab} \delta (\hbar \Omega - \epsilon_{ba}) \notag \\
		&\times \left[ \left( R_{ab;+}^\mu -R_{ab;-}^\mu  \right) \left(  g^{xx}_{ab} + g^{yy}_{ab}\right) -  \left( R_{ab;+}^\mu + R_{ab;-}^\mu  \right) \Omega^{xy}_{ab}  \right], 	
	\end{align}
Owing to the spin degree of freedom, the contribution from the band-resolved Berry curvature is not canceled out in contrast to the formula for the spinless fermions [Eq.~\eqref{gyration_current_decomposition_no_BCV}]. 

\begin{figure*}[htbp]
			\centering
			\begin{tabular}{ccc}
			\includegraphics[height=30mm,clip]{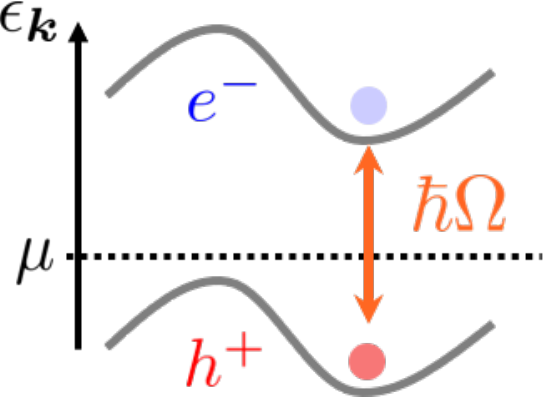} \hspace*{5mm}&
			\includegraphics[height=30mm,clip]{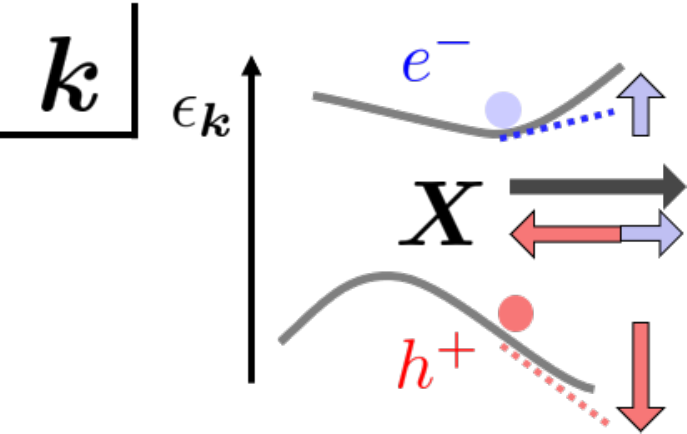} \hspace*{10mm}&
			\includegraphics[height=30mm,clip]{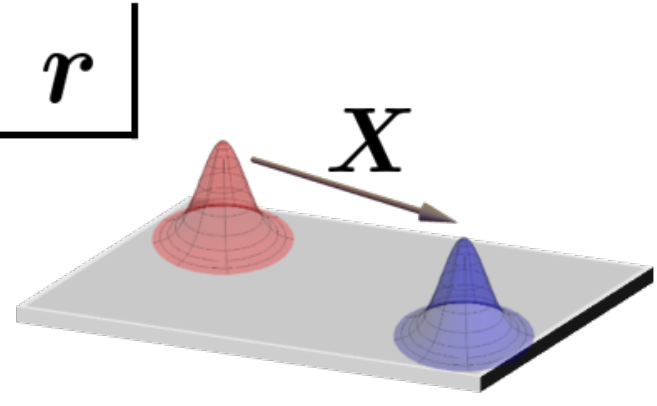} \\
			(a)&
			(b)&
			(c)
			\end{tabular}
		\caption{Schematic picture of the two processes causing the photocurrent response, (a) electron-hole pair creation and (b,c) alternating rectification of paired charges by the director. There are two kinds of the director. (b) The group velocity difference for the injection currents (momentum space picture). (c) The positional shift of wave-packets for the shift and gyration currents (real space picture).}
		\label{Fig_schmatics_photocurrent}
		\end{figure*}

Combining these findings with the known results of photocurrent allowed in insulators, we notice that the photocurrent response arises from the two processes; particle-hole pair creation and ``director" of created charges. The particle-hole creation is determined by the Pauli blockade effect $f_{ab} \delta \left( \hbar \Omega - \epsilon_{ab} \right)$ and the dipole-transition amplitude $T^{\nu\lambda}$ given by the product of the interband Berry connections. The other is the director $X^\mu$ which rectifies the created particles and holes. The overall formula is given by
		\begin{equation}
		\sigma^{\mu;\nu\lambda}\propto \int d \bk X^\mu T^{\nu\lambda}f_{ab} \delta \left( \hbar \Omega - \epsilon_{ab} \right),
		\end{equation}
where $X^\mu$ and $T^{\nu\lambda}$ are different between each photocurrent response.
It is known that the particle-hole excitation determines the linear optical (absorptive) response~\cite{Grosso2013Book}. Thus, the photocurrent response can be intuitively understood as follows; electron-hole pairs are created under irradiating lights as in the linear optical response, and then the director rectifies created pairs to produce an electric current (Fig.~\ref{Fig_schmatics_photocurrent}). Note that the director arises from the geometric property of electrons while it is the internal electric field in the case of the prototypical photocurrent response in the ferroelectric materials and p-n junction. In the case of the electric injection current, for instance, the transition amplitude and director are Berry curvature $T^{\nu\lambda} = \Omega^{\nu\lambda}$ and group velocity difference $X^\mu = \Delta^\mu$, respectively. The set $\left( X^\mu, T^{\nu\lambda} \right)$ for each photocurrent is summarized in Table~\ref{Table_photocurrent_ingredients}.

		\begin{table}[htbp]
		\caption{Director $X^\mu$ and dipole-transition amplitude $T^{\nu\lambda}$ for the photocurrent responses allowed in insulators (see also Table~\ref{Table_photocurrent_circular_linear_classification}). The directors $\Delta^\mu$, $R^\mu$, and $R^\mu_\pm$ are group velocity difference, shift vector, and chiral shift vector, respectively. The transition amplitude is characterized by the quantum metric $g^{\nu\lambda}$ and Berry curvature $\Omega^{\nu\lambda}$.}
		\label{Table_photocurrent_ingredients}
		\centering
		\begin{tabular}{ccc}\hline
			&$X^\mu$&$T^{\nu\lambda}$\\ \hline
			electric injection current&$\Delta^\mu$&$\Omega^{\nu\lambda}$\\
			shift current&$R^\mu$&$g^{\nu\lambda}$\\
			magnetic injection current&$\Delta^\mu$&$g^{\nu\lambda}$\\
			gyration current&$R^\mu_\pm$&$g^{\nu\lambda},\Omega^{\nu\lambda}$\\\hline
		\end{tabular}
		\end{table}

\subsection{Model study of gyration current}\label{Sec_Rashba_Dressel_model}
In this section, we present a microscopic calculation of the gyration current in a spinful model. The \PT{}-preserved but \Pa{}-broken system is realized by the antiferromagnetic order in locally-noncentrosymmetric systems.

The locally-noncentrosymmetric system hosts crystalline sublattices whose site-symmetry lacks the \Pa{}-symmetry while the global \Pa{}-symmetry is preserved by interchanging the sublattice. The prototypical examples are the honeycomb lattice and bilayer system. Such peculiar crystal symmetry gives rise to the sublattice-dependent anti-symmetric spin-orbit coupling (sASOC)~\cite{Kane2005,Yanase2014zigzag,Zelezny2014NeelorbitTorque,Zhang2014HiddenSpin}. In many cases, effects of the sASOC do not appear in macroscopic phenomena while the spin- and momentum-resolved spectroscopy can capture a fingerprint of the sASOC~\cite{Gotlieb2018HiddenSpinInCuprate}. On the other hand, a sublattice-dependent order unveils the sASOC in the way that a coupling between the sASOC and order parameter gives rise to nontrivial electronic structures and cross-correlated responses~\footnote{Refs.~\cite{Watanabe2018grouptheoretical,Hayami2018Classification}, and references therein.}. For instance, the combination of the sASOC with an antiferromagnetic order leads to an asymmetric band dispersion, which is an essential ingredient in the Drude term. Note that such parity-breaking magnetic systems exist in a broad range of compounds~\cite{Gallego2016,Watanabe2018grouptheoretical,Watanabe2018}.

The adopted Hamiltonian is modeled after such parity-violating magnets. A two-dimensional rectangular lattice system consists of two sublattices labeled as A and B. Owing to the locally-noncentrosymmetric property, the site-symmetry is denoted by the noncentrosymmetric point group $C_{2v}$ ($mm2$), while the global symmetry is centrosymmetric labeled by $D_{2h}$ ($mmm$). In the point group $C_{2v}$, the Rashba-type ASOC and (anisotropic) Dresselhaus-type ASOC are allowed~\cite{Manchon2019spin-orbit-torque_review}. Thus, the system hosts these types of ASOC in the sublattice-dependent way as the sASOC. Using the tight-binding approximation, the Bloch Hamiltonian is given by
		\begin{equation}
 		H(\bm{k}) =\begin{pmatrix}
					\epsilon_0(\bm{k}) + \bm{g}_{\rm A}(\bm{k})  \cdot \bm{\sigma}& V_{\rm AB} (\bm{k})\\
					 V_{\rm AB} (\bm{k})&\epsilon_0(\bm{k})   +\bm{g}_{\rm B}(\bm{k})  \cdot \bm{\sigma}
					 \end{pmatrix}, \label{Rashba_Dresselhaus_hamiltonian}
 		\end{equation} 
where $\bm{\sigma}$ and $\bm{\tau}$ are Pauli matrices representing the spin and sublattice degrees of freedom, respectively. The components are defined as
		\begin{align}
		&\epsilon_0 (\bm{k})= -t \left(	\cos{k_x} +\cos{ k_y }	\right),\label{intrasub_kinetic}\\
		&V_\text{AB}(\bm{k})= -2\tilde{t}  	\cos{ \frac{k_x}{2}  } \cos{ \frac{k_y}{2} },\\
		&\bm{g}_\text{A}(\bm{k}) = \bm{g}_0 (\bm{k}) + \bm{h}_\text{AF} =
			\begin{pmatrix}
			h^x_\text{AF} - \alpha_\text{R}  \sin{k_y} + \alpha_\text{D} \sin{k_y} \\
			h^y_\text{AF} + \alpha_\text{R}  \sin{k_x} + \alpha_\text{D} \sin{k_x}\\
			h^z_\text{AF}
			\end{pmatrix},
		\end{align}
and $\bm{g}_\text{B}(\bm{k}) = - \bm{g}_\text{A}(\bm{k})$. The parameters $t=1.0$ and $\tilde{t} = 0.5$ are intra-sublattice and inter-sublattice hopping parameters, respectively. Importantly, we introduce the Rashba-type and Dresselhaus-type sASOC parameterized by $\alpha_\text{R} = 0.2$ and $\alpha_\text{D} = 0.4$, respectively. In the specific case that $|\alpha_\text{R}|=|\alpha_\text{D}|$, the gyration current response vanishes since the emergent symmetry may be present~\cite{Bernevig2006}. We take the molecular field for the antiferromagnetic order as $\bm{h}_\text{AF} = (1.6,0,0)$, which represents $x$-collinear antiferromagnetic order. The doubly-degenerate energy spectrum for Eq.~\eqref{Rashba_Dresselhaus_hamiltonian} is obtained as
		\begin{equation}
		\epsilon_{\bm{k}\pm} = \epsilon_0 (\bm{k}) \pm \sqrt{V_{\rm AB}(\bm{k})^2 + \bm{g}(\bm{k})^2 }. \label{energyspectrum}
		\end{equation}
Mainly owing to the large molecular field $\bm{h}_\text{AF}$, two degenerate bands are separated by the energy gap, $\delta \epsilon = 2 \sqrt{V_{\rm AB}(\bm{k})^2 + \bm{g}(\bm{k})^2 }$.

The point group symmetry is denoted by $mm'm$ lacking the \Pa{}-symmetry in the antiferromagnetic state. Indeed, the antiferromagnetic order parameter is characterized by the odd-parity irreducible representation $B_{2u}$ of the point group $D_{2h}$. According to the reduced symmetry, we have
		\begin{equation}
  		\kappa^{xz}_\text{gyro} \neq 0,~\kappa^{yz}_\text{gyro} = 0. \label{gyration_current_symmetry_in_RD_system}
  		\end{equation}  
Note that we can only take the index $\nu =z$ in $\kappa^{\mu\nu}_\text{gyro}$ because of the absence of the $k_z-$dispersion in the two-dimensional model. A lot of well-known magnetoelectric insulators such as Li\textit{T}PO$_4$ (\textit{T} = Fe,~Co,~Ni)~\cite{Vaknin2002LiCoPO4,VanAken2007LiCoPO4toroidicDomains,Fogh2017LiCoPO4,Rousse2003LiFePO4,Petersen2015LiFePO4,Kornev2004LiCo_NiPO4} are characterized by the same irreducible representation and allow the gyration current response in Eq.~\eqref{gyration_current_symmetry_in_RD_system}.

In addition to the gyration current response function, we calculate the joint density of states $J(\Omega)$ in Eq.~\eqref{joint_density_of_states} and the attenuation coefficient $\varepsilon_\text{att}$ given by~\cite{Sipe2000secondorder,AzpirozSouza2018,NastosSipe2007LinearResponse}
		\begin{align}
		&\varepsilon^{\mu\nu}_\text{att} = i\pi q^2 \int \frac{d\bm{k}}{\left( 2\pi \right)^2} \sum_{a\neq b} \mathcal{A}^\mu_{ab} \mathcal{A}^\nu_{ba} f_{ab} \delta (\hbar \Omega -\epsilon_{ba}),\label{attenuation_coefficient}\\
			&=i\pi q^2 \int \frac{d\bm{k}}{\left( 2\pi \right)^2} \sum_{a\neq b} \left( g^{\mu\nu}_{ab} -\frac{i}{2} \Omega^{\mu\nu}_{ab} \right)  f_{ab} \delta (\hbar \Omega -\epsilon_{ba}),
		\end{align}
which is derived from the absorptive part of the expectation value $\trace{q \bm{r}_\text{e} P^{(1)}}$ with the interband position operator $\bm{r}_\text{e}$ and the first-order perturbed density matrix $P^{(1)}$. Under the linearly-polarized light along the $\mu$-direction, the attenuation coefficient is solely determined by the band-resolved quantum metric $g^{\mu\nu}_{ab}$. Thus, the comparison between the shift current coefficient $\sigma^{\mu;\nu\nu}_\text{shift}$ and the symmetric component of the attenuation coefficient $\varepsilon^{\mu\nu}_\text{att}$ is informative~\cite{Sturman1992Book,AzpirozSouza2018,YoungRappe2012_FirstPrincipleBTO,YoungRappe2012BiFeO3}. On the other hand, the attenuation of the circularly-polarized light arises from both of the band-resolved quantum metric and Berry curvature~\cite{Gao2020TunablePGE,Souze2008Dichroic}. We define the attenuation coefficients of the left-handed ($+$) and right-handed ($-$) circularly-polarized lights as
		\begin{align}
		&\varepsilon^{\pm}_\text{att} =  \frac{1}{2} \left( \varepsilon^{xx}_\text{att} + \varepsilon^{yy}_\text{att}  \right) \mp \frac{i}{2} \left( \varepsilon^{xy}_\text{att} - \varepsilon^{yx}_\text{att}  \right),\\
			&= i\pi q^2 \int \frac{d\bm{k}}{\left( 2\pi \right)^2} \sum_{a\neq b}\left[  \frac{1}{2} \left( g^{xx}_{ab} +g^{yy}_{ab}  \right) \mp \frac{1}{2} \Omega^{xy}_{ab}  \right]  f_{ab} \delta (\hbar \Omega -\epsilon_{ba}), \label{CP_attenuation}\\
			&= i\pi q^2 \int \frac{d\bm{k}}{\left( 2\pi \right)^2} \sum_{a\neq b}\frac{1}{2} |\mathcal{A}^\pm_{ab}|^2  f_{ab} \delta (\hbar \Omega -\epsilon_{ba}). \label{circular_attenuation}
		\end{align}
In the \T{}-/\PT{}-symmetric systems, the band-resolved Berry curvature does not contribute to the attenuation coefficients in Eq.~\eqref{CP_attenuation} due to the Kramers degeneracy. Thus, in the numerical calculation, we calculate $\varepsilon_\text{att} = \varepsilon^{xx}_\text{att}/2 + \varepsilon^{yy}_\text{att}/2$ and take a dimensionless value defined by $\varepsilon_\text{r} =\varepsilon_\text{att}/(\varepsilon_0 l) $, where $\varepsilon_0$ and $l$ are the vacuum permittivity and thickness of the system, respectively~\cite{AzpirozSouza2018}. 

\begin{figure}[htbp]
		\centering 
		\includegraphics[width=80mm,clip]{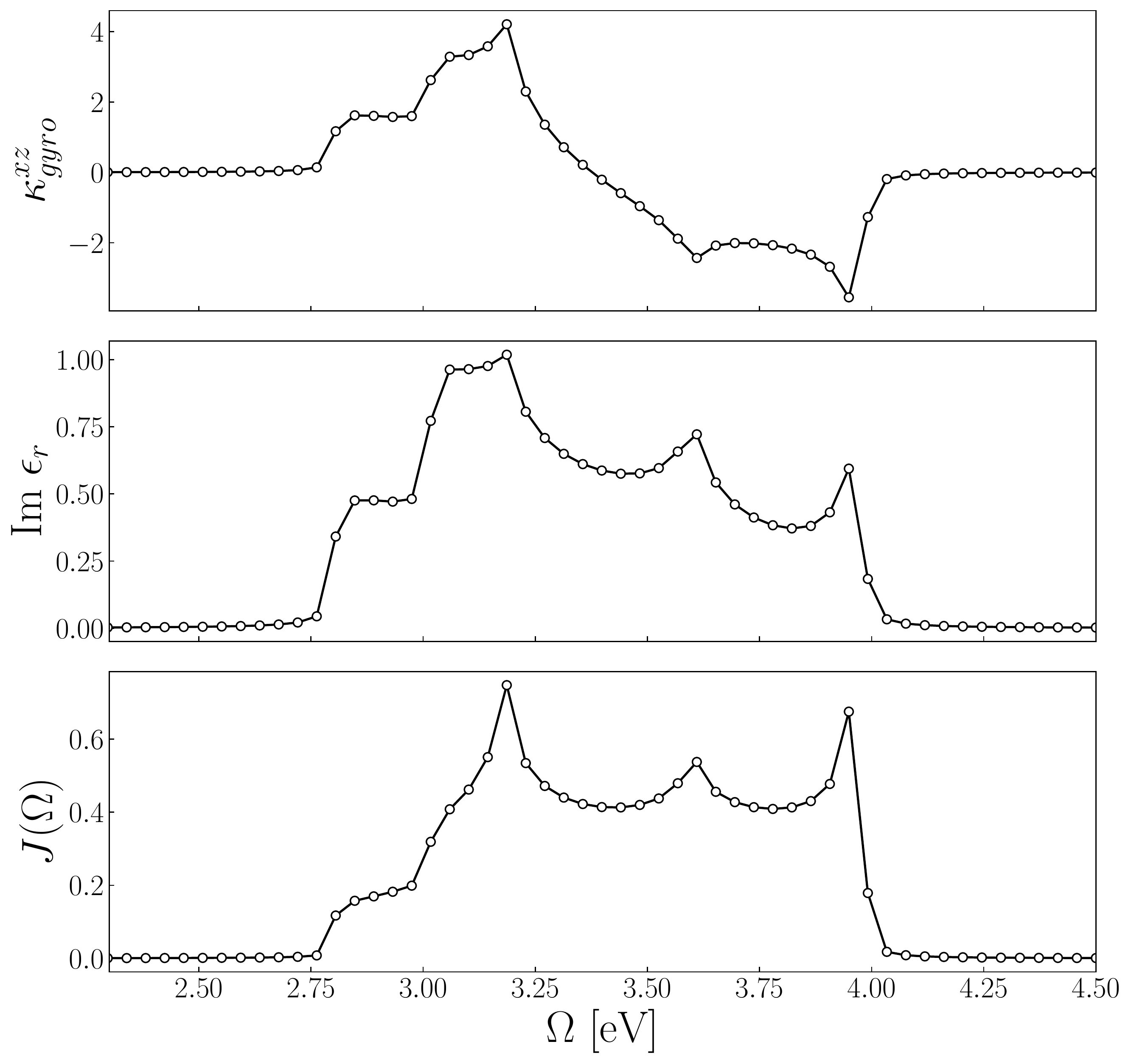}
		\caption{Frequency dependence of (upper panel) the gyration current coefficient $\kappa^{xz}_\text{gyro}$ [$\mu$A$\cdot$V$^{-2}$], (middle panel) dimensionless attenuation coefficient $\varepsilon_r$, and (lower panel) joint density of states $J (\Omega)$ [eV$^{-1}$].  We adopted $q = \mr{1.60 \times 10^{-19} }{[C]}$, $\varepsilon_0 = \mr{8.85 \times 10^{-12} }{[F\cdot m^{-1}]}$, $l = \mr{1}{[nm]}$, and $|t| =\mr{1}{[eV]}$. }
		\label{FIG_RD_model_comparison} 
		\end{figure}

We show the numerically-calculated gyration current coefficient $\kappa^{\mu\nu}_\text{gyro}$, attenuation coefficient $\varepsilon_\text{r}$, and joint density of states $J(\Omega)$~\footnote{The dimension of gyration current coefficient is [A$\cdot$V$^{-2}\cdot $m] in two-dimensional systems. Taking the thickness of the system $l$, the coefficient in three dimension is obtained as $\kappa_\text{gyro} (\text{3d}) = \kappa_\text{gyro} (\text{2d})/l$~\cite{Eangel2017,Cook2017design}. In this work, the thickness and lattice constant are assumed to be the same for simplicity.} in Fig.~\ref{FIG_RD_model_comparison}. For numerics, we approximate the delta function in Eqs.~\eqref{joint_density_of_states}, \eqref{gyration_current_second_tensor_form_U2}, and \eqref{circular_attenuation} by the Lorentian function. This treatment corresponds to taking into account a phenomenological scattering rate $\gamma = 0.01$. We assume the absolute zero temperature ($T=0$) and fix the chemical potential between the two bands in Eq.~\eqref{energyspectrum}. Thus, the system in the insulating state satisfies $f (\epsilon_{\bm{k}a})  = 0$ for the upper band ($a =+$) and $f (\epsilon_{\bm{k}a})  = 1$ for the lower band ($a =-$). 

Figure~\ref{FIG_RD_model_comparison} plots the frequency dependence. We see that the three quantities mostly share the peak positions. Thus, it is indicated that the frequency dependence of the gyration current coefficient is roughly determined by the joint density of states. This is consistent with the conventional understanding of the optical conductivity~\cite{Grosso2013Book}. A large joint density of states may be found in low-dimensional magnetoelectric materials such as those crystalize in a pyroxene structure~\cite{Watanabe2018grouptheoretical,Jodlauk2007pyroxene}. On the other hand, in the presence of a geometrically nontrivial electronic structure, the gyration current may show strong enhancement which cannot be attributed to the joint density of states. As an example, we investigate the gyration current in a topologically nontrivial antiferromagnet in the next subsection. 

\subsection{Enhanced gyration current in topological materials}\label{Sec_gyration_in_topological_materials}
Dirac and Weyl electrons with gapless band dispersions give rise to various nontrivial phenomena. For instance, geometric properties of such electronic structure lead to unconventionally large nonlinear responses such as nonlinear Hall effect~\cite{Sodemann2015,Xu2018BCD_switchable,Ma2019BCD_experiment_WTe2}, higher harmonic generations~\cite{Wu2016SHG_in_TaAs,Parker2019}, injection current~\cite{DeJuan2016,Ishizuka2016WeylPGE,ChanPALee2017PGEinWeyl,Raguchi2016PCMinWeyl,Yang2017DivergentBPVEinWeyl,Ma2017CPGE_TaAs,Flicker2018Multifoldfermions,Chang2019Fermiarc_photocurrent_theory,Kastl2015UltrafastPGE_Bi2Se3}, and shift current~\cite{Morimoto2016topological,Osterhoudt2019}. Based on these findings, we investigate the possibility of the giant gyration current response in topological materials.

\begin{figure}[htbp]
		\centering 
		\includegraphics[width=85mm,clip]{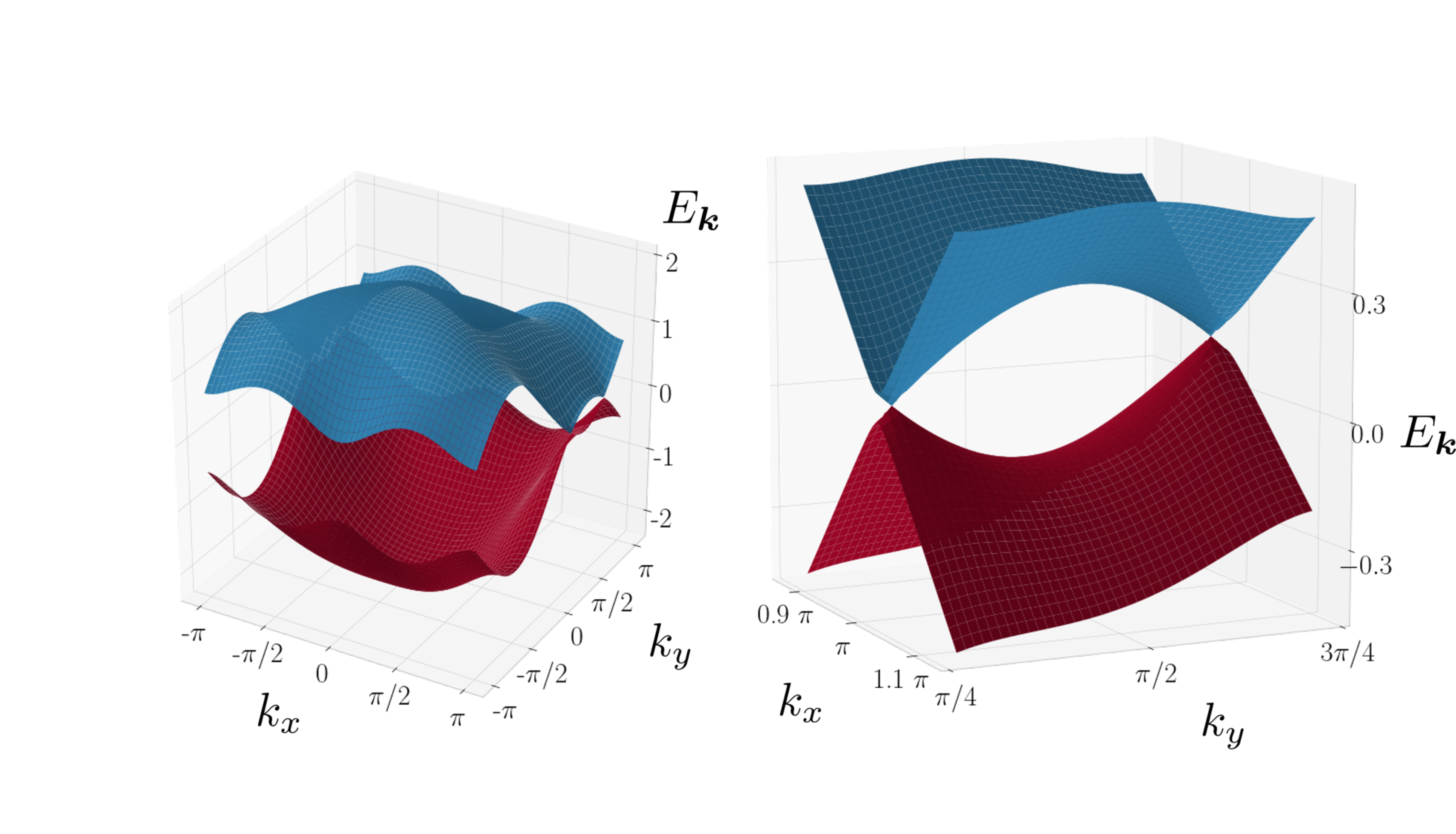}
		\caption{Band structure of the Hamiltonian [Eq.~\eqref{Rashba_Dresselhaus_hamiltonian} with Eq.~\eqref{Rashba_Hamiltonian_parameters}]. (Left panel) Dispersion over all the Brillouin zone. (Right panel) Enlarged view around the gapless Dirac points.}
		\label{FIG_R_dispersion} 
		\end{figure}

The model Hamiltonian is obtained by taking the parameters in Eq.~\eqref{Rashba_Dresselhaus_hamiltonian} as
		\begin{equation}
  		t = 0.08,~ \tilde{t} =1,~ \alpha_\text{R} = 0.8,~\alpha_\text{D} = 0,~\bm{h}_\text{AF} = (0.6,0,0). \label{Rashba_Hamiltonian_parameters}
  		\end{equation}  
This model has been proposed as an effective two-dimensional model Hamiltonian of tetragonal CuMnAs~\cite{Smejkal2017Electric}. We plot the band dispersion of the Hamiltonian in Fig.~\ref{FIG_R_dispersion}. Interestingly, gapless points appear along the high-symmetry line ($k_x = \pi$). Appearance of the gapless points is due to the facts that the sASOC overwhelms the molecular field and that the inter-sublattice hoppings are forbidden by the mirror symmetry denoted by $ \{M_x | [1/2,0,0] \}$~\cite{Liang2016Nonsymmorphic,Smejkal2017Electric,Sumita2016,Ishizuka2018OddparitySC}. The coordinates of the gapless points are analytically obtained as $\bk = (\pi,\pi/2 \pm k_0)$ with $k_0 = \text{arccos} (h_\text{AF}/\alpha_\text{R}) \in (0,\pi/2]$. Here we denoted $\bm{h}_\text{AF} = h_\text{AF}\hat{x}$, and adopt the energy unit $|\tilde{t} |=\mr{1}{[eV]}$ for a quantitative estimation.

To calculate the gyration current arising from the gapless band electrons, we analyze an effective Dirac Hamiltonian given by
		\begin{equation}
		H (\bm{k};s_z) = v_0 k_y  + a_1 k_y \sigma_x - a_2 k_x \sigma_y + w k_x \tau_x + \Delta, \label{effective_Dirac_Hamiltonian}
 		\end{equation}
where the coefficients are obtained from the microscopic parameters as
		\begin{align}
		v_0 = t \cos{k_0},~a_1 = \alpha_\text{R} \sin{k_0} \,s_z,~a_2 = \alpha_\text{R} ,\notag \\
		w = \tilde{t} \cos{\left( \frac{\pi/2 + s_z k_0}{2}  \right)}, ~\Delta =t \sin{k_0} \,s_z.
		\end{align}
We introduced the label $s_z = \uparrow,\downarrow$ representing the Dirac nodes at $(\pi,\pi/2 + k_0)$ and $(\pi,\pi/2 - k_0)$, respectively. Note that the $v_0$ term gives rise to tilting of the Dirac cones along the $y$-axis, whereas $\Delta$ gives the opposite energy shift to the two Dirac nodes. Below we show that the tilting is important to enhance the gyration current. 

Here, we take one of the Dirac nodes and calculate contribution to the gyration current coefficient. Introducing the polar coordinate by $\rho \sin{\theta} = |a_1| k_y$ and $\rho \cos{\theta} =\left( a_2^2 + w^2 \right)^{1/2} k_x $, we write the energy spectrum of Eq.~\eqref{effective_Dirac_Hamiltonian} as
		\begin{equation}
		\epsilon_{\bm{k}\pm ;s_z} = \rho  \left( \frac{v_0}{ |a_1|}\sin{\theta} \pm 1  \right) + \Delta.
		\end{equation}
Owing to the double degeneracy, summation over the band indices can be computed with putting aside the energy-related term, $f_{ab} \delta (\hbar \Omega  -\epsilon_{ba} )$ in Eq.~\eqref{gyration_current_decomposition_U2_z_CP}. When we take the frequency of light as $\Omega >0$ and assume the absolute zero temperature $T=0$, the summation is evaluated as
		\begin{align}
 		& \sum_{a = - }\sum_{b = +} \text{Re}\, \left( i \left[ \ud_\mu  \mathcal{A}^+ \right]_{ab} \mathcal{A}^-_{ba}  - i\left[ \ud_\mu  \mathcal{A}^- \right]_{ab} \mathcal{A}^+_{ba}  \right) \notag \\
 		&= \frac{1}{\rho^3} a_1^2  \left(  a_2^2 +w^2 \right) \sin{\theta}. \label{gyration_component_Dirac_dispersion}
 		\end{align}
The summation was taken over the lower degenerate bands for $a$ and over the upper degenerate bands for $b$, respectively. We notice that the gyration current is totally canceled out if the tilting parameter is zero, since the energy dispersion is symmetric under $k_y \rightarrow -k_y$ when $v_0=0$. Thus, the tilting parameter is an essential ingredient for the gyration current response.

\begin{figure}[htbp]
		\centering 
		\begin{tabular}{c}
		\includegraphics[height=30mm,clip]{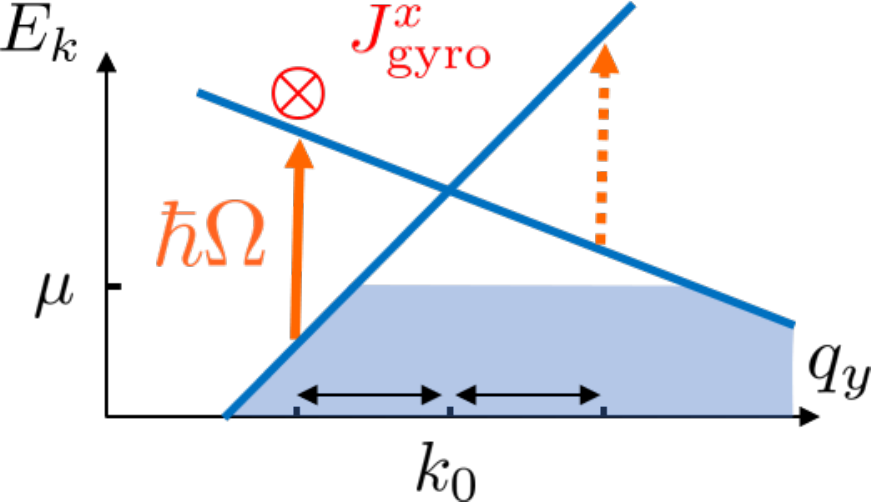}\\
		(a) \vspace{3mm}\\
		\includegraphics[height=30mm,clip]{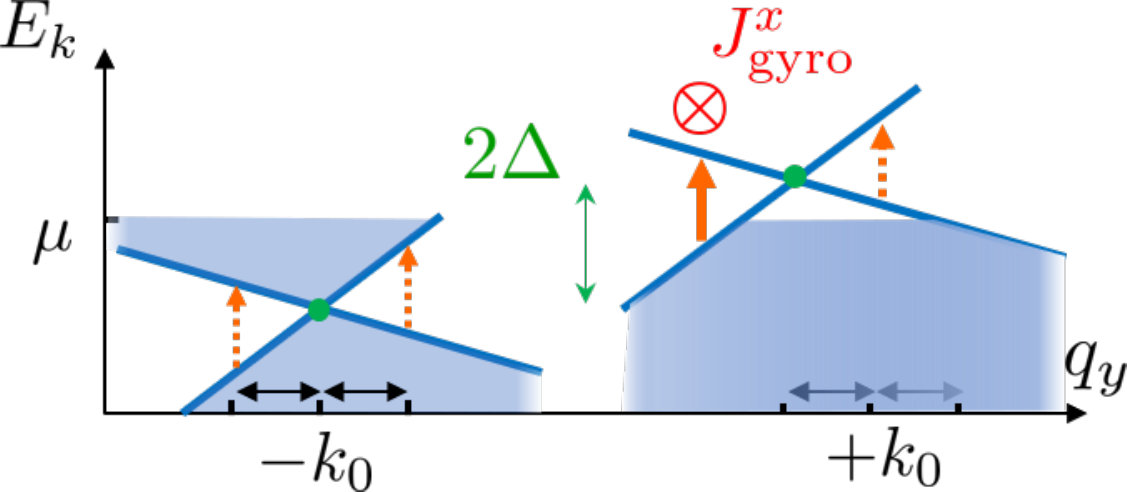}\\
		(b)
		\end{tabular}
		\caption{Mechanism of the enhanced gyration current response in the tilted Dirac system. A coordinate $q_y = k_y -\pi/2$ is introduced. (a) Contributions from $q_y=k_0 \pm k$ are not canceled out because of the tilting of a single Dirac cone. A dotted arrow represents the transition prohibited by Pauli blockade. (b) The opposite energy shift $\pm |\Delta|$ of the nodes prevents cancellation between two Dirac nodes.}
		\label{FIG_tilting} 
		\end{figure}

\begin{figure}[htbp]
		\centering 
		\includegraphics[width=70mm,clip]{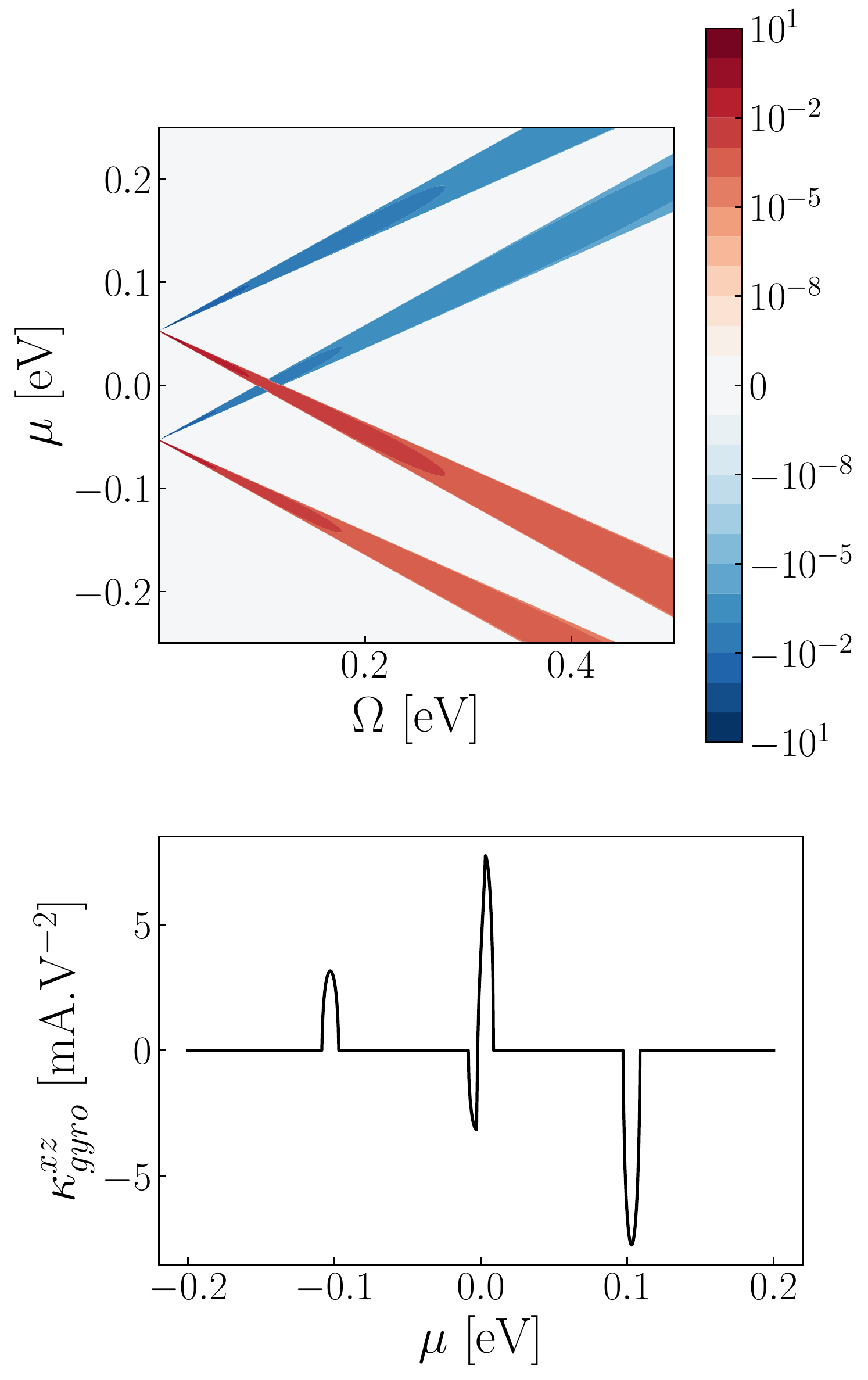}
		\caption{(Upper panel) Chemical potential and frequency dependence of the gyration current coefficient in the unit [A$\cdot$V$^{-2}$]. (Lower panel) Chemical potential profile of the gyration current coefficient for $\hbar\Omega = \mr{100}{[meV]}$.}
		\label{FIG_R_chemi_freq_dep_gyration} 
		\end{figure}

After some simple algebra, we obtain the analytical expression for the gyration current coefficient as
		\begin{align}
		&\kappa^{xz}_\text{gyro} (\Omega) \notag \\
		&= \sum_{s_z =\uparrow,\downarrow} \frac{2q^3}{\pi \hbar^3 \Omega^2} \left( a_2^2 +w^2 \right)^{1/2} \text{sgn}(v_0) \notag \\
		&\times \text{Re} \left[ \sqrt{1 - \frac{a_1^2}{v_0^2} \left(  \frac{\mu + \Delta}{\hbar \Omega/2 } +1  \right)^2 } - \sqrt{1 - \frac{a_1^2}{v_0^2} \left(  \frac{\mu + \Delta}{\hbar\Omega/2 } -1  \right)^2 } \right].\label{gyration_current_analytical_expression}
		\end{align}
Differences in contributions from the two Dirac nodes can be found in the energy shift of the Dirac nodes $\Delta$ and in $w (s_z =\uparrow) \neq w (s_z =\downarrow)$. Otherwise, Dirac electrons around $(\pi,\pi/2 \pm k_0)$ equally contribute to the gyration current response. As a result, the tilting parameters $v_0$ and the energy shift $\Delta$ play two important roles as illustrated in Fig.~\ref{FIG_tilting}. Firstly, the tilting of each Dirac cone due to $v_0$ prevents the gyration current from compensation of the contributions from $\pm k_y$. Secondly, cancellation between the gyration current from the two Dirac cones is suppressed when the opposite potential shift $\Delta$ sufficiently separates the Dirac nodes. Supposing a small potential difference $\Delta$, the gyration current is partially compensated in the low-frequency regime as shown in the lower panel of Fig.~\ref{FIG_R_chemi_freq_dep_gyration}. Therefore, Dirac nodes separated along the energy axis are favorable for a divergent photocurrent response in the low-frequency regime. Consequently, for an enhanced gyration current response, it is important to hunt for materials hosting strongly tilted gapless dispersions such as the type II Dirac materials~\cite{Chang2017TypeIIDirac_VAl3,Noh2017TypeIIDriac_exp_PdTe2}. 

We demonstrate the impact of tilting by taking the Dirac node labeled by $s_z =\uparrow$. For a fixed frequency $\Omega$, the gyration current appears in the region given by 
		\begin{equation}
		- \frac{\hbar \Omega}{2} \left( \left|\frac{v_0}{a_1}\right| +1 \right) \leq \mu -t \sin{k_0} \leq \frac{\hbar \Omega}{2} \left( \left|\frac{v_0}{a_1}\right| -1 \right),\label{lower_active_band}
		\end{equation}
for $\mu < t \sin{k_0}$, and
		\begin{equation}
		- \frac{\hbar \Omega}{2} \left( \left|\frac{v_0}{a_1}\right| -1 \right) \leq \mu -t \sin{k_0} \leq \frac{\hbar \Omega}{2} \left( \left|\frac{v_0}{a_1}\right| +1 \right), \label{upper_active_band}
		\end{equation}
for $\mu > t \sin{k_0}$. For the parameters in Eq.~\eqref{Rashba_Hamiltonian_parameters}, $\left|v_0/a_1\right| =0.11 <1$. Thus, the chemical potential has energy windows where the gyration current response is finite. The width $ \delta \Omega_\text{I} =  \Omega \left|v_0 /a_1\right|$ increases in proportion to the frequency $\Omega$, while it vanishes in the non-tilted system ($v_0 =0$). When the chemical potential lies in the window, the gyration current is extensively enhanced as $O(\Omega^{-2})$ in the low-frequency regime.

\begin{figure}[htbp]
		\centering 
		\includegraphics[width=80mm,clip]{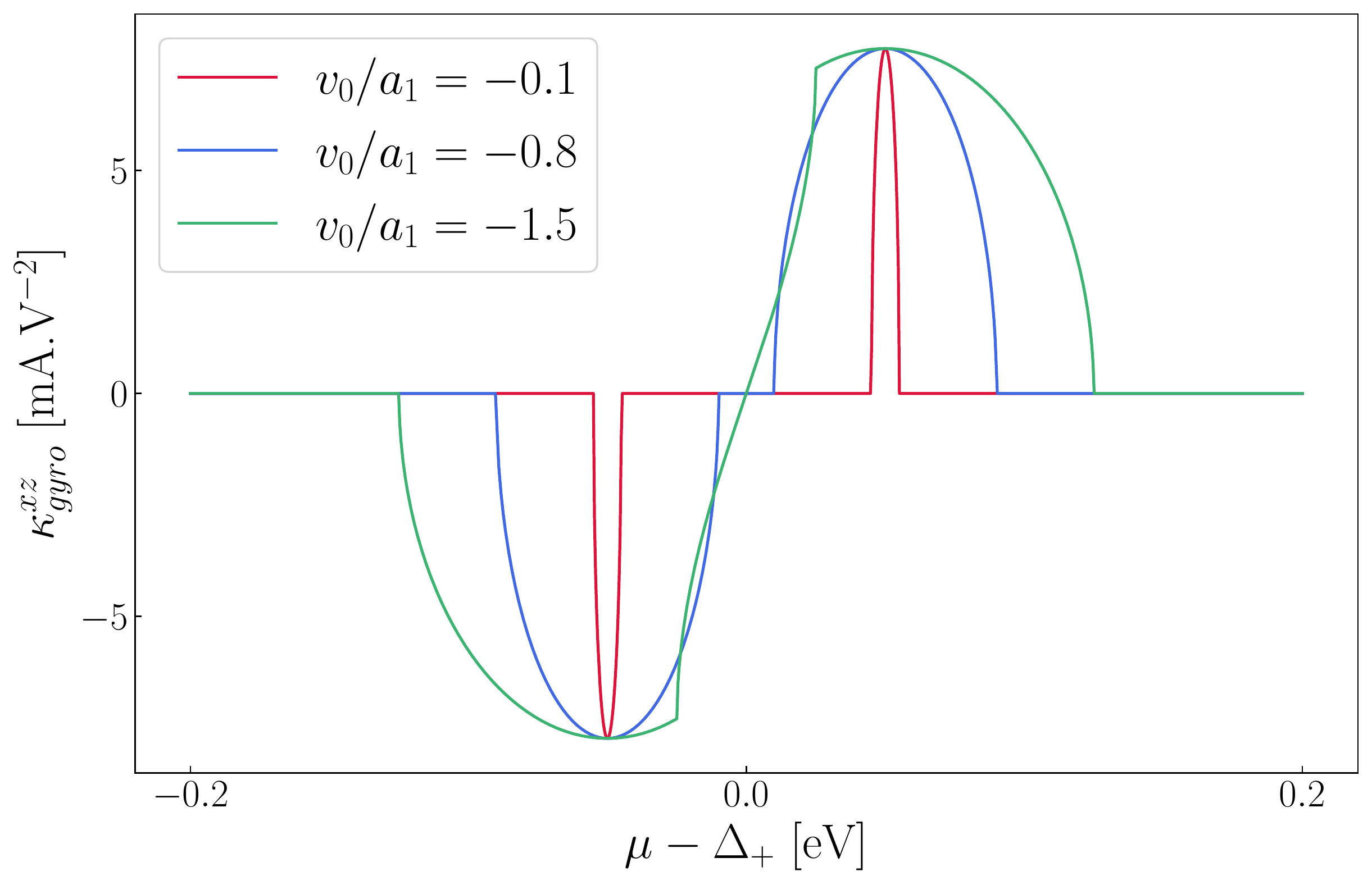}
		\caption{Chemical potential dependence of the gyration current coefficient for the frequency of light $\hbar \Omega = \mr{100}{[meV]}$. Several plots are shown by changing $v_0$. The ratio $v_0/a_1 = -0.1$, $-0.8$, and $-1.5$ represent the type-I Dirac (red line), highly-tilted type-I Dirac (blue line), and type-II Dirac systems (green line). We introduced $\Delta_+ = t \sin{k_0} $ for the energy shift of the Dirac node $s_z = \uparrow $.}
		\label{FIG_R_chemi_freq_dep_gyration_a1sweep} 
		\end{figure}

When the tilting parameter $v_0$ increases, the system changes from a type-I Dirac system ($\left|v_0/a_1\right| <1$) to a type II Dirac system ($\left|v_0/a_1\right| >1$). In the type-II Dirac system, the width of the energy window reaches as large as $\delta \Omega_\text{II} \geq \Omega $. On the other hand, the upper and lower energy windows given in Eqs.~\eqref{lower_active_band} and \eqref{upper_active_band} overlap with each other, and hence the gyration current is partially canceled out. The tilting parameters do not influence the maximal value of the gyration current coefficient as shown in Fig.~\ref{FIG_R_chemi_freq_dep_gyration_a1sweep}, because the Berry connection itself is not relevant to the trace of the Dirac Hamiltonian in Eq.~\eqref{effective_Dirac_Hamiltonian}.

On the basis of the analytical formula in Eq.~\eqref{gyration_current_analytical_expression}, we plot $(\mu,\Omega)$ dependence of the gyration current coefficient by taking both of the two Dirac nodes into account (Fig.~\ref{FIG_R_chemi_freq_dep_gyration}). It is clearly shown that the energy windows of two nodes grow from the offset energies given by $\Delta$ and overlap near $\hbar\Omega \sim 2 |\Delta|\sim \mr{0.1}{[eV]}$. In the overlapped region, the total gyration current coefficient is decreased by partial cancellation. Interestingly, the gyration current shows divergent behavior in the low-frequency regime $\Omega \ll 1$. Taking an available low-frequency light in the Terahertz regime $\hbar\Omega = \mr{1}{[meV]}$, the energy window of each node is evaluated as
		\begin{align}
		& 0.44 \leq \left|\mu - 52 \right| \leq 0.56 & \text{for  } s_z = \uparrow, \\
		& 0.44 \leq \left|\mu + 52 \right| \leq 0.56 & \text{for  } s_z = \downarrow, 
		\end{align}   
where the unit [meV] is abbreviated. The gyration current coefficient is estimated as large as $\left|\kappa^{xz}_\text{gyro}\right| \sim \mr{10}{[A\cdot V^{-2}]}$. 

Note that the divergent response shown in Fig.~\ref{FIG_R_chemi_freq_dep_gyration} is also found in our calculations for the original tight-binding Hamiltonian in Eqs.~\eqref{Rashba_Dresselhaus_hamiltonian} and \eqref{Rashba_Hamiltonian_parameters}. Thus, the effective Dirac model picks up the photocurrent response well in the low-frequency regime ($\hbar \Omega \lesssim \mr{100}{meV}$ in this model).

\begin{figure}[htbp]
		\centering 
		\includegraphics[width=80mm,clip]{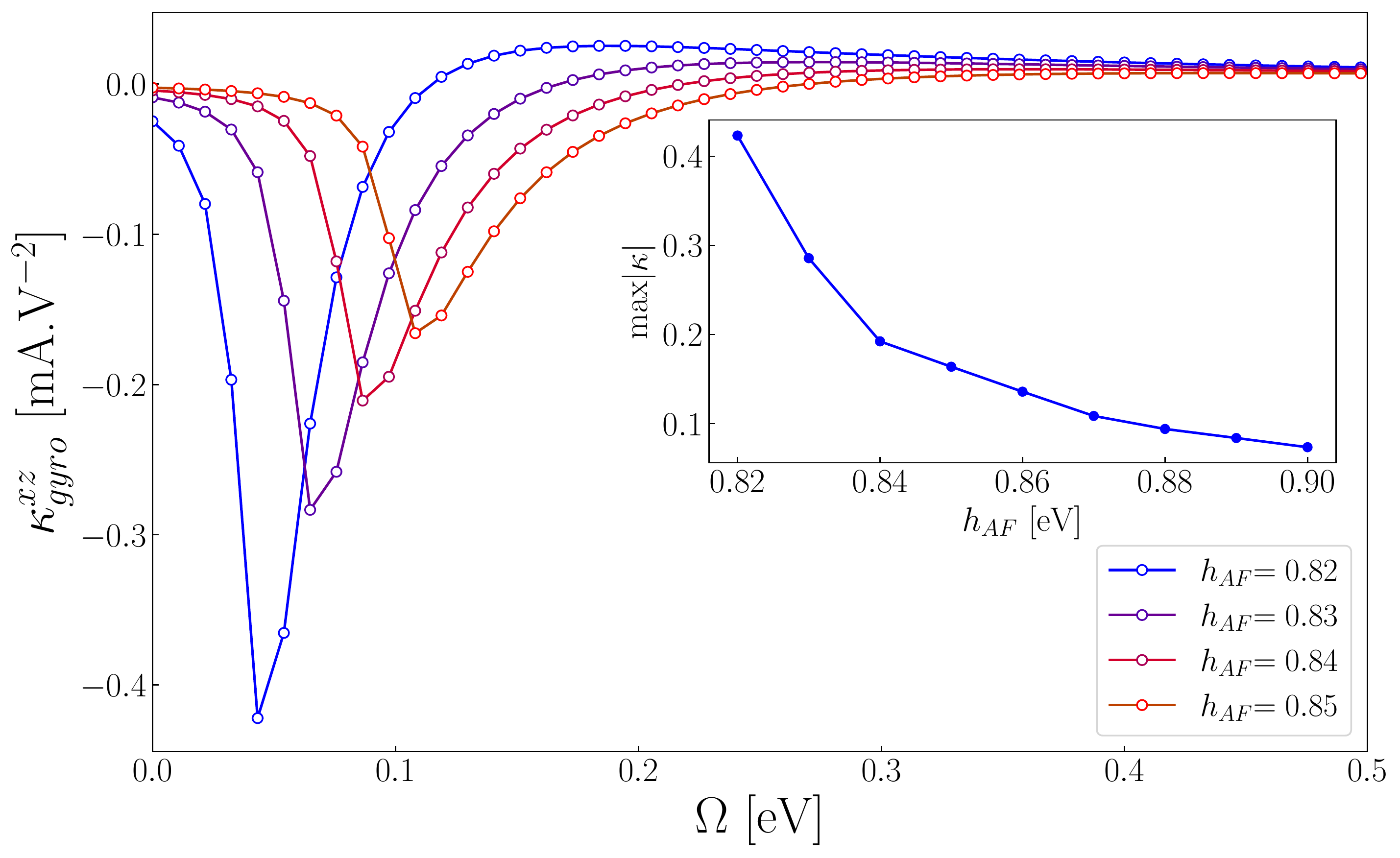}
		\caption{Frequency dependence of the gyration current coefficient with changing the molecular field $h_\text{AF}$. The other parameters are the same as Eq.~\eqref{Rashba_Hamiltonian_parameters}. The inset plots the maximum magnitude of $\kappa^{xz}_\text{gyro}$ as a function of $h_\text{AF}$.}
		\label{FIG_R_freq_dep_gyration_with_hAFsweep} 
		\end{figure}

Here we move on to a slightly-gapped Dirac system which is realized when $|h_\text{AF}| \geq |\alpha_\text{R}|$. We numerically calculate the gyration current response and find that the massive Dirac dispersion is also responsible for an enhanced gyration current. When the molecular field increases so as to surpass the sASOC, the two Dirac nodes merge at $(k_x,k_y) = (\pi,\pi/2)$ and then turn into the massive Dirac dispersion. Figure~\ref{FIG_R_freq_dep_gyration_with_hAFsweep} shows the numerical results of Eq.~\eqref{gyration_current_decomposition_U2_z_CP} with the discretized Brillouin zone mesh $N=1500^2$ and the phenomenological scattering rate $\gamma = \mr{0.01}{[eV]}$. We assume an insulating state at the zero temperature $T=0$, that is, the chemical potential is positioned in the energy gap. Such electronic structure may be realized in MnBi$_2$Te$_4$ thin films consisting of the double septuple layers~\cite{Shiqiao2020MnBi2Te4}. Interestingly, we see a large gyration current coefficient $\left|\kappa^{xz}_\text{gyro}\right| \sim \mr{100}{[\mu A\cdot V^{-2}]}$  for a relatively high frequency $\hbar \Omega \sim \mr{100}{[meV]}$ of light~\cite{Ma2017CPGE_TaAs,Osterhoudt2019}. The coefficient is therefore expected to be an order of magnitude larger than the photoconductivity of typical semiconductors such as GaAs~\cite{AzpirozSouza2018}.

The enhanced photocurrent response may be attributed to two reasons. One is that the quadratic band edge at $(k_x,k_y) =(\pi,\pi/2)$ forms a generalized van Hove singularity [see Eq.~\eqref{generalized_van_Hove_singularity}]. The van Hove singularity gives rise to a large joint density of states $J(\Omega)$ leading to an enhanced gyration current, while this factor is absent in the linear and gapless Dirac system. The other reason is that the geometric quantity is still large in a slightly-gapped regime. As the antiferromagnetic molecular field $h_\text{AF}$ increases and geometric quantity becomes smaller, the maximum value of the gyration current coefficient is suppressed [inset of Fig.~\ref{FIG_R_freq_dep_gyration_with_hAFsweep}]. The exchange splitting due to the antiferromagnetic order grows as the temperature is lowered, and the gyration current is therefore expected to show a drastic temperature dependence. This nontrivial temperature dependence is a striking property of the photocurrent in magnetically-parity-violating systems.

\begin{figure*}[t]
		\centering 
		\begin{tabular}{cc}
		\includegraphics[height=65mm,clip]{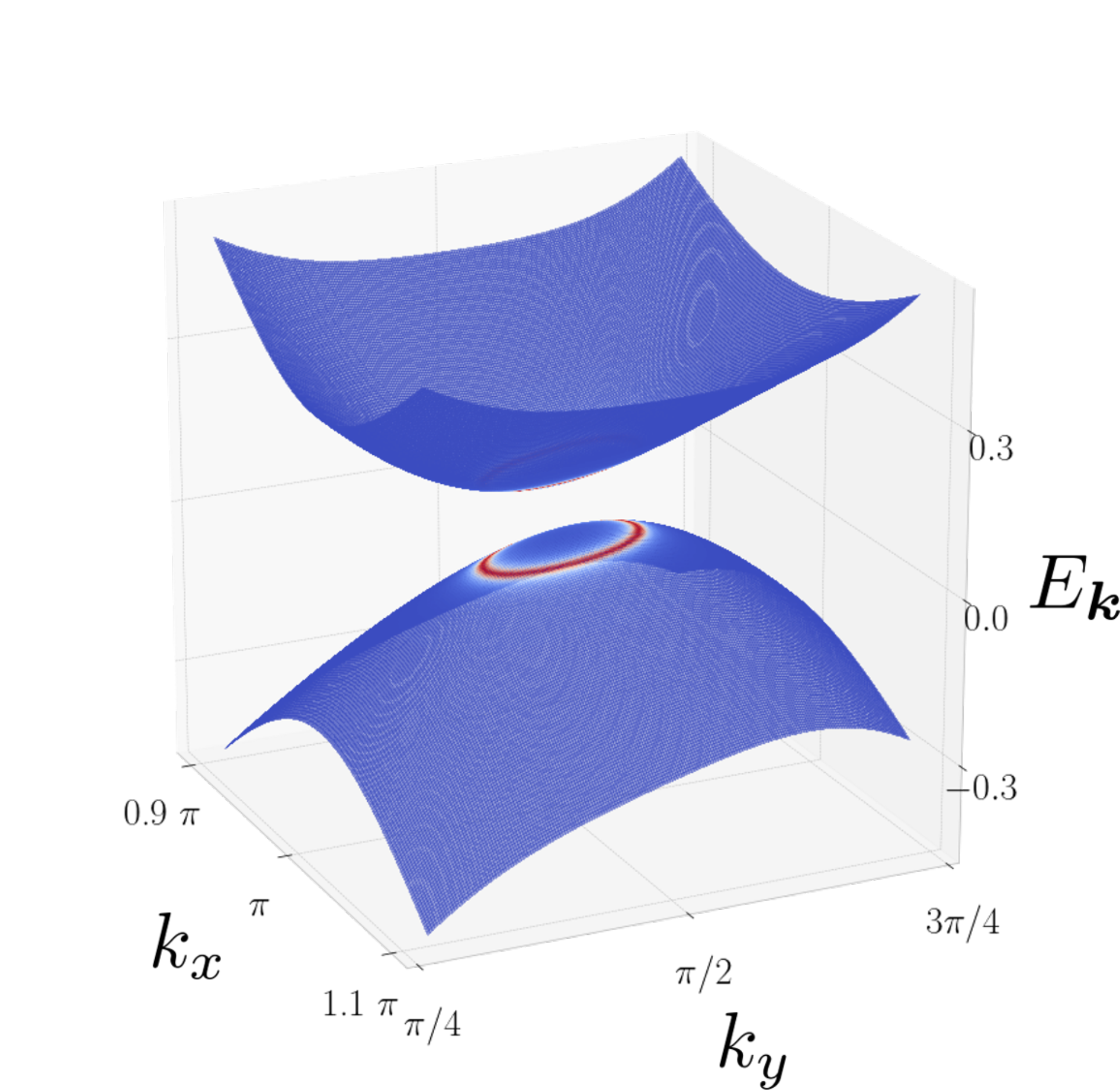} \hspace*{10mm}&
		\includegraphics[height=65mm,clip]{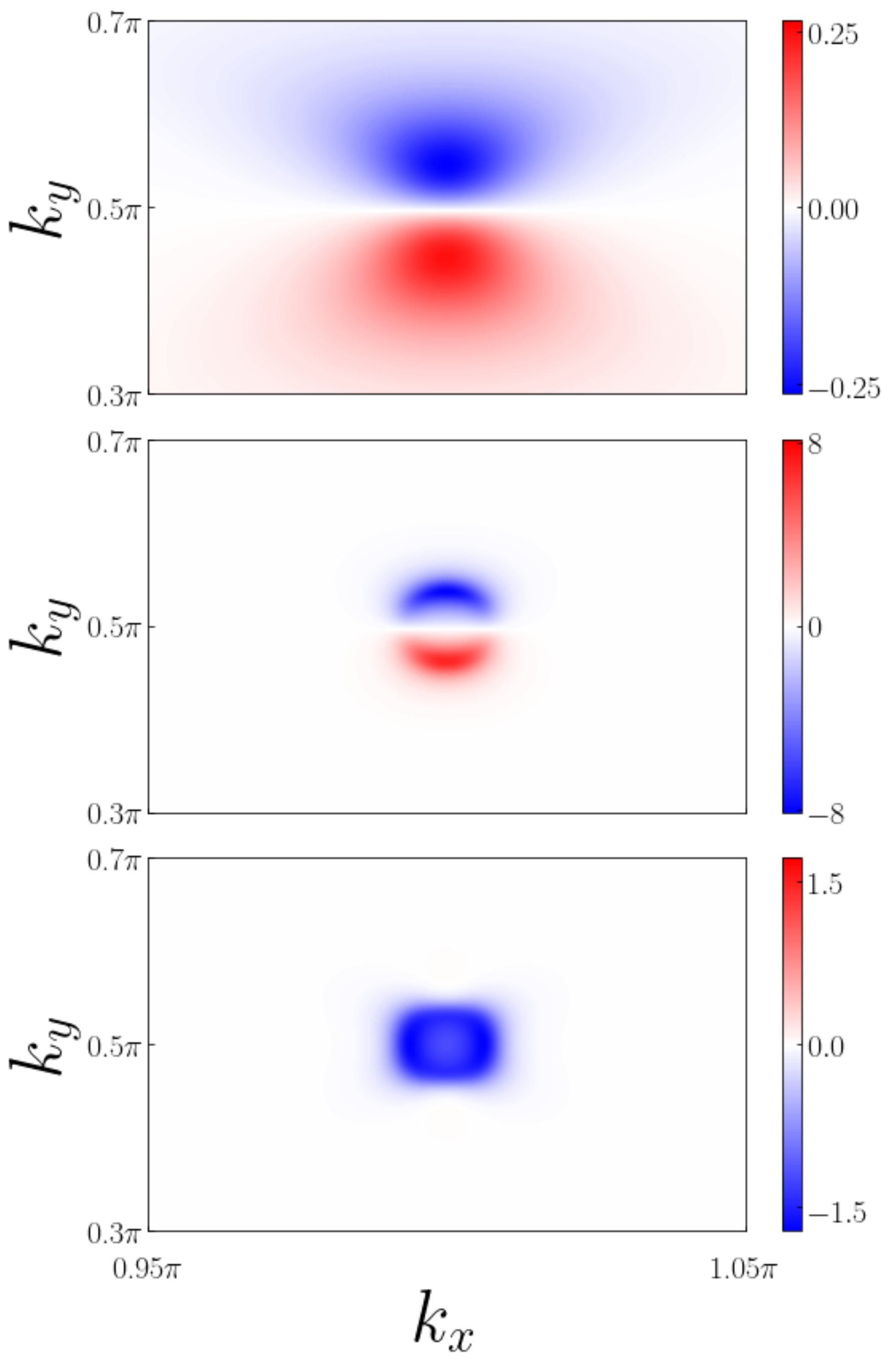}\\
		(a)&
		(b)
		\end{tabular}
		\caption{(a) Energy dispersion in the vicinity of $(k_x,k_y) = (\pi, \pi/2)$ in the slightly-gapped regime ($h_\text{AF} = 0.85$). The red-colored region represents the momentum where interband transitions are allowed with the frequency $\hbar \Omega = \mr{120}{[meV]}$ and scattering rate $\gamma = \mr{10}{[meV]}$. (b) The momentum-resolved distributions are shown for  $G^{xz} (k_x,k_y)$ (upper panel), $\bar{\kappa}^{xz}_\text{gyro} (k_x,k_y) $ (middle panel), and $\bar{\kappa}^{xz}_\text{gyro} (k_x,k_y) + \bar{\kappa}^{xz}_\text{gyro} (k_x,\pi-k_y) $ (lower panel).}
		\label{FIG_profile} 
		\end{figure*}

To discuss geometric properties of the system in more details, we introduce a quantity defined by
		\begin{align}
		 &G^{\mu z} (k_x,k_y)\notag \\
		 & = \frac{\pi q^3}{\hbar}  \sum_{a:\text{occ.}} \sum_{b:\text{unocc.}} \text{Re}\, \left( i \left[ \ud_\mu  \mathcal{A}^+ \right]_{ab} \mathcal{A}^-_{ba}  - i\left[ \ud_\mu  \mathcal{A}^- \right]_{ab} \mathcal{A}^+_{ba}  \right),\label{gyration_current_profile_function}
		\end{align}
which is indeed a part of integrand in Eq.~\eqref{gyration_current_decomposition_U2_z_CP}. We also consider a momentum-resolved gyration current coefficient defined by
		\begin{equation}
		\bar{\kappa}^{\mu z}_\text{gyro} (k_x,k_y)  = G^{\mu z} (k_x,k_y) \delta (\hbar \Omega - \delta \epsilon),
		\end{equation}
where $\delta \epsilon$ is the energy gap. Since both of $G^{xz}$ and $\bar{\kappa}^{xz}_\text{gyro} (\bm{k})$ show a dipolar profile [Fig.~\ref{FIG_profile}~(b)] around the massive Dirac point at $(k_x,k_y) = (\pi, \pi/2)$, the total gyration current coefficient is seemingly canceled out by integration over $(k_x ,k_y)$. However, the cancellation is actually prevented by an asymmetric energy dispersion along the $k_y$ axis. The inter-sublattice hopping $V_\text{AB} \left( \bm{k} \right)$ gives rise to the asymmetry of the energy gap between the momentum $(k_x, k_y)$ and $(k_x,\pi-  k_y)$, and hence makes the net gyration current uncompensated. Uncompensation can be seen in the distribution of the symmetrized gyration current coefficient defined by $\bar{\kappa}^{xz}_\text{gyro} (k_x, k_y) + \bar{\kappa}^{xz}_\text{gyro} (k_x, \pi-k_y)$ [Fig.~\ref{FIG_profile}~(b)].
Thus, the microscopic origin of the enhanced gyration current response is different between massive and massless Dirac systems. In the former the asymmetric band gap due to the inter-sublattice hopping plays an important role, while in the latter cancellation is prevented by the combination of tilting in the Dirac nodes and Pauli blockade (See Fig.~\ref{FIG_tilting}).

At the end of this section, we comment on the extrinsic effect due to impurity scattering. In the presence of the metallic conductivity, the impurity effect may overwhelm the intrinsic terms as in the case of the anomalous Hall effect~\cite{Nagaosa2010}. Theoretical works have reported that such extrinsic contributions play an important role in \T{}-symmetric metals~\cite{Du2019Disorder,Isobe2020Rectification}, which may smear the topological enhancement of the intrinsic photocurrents such as the injection current. On the other hand, in the \PT{}-symmetric systems, the extrinsic effects are strongly suppressed~\cite{Watanabe2020NLC} except for the trivial correction such as smearing resonant behavior [See Eq.~\eqref{electric_injection_with_scattering}]. Thus, we expect that the enhancement of the gyration current in topological materials is robust to the admixture with other contributions, in contrast to the intrinsic CP-photocurrent in the \T{}-symmetric systems.

\section{Summary and Discussion}\label{Sec_discussion_summary}

In this work, we systematically investigated the second order photocurrent and uncovered new types of photocurrent, named intrinsic Fermi surface effect and gyration current. Our formalism is based on the well-established perturbative calculations~\cite{Sipe1993,Aversa1995,Sipe2000secondorder}, and presents formulas unifying the \PT{}-symmetric parity-violating system (magnetic parity-violation) and the \T{}-symmetric parity-violating one. We showed that the \T{}- and \PT{}-symmetry play contrasting roles in the classification of photocurrent responses. The symmetry determines which the linearly-polarized light or circularly-polarized light generates the photocurrent via the injection current, intrinsic Fermi surface effect, and shift current. Our formulation also identifies the geometric quantities which give rise to these photocurrent responses. Making use of the result of classification, we found the chiral photocurrent arising from the gyration current in the \PT{}-symmetric systems, which is the counterpart of the shift current in the \T{}-symmetric systems~\cite{Sturman1992Book,Kraut1981Photovoltaiceffect,Sipe2000secondorder}.

We also elucidated that the gyration current is enhanced in the topological systems. On the basis of the minimal model for the \PT{}-symmetric and topologically nontrivial antiferromagnet CuMnAs, we derived analytical expressions revealing a divergent gyration current in the low-frequency regime. In particular, tilting of the gapless Dirac dispersion is an essential ingredient for the enhanced gyration current. As shown in Fig.~\ref{FIG_R_freq_dep_gyration_with_hAFsweep}, massive Dirac systems may also cause an enhanced gyration current due to relatively large joint density of states and quantum geometric quantity. We expect that the experimental detection of the enhanced chiral photocurrent is promising because the gyration current is not admixed with other chiral photocurrents (See Table~\ref{Table_photocurrent_circular_linear_classification}) and because extrinsic contributions from the impurity scattering~\cite{Du2019Disorder,Isobe2020Rectification} play a minor role in generating the photocurrent in the \PT{}-symmetric systems.

More elaborate investigations of the gyration current in various topological materials are desirable, although this work focuses on two-dimensional Dirac electron systems. 
Recent studies have clarified that some magnetic space groups can ensure the multi-fold degeneracy at high-symmetry points in the Brillouin zone~\cite{Cano2019}. For instance, Cu$_3$TeO$_6$ undergoes the parity-violating magnetic order and may possess six-fold degenerate electrons at the Brillouin zone corner~\cite{Cano2019,Kangkang2017CuTeO6_theory,Bao2018Cu3TeO6_experiment,Yao2018Cu3TeO6_experiment}. Although this compound is insulating and the degenerate states do not lie near the Fermi energy, related compounds may be a potential candidate to realize a giant gyration current response. Alternatively, the photocurrent may be enhanced by large joint density of states. Some of the magnetoelectric materials show the low-dimensional behavior and thus may be potential candidates for a good photocurrent generator~\cite{Jodlauk2007pyroxene,Tokura2014Multiferroics}.    

This work completes all the photocurrent responses of the band electrons. On the other hand, electron correlation effect may enrich the photocurrent phenomena~\cite{Morimoto2018Nonreciprocal_ElectronCorrelation}. Indeed, it has been shown that the strong correlation influences the frequency dependence of photocurrent responses~\cite{Morimoto2016topological}.
Furthermore, it has been proposed that the photocurrent can be generated through the bosonic excitations such as electromagnon and exciton in correlated systems~\cite{Morimoto2016Excitonic_nonlinear,Morimoto2019shiftcurrent_electromagnon,Liu2020Cirularphoto_perovskate}. Thus, interplay of correlation effects and topological electronic structures in the photocurrent generation is desired to be clarified in future works. Moreover, relaxation of photo-excited electrons should be elaborated for more accurate description of photocurrent responses. Throughout this work, we take into account the relaxation within the relaxation-time approximation as in Eq.~\eqref{electric_injection_with_scattering}. Although this assumption may be reasonable in the optical regime where the inverse relaxation time $\tau^{-1}$ is much smaller than the frequency of light $\Omega \gg \tau^{-1}$, the enhanced photocurrent response we interested in may be in the low-frequency regime where $\Omega \ll \tau^{-1}$. Thus, it remains a task to clarify how the photocurrent responses are influenced by the self-energy and vertex correction arising from impurity scatterings. The gyration current, however, may not be significantly changed by scattering because its counterpart, the shift current, is quite invulnerable to the impurities~\cite{Hatada2020} and the extrinsic contributions beyond the relaxation-time approximation are strongly suppressed by the \PT{}-symmetry~\cite{Watanabe2020NLC}. Recent experiments have used ultrafast spectrometry and successfully observed dynamics of the photo-electrons~\cite{Kastl2015UltrafastPGE_Bi2Se3,Braun2016surface_shift_current_experiment,Ogawa2017SbSI,Sotome2019Sn2P2S6,Sirica2019,Liu2020Cirularphoto_perovskate,Gao2020ChiralTerahertz}. These previous experiments worked on the photocurrent responses in the \T{}-symmetric systems. On the other hand, because the antiferromagnetic magnon excitations are present, the time-resolved dynamics of photo-electrons in the \PT{}-symmetric magnetic systems may show relaxation distinct from the nonmagnetic systems. The relaxation process of the photo-induced electrons may be an important key to realize high-performance photo-electric devices.

Interest on the gyration current will be shared in a vast range of the field in condensed matter physics, such as optoelectronic, multiferroics, spintronics, and topological science. In particular, the gyration current coefficient is sensitive to the parity-violating magnetic order. Thus, it may enable us to observe domain states via optical probes and to realize a magnetically-switchable photocurrent response~\cite{Zhang2019switchable}. We expect that further studies of the gyration current will be beneficial not only for fundamental research clarifying magnetic compounds but also for applications to multi-functional devices where the light, spins, and electrons are closely correlated with each other. 

\textbf{Note added}---
Recently a theoretical work on the same topic is conducted by J.~Ahn and N.~Nagaosa~\cite{AhnNagaosa}. They also successfully show the \T{}-/\PT{} correspondence and propose the enhanced photocurrent responses in topological materials. Their results are consistent with ours although each work has been done in a completely independent way. We sincerely thank J.~Ahn and N.~Nagaosa for sending the manuscript before submission and agreeing to the simultaneous submission.

\textbf{Acknowledgments}---

The authors are grateful to A.~Daido and F.~de Juan for valuable comments. Especially, the authors thank F.~de Juan for letting us making aware of Ref.~\cite{Fernando2020Difference} which is relevant to this work. This work was supported by JSPS KAKENHI (Grants No. JP15H05884, No. JP18H04225, No. JP18H05227, No. JP18H01178, and No. 20H05159). H.W. is a JSPS research fellow and supported by JSPS KAKENHI (Grant No.~18J23115).

\appendix

\section{Symmetry considerations of \T{}- and \PT{}-symmetries} \label{AppSec_Symmetry_requirement}
In this section, we introduce basic transformation properties under the anti-unitary operations such as \T{}- and \PT{}-symmetries. Let us consider an anti-unitary symmetry described by an operator $a = \theta g$, where $\theta$ and $g$ are the time-reversal operation and a unitary symmetry operation, respectively. In particular, we take the space-inversion operation $ g = I$ for the \PT{}-symmetry, whereas $g=1$ for the \T{}-symmetry. Bloch states at the momentum $\bm{k}$ are related to those at $-g^{-1} \bm{k}$. The transformation property is given by
		\begin{equation}
		a \ket{u_a (\bm{k})} = \ket{u_b (-g^{-1} \bm{k})} w_{ba} (\bm{k}),\label{anti_unitary_transformation}
		\end{equation}
where the matrix $\hat{w}(\bm{k})$ is unitary. In the following, we describe the basic transformation properties in spinless and spinful systems.

\subsection{spinless system}
In the spinless system, the time-reversal operation is expressed by the complex conjugation operator, $\theta  = \mathcal{K}$. Then, the unitary matrix can be taken as the scalar $\hat{w} (\bk) = 1$ when $g^2 =1$. Owing to the equation $\theta^2 = \mathcal{K}^2 = 1$, we obtain the formula
		\begin{equation}
		\Braket{u_a (\bk) |u_b (\bk) } = \Braket{\theta u_b (\bk) | \theta u_a (\bk) } = \Braket{u_b (-\bk) |u_a (-\bk) }.   
		\end{equation}
Thus, the \T{}-symmetry gives constraint on the Berry connection
		\begin{equation}
		\xi^\mu_{ab} (\bk) = \xi^\mu_{ba} (-\bk).\label{App_T_sym_Berry_connection}
		\end{equation}
Similarly, we obtain
		\begin{equation}
		\xi^\mu_{ab} (\bk) = -\xi^\mu_{ba} (\bk), \label{App_PT_sym_Berry_connection}
		\end{equation}
for the \PT{}-symmetry.

In general, the $\hat{w} (\bk)$ can take an arbitrary phase factor due to the U(1)-gauge degree of freedom. For instance, for a gauge $\theta \ket{u_a (\bm{k})} = \ket{u_a (-\bm{k})} \exp{[-i\phi_a (\bk)]} $, the Berry connection satisfies the relation
		\begin{equation}
 		\xi^\mu_{ab} (\bk) = \xi^\mu_{ba}(-\bk)e^{-i \left[ \phi_a (\bk) - \phi_b (\bk) \right]}.
 		\end{equation}
The formulas in Eqs.~\eqref{quantum_geometric_tensor_T_sym_spinless} and \eqref{BCderiv_BC_prod_T_sym}, however, are irrelevant to the choice of the matrix $\hat{w} (\bk)$. This is consistent with the fact that the obtained formulas for photocurrent responses are U(1)-gauge invariant.

\subsection{spinful system}
In the presence of the spin degree of freedom, the matrix $\hat{w}(\bm{k})$ is at least two-dimensional and has no diagonal component according to the Kramers theorem. When $g^2 =1$, the unitary matrix $\hat{w}(\bm{k})$ is written as
		\begin{equation}
		\hat{w} (\bm{k}) = 
				\begin{pmatrix}
				0 & e^{-i\theta_{\bm{k}}}\\
				e^{-i\phi_{\bm{k}}} & 0
				\end{pmatrix},
		\end{equation}
where $\theta_{\bm{k}}$ and $\phi_{\bm{k}}$ denote real-valued functions of $\bm{k}$. Owing to the Kramers theorem,
		\begin{align}
		- \ket{u_a (\bm{k})} = a^2 \ket{u_a (\bm{k})} 
				&= a \left[   \ket{u_b (-g^{-1} \bm{k})} w_{ba} (\bm{k})  \right],\\
				&= \ket{u_c ( \bm{k})} w_{ba}^\ast (\bm{k}) w_{cb} (-g^{-1}\bm{k}),
		\end{align}
leads to the relation
 		\begin{equation}
		\theta (\bm{k}) =  \phi (-g^{-1}\bm{k}) + \pi.
 		\end{equation}
Therefore, we obtain the unitary matrix
		\begin{equation}
		\hat{w} (\bk) = 
				\begin{pmatrix}
				0 & e^{-i\theta (\bm{k})}\\
				-e^{-i\theta (-g^{-1}\bm{k}) } & 0
				\end{pmatrix},\label{sewing_matrix}
		\end{equation}
which describes the transformation property between doubly-degenerate states. Especially, when we take the gauge so as to satisfy $\theta (\bm{k}) \equiv 0$, the corresponding unitary matrix represents the well-known transformation property between the Kramers doublet
		\begin{equation}
  		a \ket{u_\pm (\bm{k})} = \pm\ket{u_\mp (-g^{-1} \bm{k}) },
  		\end{equation}  
where the subscript $\pm$ denotes the Kramers degree of freedom. Although the discussion can be generalized to other anti-unitary operations satisfying $g^2 \neq 1$, the above discussion sufficiently describes the \T{}- and \PT{}-symmetries.

Here, we proceed to analyze the \PT{}-symmetry. For the \PT{}-symmetry, $\bm{k} = - g\bm{k}$, and thus, we have the Kramers doublet labeled by $\sigma =\pm$ as $\ket{u_{a \sigma}(\bm{k})}$, where $a$ denotes the band index. Below we abbreviate the momentum dependence unless explicitly stated. The Kramers doublet is related to each other by Eq.~\eqref{anti_unitary_transformation}, and the unitary matrix $\hat{w}\left( \bm{k} \right)$ is 
		\begin{equation}
		\hat{w}\left( \bm{k} \right) = i \sigma_y e^{-i\theta} .
		\end{equation}
Note that we take into account a band-independent phase factor $\theta$.

First, we show the proof of the formula 
		\begin{equation}
		\mathcal{A}^\mu_{ab}(\bm{k}) \mathcal{A}^\nu_{ba}(\bm{k}) =   \mathcal{A}^\mu_{\bar{b}\bar{a}} (\bm{k})\mathcal{A}^\nu_{\bar{a}\bar{b}}(\bm{k}),\label{app_BC_product_PT_symmetry}
		\end{equation}
where $\mathcal{A}_{ab}^\mu$ is the interband component of the U(2) Berry connection in Eq.~\eqref{U2_BC_decomposition}, and $(s,\bar{s})$ labels a Kramers pair. The transformation property of the Berry connection is obtained as
 		\begin{align}
 		\xi^\mu_{a\sigma;b\tau} 
 			&= i\Braket{u_{a\sigma} |\partial_\mu u_{b\tau}},\\ 
 			&= i\Braket{ a \left( \partial_\mu u_{b\tau} \right)  |a \left(  u_{a\sigma}  \right)},\\ 
 			&= i \left[ \partial_\mu \left( \ket{ u_{b\tau'} } w_{\tau'\tau}  \right)  \right]^\ast  \ket{u_{a\sigma'} } w_{\sigma'\sigma} ,\\
 			&= i \left( \Braket{\partial_\mu u_{b\tau'} | u_{a\sigma'} } + i \partial_\mu \theta \Braket{u_{b\tau'} | u_{a\sigma'} }     \right) \notag \\
 			&~~~\times \left( i\sigma_y \right)^\dagger_{\tau\tau'}  \left( i\sigma_y \right)_{\sigma'\sigma},\\
 			&= \left( - \xi^\mu_{b\tau';a\sigma'} -\partial_\mu \theta \delta_{ab}\delta_{\tau'\sigma'}      \right) \left( i\sigma_y \right)^\dagger_{\tau\tau'}  \left( i\sigma_y \right)_{\sigma'\sigma}.\label{app_BC_PT_transformation}
		 \end{align}
Taking different band indices $a\neq b$ and applying Eq.~\eqref{app_BC_PT_transformation} to the product $\mathcal{A}^\mu_{ab} \mathcal{A}^\nu_{ba}$, we obtain Eq.~\eqref{app_BC_product_PT_symmetry}.

Similarly, we can derive the formula
 		\begin{equation}
		\left[ \ud_\mu (\bm{k}) \mathcal{A}^\nu (\bm{k}) \right]_{ab} \mathcal{A}^\lambda_{ba} (\bm{k}) = \left[ \ud_\mu (\bm{k}) \mathcal{A}^\nu (\bm{k}) \right]_{ba} \mathcal{A}^\lambda_{\bar{a}\bar{b}} (\bm{k}), \label{app_BC_BCderiv_PT_symmetry}
		\end{equation}
in which $\ud_\mu$ indicates the U(2)-gauge covariant derivative shown in Eq.~\eqref{app_U2_gauge_covariant_derivative}. For the band indices satisfying $\epsilon_{\bm{k}a}\neq \epsilon_{\bm{k}b}$, the covariant derivative of the Berry connection is transformed as 
		\begin{align}
		&\left[ \ud_\mu (\bm{k}) \xi^\nu (\bm{k}) \right]_{a\sigma;b\tau}\notag \\
			&=  \partial_\mu \xi^\nu_{a\sigma;b\tau} - i \left( \xi^\mu_{a\sigma;a\sigma} - \xi^\mu_{b\tau;b\tau} \right)\xi^\nu_{a\sigma;b\tau} \notag \\
			&~~~ - i \left( \xi^\mu_{a\sigma;a\bar{\sigma}}\xi^\nu_{a\bar{\sigma};b\tau} - \xi^\nu_{a\sigma;b\bar{\tau}} \xi^\mu_{b\bar{\tau};b\tau} \right) ,\\
			&= \Bigl[ - \partial_\mu \xi^\nu_{b\bar\tau';a\sigma'}  - i \left(  \xi^\mu_{a\bar{\sigma};a\bar{\sigma}} + \partial_\mu \theta   -\xi^\mu_{b\bar{\tau};b\bar{\tau}} -\partial_\mu \theta  \right) \xi^\nu_{b\tau';a\sigma'}  \notag \\
			&~~- i \xi^\mu_{a\bar{\sigma}';a\sigma'} \xi^\nu_{b \tau';a\bar{\sigma}'}+ i \xi^\nu_{b \bar{\tau}';a\sigma'} \xi^\mu_{b\tau';b\bar{\tau}'}\Bigr] \times \left( i\sigma_y \right)^\dagger_{\tau\tau'}  \left( i\sigma_y \right)_{\sigma'\sigma},\\
			&= - \left[ \ud_\mu \xi^\nu \right]_{b\tau';a\sigma'} \left( i\sigma_y \right)^\dagger_{\tau\tau'}  \left( i\sigma_y \right)_{\sigma'\sigma}.
		\end{align}
Combining this equation with Eq.~\eqref{app_BC_PT_transformation}, we obtain Eq.~\eqref{app_BC_BCderiv_PT_symmetry}. This equation is essential for the derivation of the gyration current formula in the main text. Note that a similar analysis can be conducted in the case of \T{}-symmetric spinful systems.

\section{U(2)-gauge description of photocurrent responses in \PT{}-symmetric systems}\label{Sec_app_U2gauge}
In this section, we show derivation of the photocurrent formulas in the \PT{}-symmetric spinful systems. Previous theoretical studies considered non-degenerate Bloch states and characterized intraband effects by the diagonal component of the Bloch basis~\cite{Aversa1995,Sipe2000secondorder}. This assumption is reasonable in the spinless system or in the \PT{}-violated spinful system. Indeed, much attention have been paid to the \Pa{}-broken nonmagnetic systems, and hence the U(1)-covariant formulation is sufficient to obtain gauge-invariant expressions. On the other hand, the \PT{}-symmetric and spinful systems have the Kramers degeneracy in the band structure at each $\bk$. Thus, we have to carefully proceed to the perturbative calculations with the use of the U(2)-covariant derivative as follows.

Using the U(2) intraband position operator in Eq.~\eqref{U2_intra_position_operator}. the U(2)-gauge covariant derivative is defined by $\ud_\mu = -i r_i^\mu$. The derivative acts on the physical quantities in the Bloch representation $O_{ab}$ as
		\begin{equation}
		[\ud_\mu O]_{ab} = \partial_\mu O_{ab} - i \left( \sum_c \alpha^\mu_{ac} O_{ca} - \sum_c O_{ac}\alpha^\mu_{cb}\right).\label{app_U2_gauge_covariant_derivative}
		\end{equation}
We can check that $[\ud_\mu O]_{ab}$ is U(2)-covariant by taking the U(2)-gauge transformation $\ket{u_a (\bk)} \rightarrow \ket{u_b (\bk)} U_{ba}$, where the summation of the band index $b$ is taken over the Kramers pair, $\epsilon_{\bk b}  =\epsilon_{\bk a}$. The U(1) quantum metric and Berry curvature are defined by
		\begin{align}
		&g^{\mu\nu}_{a} = \frac{1}{2}\sum_\sigma \sum_{b}  \left(  \mathcal{A}^\mu_{a\sigma;b} \mathcal{A}^\nu_{b;a\sigma} + \mathcal{A}^\nu_{a\sigma;b} \mathcal{A}^\mu_{b;a\sigma}  \right), \\
		&\Omega^{\mu}_{a} = \frac{i}{2} \sum_\sigma \sum_{b} \epsilon_{\mu\nu\lambda} \left(  \mathcal{A}^\nu_{a\sigma;b} \mathcal{A}^\lambda_{b;a\sigma} - \mathcal{A}^\lambda_{a\sigma;b} \mathcal{A}^\nu_{b;a\sigma}  \right),
		\end{align}
where we explicitly show the Kramers degree of freedom $\sigma$ for the $a$-th band. Accordingly, we define the band-resolved U(1) quantum metric and Berry curvature as
		\begin{align}
		&g^{\mu\nu}_{ab} = \frac{1}{2}\left(  \mathcal{A}^\mu_{ab} \mathcal{A}^\nu_{ba} + \mathcal{A}^\nu_{ab} \mathcal{A}^\mu_{ba}  \right),\label{app_bandresolved_quantum_metric_u1_sector}\\ 
		&\Omega^{\mu\nu}_{ab} = i \left(  \mathcal{A}^\mu_{ab} \mathcal{A}^\nu_{ba} - \mathcal{A}^\nu_{ab} \mathcal{A}^\mu_{ba}  \right). 
		\end{align}

Following the U(2)-covariant decomposition of the position operator, we divide the nonlinear optical conductivity into four parts [see Eq.~\eqref{second_order_nonlinear_conductivity}]  as
		\begin{widetext}
		\begin{align}
		\sigma^{\mu;\nu \lambda}_\text{ii} \left( \omega; \omega_1,\omega_2 \right)
			&= \frac{q^3}{2}\int \frac{d\bm{k}}{\left( 2\pi \right)^d} \sum_a -v^\mu_{aa} d_{aa}^{\,\omega} d_{aa}^{\,\omega_2}  \partial_\nu \partial_\lambda f(\epsilon_{\bm{k}a}) + \left[ \left( \nu, \omega_1 \right) \leftrightarrow \left( \lambda, \omega_2 \right) \right], \label{second_order_conductivity_ii_spinful} \\
		\sigma^{\mu;\nu \lambda}_\text{ei} \left( \omega; \omega_1,\omega_2 \right)
			&=\frac{q^3}{2}\int \frac{d\bm{k}}{\left( 2\pi \right)^d} \sum_{a,b}   -i v^\mu_{ab} d_{ba }^{\,\omega}   d_{aa}^{\,\omega_2}   \mathcal{A}^\nu_{ba } \partial_\lambda f_{ba} + \left[ \left( \nu, \omega_1 \right) \leftrightarrow \left( \lambda, \omega_2 \right) \right], \label{second_order_conductivity_ei_spinful}\\
		\sigma^{\mu;\nu \lambda}_\text{ie} \left( \omega; \omega_1,\omega_2 \right)
			&= \frac{q^3}{2}\int \frac{d\bm{k}}{\left( 2\pi \right)^d} \sum_{a,b} i  v^\mu_{ab}  d_{ ba}^{\,\omega}  \Biggl[ \partial_\nu \left( d_{ba}^{\,\omega_2}  f_{ab} \mathcal{A}^\lambda_{ba } \right)
			- i \left( \sum_c \alpha^\nu_{bc} d^{\omega_2}_{ca}  \mathcal{A}^\lambda_{ca} f_{ac}   - \sum_c \alpha^\nu_{ca} d^{\omega_2}_{bc}  \mathcal{A}^\lambda_{bc} f_{cb}  \right)    \Biggr] \notag \\
			&~~~~~ + \left[ \left( \nu, \omega_1 \right) \leftrightarrow \left( \lambda, \omega_2 \right) \right],\label{second_order_conductivity_ie_spinful} \\
 		\sigma^{\mu;\nu \lambda}_\text{ee} \left( \omega; \omega_1,\omega_2 \right)
			&= \frac{q^3}{2}\int \frac{d\bm{k}}{\left( 2\pi \right)^d} \sum_{a,b} v^\mu_{ab}  d_{ ba}^{\,\omega}  \left( \sum_c d_{ca }^{\,\omega_2}  \mathcal{A}^\nu_{bc} \mathcal{A}^\lambda_{ca}  f_{ac} - \sum_c d_{bc}^{\,\omega_2}   \mathcal{A}^\nu_{ca} \mathcal{A}^\lambda_{bc} f_{cb}  \right) +
			 \left[ \left( \nu, \omega_1 \right) \leftrightarrow \left( \lambda, \omega_2 \right) \right].\label{second_order_conductivity_ee_spinful}
		\end{align}  
		\end{widetext}
The component $\sigma_\text{ii}$ has the same form as that in the U(1)-covariant representation, whereas the remaining components are modified. Thus, we investigate the photocurrent responses arising from the components other than $\sigma_\text{ii}$ in the following subsections. Note again that we consider systems preserving the \PT{}-symmetry.

\subsection{Berry curvature dipole term}
Under the condition Eq.~\eqref{photocurrent_frequency_condition}, the component $\sigma_\text{ei}$ is recast as
		\begin{align}
		&\sigma^{\mu;\nu \lambda}_\text{ei} \notag \\
			&=\frac{q^3}{2\hbar^2 \Omega}\int \frac{d\bm{k}}{\left( 2\pi \right)^d} \sum_{a,b}  \mathcal{A}^\mu_{ab}  \mathcal{A}^\nu_{ba } \partial_\lambda f_{ba} + \left[ \left( \nu, -\Omega \right) \leftrightarrow \left( \lambda, \Omega \right) \right],\\
			&=\frac{-iq^3}{2\hbar^2 \Omega}\int \frac{d\bm{k}}{\left( 2\pi \right)^d} \sum_{a\neq b} \Omega^{\mu\nu}_{ab} \partial_\lambda f(\epsilon_{\bm{k}a}) + \left[ \left( \nu, -\Omega \right) \leftrightarrow \left( \lambda, \Omega \right) \right],\label{app_U2_BCD_bare}
		\end{align}
Applying Eq.~\eqref{app_BC_product_PT_symmetry} to Eq.~\eqref{app_U2_BCD_bare} and using the relation $\epsilon_{\bk a } = \epsilon_{\bk \bar{a} }$, we find that the photocurrent $\sigma_\text{ei}$ vanishes. Thus, the Berry curvature dipole term is forbidden due to the \PT{}-symmetry as in the spinless system.  

\subsection{injection current}
We consider a part of the $\sigma_\text{ee}$ term derived from the diagonal component of the velocity operator $v^\mu_{ab}$ in Eq.~\eqref{second_order_conductivity_ee_spinful}. The corresponding contribution $\sigma_\text{ee;d}$ is given by
		\begin{align}
		&\sigma^{\mu;\nu \lambda}_\text{ee;d} \left( \omega; \omega_1,\omega_2 \right) \notag \\
			&= \frac{q^3}{2 \hbar  \omega}\int \frac{d\bm{k}}{\left( 2\pi \right)^d} \sum_{a,b} \Delta^\mu_{ab}  \mathcal{A}^\nu_{ab} \mathcal{A}^\lambda_{ba}  f_{ab} \left(  d_{ba }^{\,\omega_2} + d_{ab}^{\,\omega_1}  \right).
		\end{align}
Owing to the pole at $\omega =0$, it is necessary to pick up the terms in the integrand up to $O(\omega^1)$. Following the parallel discussion of Sec.~\ref{Sec_injection_current}, 
		\begin{align}
		&\sigma^{\mu;\nu \lambda}_\text{ee;d} \left( \omega; \omega_1,\omega_2 \right)\notag  \\
			&= \frac{q^3}{2 \hbar  \omega}\int \frac{d\bm{k}}{\left( 2\pi \right)^d} \sum_{a,b} \Bigl[ -2i\pi \Delta^\mu_{ab}  \mathcal{A}^\nu_{ab} \mathcal{A}^\lambda_{ba}  f_{ab}\delta \left( \hbar \omega_2 - \epsilon_{ba} \right) \notag \\
			&+\mathcal{A}^\nu_{ab} \mathcal{A}^\lambda_{ba}  f_{ab}  \left(  \partial_\mu d_{ba }^{\,\omega_1} \right)_{| \omega_1  = -\omega_2}(\omega_1+\omega_2) \Bigr] + O((\omega_1+\omega_2)^2).\label{cond_U2_eed_expanded}
		\end{align}

Under the photocurrent condition [Eq.~\eqref{photocurrent_frequency_condition}], the first term in Eq.~\eqref{cond_U2_eed_expanded} is obtained as
		\begin{align}
 		&\sigma^{\mu;\nu \lambda}_\text{inj}\notag \\
				&= \lim_{\omega \rightarrow 0} \frac{- i\pi  q^3}{\hbar  \omega}\int \frac{d\bm{k}}{\left( 2\pi \right)^d}\sum_{a,b} \Delta^\mu_{ab}  \mathcal{A}^\nu_{ab} \mathcal{A}^\lambda_{ba}  f_{ab} \delta (\hbar \Omega  - \epsilon_{ba} ).
		\end{align}
Using Eq.~\eqref{app_BC_product_PT_symmetry}, the expression is recast as
		\begin{align}
 		&\sigma^{\mu;\nu \lambda}_\text{inj} (\mathcal{PT})\notag \\ 
				&= \lim_{\omega \rightarrow 0} \frac{- i\pi  q^3}{2 \hbar  \omega}\int \frac{d\bm{k}}{\left( 2\pi \right)^d} \notag \\
				&~~~\times \sum_{a,b} \Delta^\mu_{ab} \left(  \mathcal{A}^\nu_{ab} \mathcal{A}^\lambda_{ba} + \mathcal{A}^\lambda_{ab} \mathcal{A}^\nu_{ba}  \right) f_{ab} \delta (\hbar \Omega  - \epsilon_{ba} ),\\
				&= \lim_{\omega \rightarrow 0} \frac{- i\pi  q^3}{\hbar  \omega}\int \frac{d\bm{k}}{\left( 2\pi \right)^d} \sum_{a\neq b} \Delta^\mu_{ab} g^{\nu\lambda}_{ab} f_{ab} \delta (\hbar \Omega  - \epsilon_{ba} ),
		\end{align}
which is symmetric under the permutation $\nu \leftrightarrow \lambda$. This corresponds to the formula for the magnetic injection current [Eq.~\eqref{magnetic_injection_current_bare}].

The LP-photoconductivity $\eta_\text{intI}$ arising from the second term in Eq.~\eqref{cond_U2_eed_expanded} is obtained as
	\begin{align}
	&\eta^{\mu;\nu \lambda}_\text{intI}\notag\\
		& = \frac{q^3}{2 \hbar }\int \frac{d\bm{k}}{\left( 2\pi \right)^d} \sum_{a,b} g^{\nu\lambda}_{ab}  f_{ab}  \partial_\mu d_{ba }^{\,\Omega},\\
		& = \frac{q^3}{2 \hbar }\int \frac{d\bm{k}}{\left( 2\pi \right)^d} \sum_{a,b} g^{\nu\lambda}_{ab}  f_{ab}  \,  \partial_\mu \text{P} \frac{1}{\hbar \Omega - \epsilon_{ba}},\label{cond_U2_eed_intI}
	\end{align}
in which we use Eq.~\eqref{app_BC_product_PT_symmetry} and the \PT{}-ensured Kramers theorem. This contribution will be combined with the remaining terms among $\sigma_\text{ee}$ as shown in the next subsection.

\subsection{gyration current}
Here, we calculate the remaining terms, that is, $\sigma_\text{ie}$ and $\sigma_\text{ee;o}$. As for the component $\sigma_\text{ie}$, we use Eq.~\eqref{photocurrent_frequency_condition} and arrange the integrand as 
		\begin{align}
		& i  v^\mu_{ab}  d_{ ba}^{\,0}  \Biggl[ \partial_\nu \left( d_{ba}^{\,\Omega}  f_{ab} \mathcal{A}^\lambda_{ba } \right) \notag \\
		&~~~~~- i \left( \sum_c \alpha^\nu_{bc} d^{\Omega}_{ca}  \mathcal{A}^\lambda_{ca} f_{ac}   - \sum_c \alpha^\nu_{ca} d^{\Omega}_{bc}  \mathcal{A}^\lambda_{bc} f_{cb}  \right)    \Biggr],\\
		&= -\mathcal{A}^\mu_{ab}   \Biggl[ \partial_\nu \left( d_{ba}^{\,\Omega}  f_{ab} \mathcal{A}^\lambda_{ba } \right) \notag \\
		&~~~~~- i \left( \sum_c \alpha^\nu_{bc} d^{\Omega}_{ca}  \mathcal{A}^\lambda_{ca} f_{ac}   - \sum_c \alpha^\nu_{ca} d^{\Omega}_{bc}  \mathcal{A}^\lambda_{bc} f_{cb}  \right)    \Biggr],\\
		&=d_{ba}^\Omega f_{ab}\mathcal{A}_{ba}  \left[  \ud_\nu \mathcal{A}^\mu \right]_{ab} - \partial_\nu \left(  d_{ba}^{\,\Omega}  f_{ab} \mathcal{A}^\mu_{ab} \mathcal{A}^\lambda_{ba }  \right).
		\end{align}
Discarding the total derivative as a surface term, the $\sigma_\text{ie}$ term is simplified as
		\begin{align}
		\sigma^{\mu;\nu \lambda}_\text{ie}
			&= \frac{q^3}{2\hbar}\int \frac{d\bm{k}}{\left( 2\pi \right)^d} \sum_{a\neq b} - \left[ \ud_\nu  \mathcal{A}^\mu\right]_{ab}  \mathcal{A}^\lambda_{ba} f_{ba} d_{ba}^{\,\Omega} \notag \\
			& + \left[ \left( \nu, -\Omega \right)\leftrightarrow \left( \lambda, \Omega \right) \right].\label{U2_ie_term}
		\end{align}

The component $\sigma_\text{ee;o}$ is written as
		\begin{align}
 		\sigma^{\mu;\nu \lambda}_\text{ee;o} 
			&= \frac{q^3}{2}\int \frac{d\bm{k}}{\left( 2\pi \right)^d} \sum_{a\neq b} v^\mu_{ab}  d_{ ba}^{\,0} \notag \\
			&~~~~~\times  \left( \sum_c d_{ca }^{\,\Omega}  \mathcal{A}^\nu_{bc} \mathcal{A}^\lambda_{ca}  f_{ac} - \sum_c d_{bc}^{\,\Omega}   \mathcal{A}^\nu_{ca} \mathcal{A}^\lambda_{bc} f_{cb}  \right) \notag \\
			&~~~~~+ \left[ \left( \nu, -\Omega \right) \leftrightarrow \left( \lambda, \Omega \right) \right].\label{cond_U2_eed_bare}
		\end{align}
Although the summation of the band indices includes the Kramers pair $(a,b) = (a,\bar{a})$, the matrix element of the velocity operator satisfies $v^\mu_{a\bar{a}} = 0 $ by taking the orthogonal Bloch states, $\ket{u_a (\bk)}$ and $\ket{u_{\bar{a}} (\bk)}$. Owing to the adiabatic parameter in $\dmat{a\bar{a}}{0} = \left( +0 \right)^{-1}$, we have $v^\mu_{a\bar{a}}\dmat{a\bar{a}}{0} = 0 $. Thus, the integrand in Eq.~\eqref{cond_U2_eed_bare} is recast as
		\begin{align}
 		&\sum_{a\neq b} v^\mu_{ab}  d_{ ba}^{\,0}  \left( \sum_c d_{ca }^{\,\Omega}  \mathcal{A}^\nu_{bc} \mathcal{A}^\lambda_{ca}  f_{ac} - \sum_c d_{bc}^{\,\Omega}   \mathcal{A}^\nu_{ca} \mathcal{A}^\lambda_{bc} f_{cb}  \right),\\
 		&= \sum_{a\neq b \neq c} i \mathcal{A}^\mu_{ab}  \left(  d_{ca }^{\,\Omega}  \mathcal{A}^\nu_{bc} \mathcal{A}^\lambda_{ca}  f_{ac} -  d_{bc}^{\,\Omega}   \mathcal{A}^\nu_{ca} \mathcal{A}^\lambda_{bc} f_{cb}  \right),\\
 		&= \sum_{a \neq c} i  d_{ca}^\Omega \mathcal{A}^\lambda_{ca} f_{ac} \left[ \mathcal{A}^\mu,  \mathcal{A}^\nu \right]_{ac},\\
 		&= \sum_{a \neq c}   d_{ca}^\Omega \mathcal{A}^\lambda_{ca} f_{ac} \left( \left[ \ud_\mu  \mathcal{A}^\nu \right]_{ac} - \left[ \ud_\nu  \mathcal{A}^\mu \right]_{ac}  \right),
		\end{align}
where we used a formula for the U(2)-covariant derivative
		\begin{equation}
		\left[ \ud_\mu  \mathcal{A}^\nu \right]_{ab} - \left[ \ud_\nu  \mathcal{A}^\mu \right]_{ab}  = i \left[ \mathcal{A}^\mu, \mathcal{A}^\nu \right]_{ab}.
		\end{equation}
As a result, we obtain
		\begin{align}
		&\sigma^{\mu;\nu \lambda}_\text{ee;o} \notag \\
		&= \frac{q^3}{2\hbar}\int \frac{d\bm{k}}{\left( 2\pi \right)^d} \sum_{a\neq b}  \left( \left[ \ud_\mu  \mathcal{A}^\nu \right]_{ab} - \left[ \ud_\nu  \mathcal{A}^\mu \right]_{ab}  \right) \mathcal{A}^\lambda_{ba}  f_{ab} d_{ba}^{\,\Omega} \notag \\
		&~~~ + \left[ \left( \nu, -\Omega \right) \leftrightarrow \left( \lambda, \Omega \right) \right].\label{U2_eeo}
		\end{align}
Summing Eqs.~\eqref{U2_ie_term} and \eqref{U2_eeo}, we obtain the photocurrent formula
		\begin{align}
		&\sigma_\text{ie+ee}^{\mu;\nu\lambda}
		= \frac{q^3}{2\hbar}\int \frac{d\bm{k}}{\left( 2\pi \right)^d} \sum_{a\neq b} d_{ba}^\Omega  f_{ab} \left[  \ud_\mu \mathcal{A}^\nu \right]_{ab}\mathcal{A}^\lambda_{ba} \notag \\
		&~~~~~~~~~+ \left[ \left( \nu, -\Omega \right) \leftrightarrow \left( \lambda, \Omega \right) \right],\label{U2_ie+ee}
		\end{align}
where both of $\mathcal{A}_{ba}^\lambda$ and $\left[  \ud_\mu \mathcal{A}^\nu \right]_{ab}$ are U(2)-covariant and hence the overall expression is U(2)-invariant. Making use of Eq.~\eqref{app_BC_BCderiv_PT_symmetry}, we obtain the final expression as
		\begin{align}
		\sigma^{\mu;\nu \lambda}_\text{ee+ie} (\mathcal{PT}) =\eta^{\mu;\nu \lambda}_\text{intII} -\frac{i}{2} \epsilon_{\nu\lambda\tau} \kappa_\text{gyro}^{\mu\tau},\label{cond_U2_eed_intII}
		\end{align}
with the photoconductivity coefficients
		\begin{align}
		&\eta^{\mu;\nu \lambda}_\text{intII}\notag \\
			&=  \frac{q^3}{2\hbar }\int \frac{d\bm{k}}{\left( 2\pi \right)^d} \sum_{a\neq b} f_{ab} \, \text{P} \frac{1}{\hbar \Omega - \epsilon_{ba}} \notag  \\
			&~~~~~\times \text{Re}\,\left(  \left[ \ud_\mu  \mathcal{A}^\nu \right]_{ab} \mathcal{A}^\lambda_{ba}  + \left[ \ud_\mu  \mathcal{A}^\lambda \right]_{ab} \mathcal{A}^\nu_{ba}  \right), \\
			&= \frac{q^3}{2\hbar }\int \frac{d\bm{k}}{\left( 2\pi \right)^d} \sum_{a\neq b} f_{ab} \, \text{P} \frac{1}{\hbar \Omega - \epsilon_{ba}} \partial_\mu g^{\nu\lambda}_{ab},
		\end{align}
for the reactive part, and 
		\begin{align}
		&\kappa^{\mu\nu}_\text{gyro}  = \frac{\pi q^3}{\hbar} \int \frac{d\bm{k}}{\left( 2\pi \right)^d} \sum_{a\neq b} f_{ab} \delta (\hbar \Omega - \epsilon_{ba}) \notag \\
		    &~~~~\times \epsilon_{\nu\lambda\tau } \text{Re}\, \left( \left[ \ud_\mu  \mathcal{A}^\lambda \right]_{ab} \mathcal{A}^\tau_{ba}\right) ,
		\end{align}
for the absorptive part, that is, gyration current. Combining Eq.~\eqref{cond_U2_eed_intI} with Eq.~\eqref{cond_U2_eed_intII}, we obtain the intrinsic Fermi surface term
\begin{align}
	&\eta_\text{IFS}^{\mu;\nu\lambda} = \eta_\text{intI} +\eta_\text{intII}^{\mu;\nu\lambda},\\
	&=\frac{q^3}{\hbar}\int \frac{d\bm{k}}{\left( 2\pi \right)^d} \sum_{a\neq b} g^{\nu\lambda}_{ab} \frac{\epsilon_{ab}}{\hbar^2 \Omega^2 - \epsilon_{ab}^2} \partial_\mu f \left( \epsilon_{\bk a} \right).\label{app_intrinsic_fermi_surface_effect_PT_symmetric}
	\end{align}
We note that the formula is determined by the band-resolved quantum metric $g^{\nu\lambda}_{ab}$ defined by Eq.~\eqref{app_bandresolved_quantum_metric_u1_sector}.

\section{Classification of photocurrent response based on magnetic point group}\label{App_Sec_point_group_list}
This section lists the noncentrosymmetric magnetic point groups preserving the \T{}- or \PT{}-symmetry.
The 122 magnetic point groups are classified into three categories; the 32 \T{}-symmetric point groups (gray group), 32 point groups whose symmetry operations are not relevant to the time-reversal operation (black group), and the others as many as 58 (black-white group, BW group). There are 21 noncentrosymmetric and \T{}-symmetric (\PT{}-symmetric) point groups in the gray (BW) group (see Table.~\ref{App_Table_magnetic_point_group}). 

For instance, let us consider the noncentrosymmetric gray group. The gray group $\bm{G}$ is described by 
		\begin{equation}
		\bm{G} = \bm{H} + \theta \bm{H}, \label{gray_group}
 		\end{equation}
with a noncentrosymmetric black group $\bm{H}$. Among the 21 groups, $20$ groups other than the case with $\bm{H} = m\bar{3}m (O)$ are piezoelectric. On the other hand, $18$ groups other than the cases $\bm{H} = \bar{6} (C_{3h}),~\bar{6}m2 (D_{3h}),~43m (T_{d})$ are gyrotropic. Similarly, we can identity the piezoelectric groups and gyrotropic groups included in the \PT{}-symmetric BW group by replacing $\theta$ with $\theta I$ in Eq.~\eqref{gray_group}. 

As we mention in the main text, piezoelectric groups allow the LP-photocurrent while gyrotropic groups allow the CP-photocurrent. Thus, by referring to Tables~\ref{Table_photocurrent_circular_linear_classification} and \ref{App_Table_magnetic_point_group}, we can systematically identify which photocurrent is allowed in a given noncentrosymmetric system. We also show candidate materials in Table~\ref{App_Table_magnetic_point_group}. Many other candidates for the \PT{}-symmetric compounds can be found in Refs.~\cite{Gallego2016,Watanabe2018grouptheoretical,Watanabe2018}.

		\begin{table}[]
			\caption{List of the photocurrent responses allowed in the \T{}-symmetric point groups (gray group) and \PT{}-symmetric BW point groups. `MPG` denotes magnetic point group and '$\mathcal{H}$' is the maximal unitary subgroup of `MPG'. The symbols $\updownarrow$ and $\circlearrowleft$ denote photocurrents induced by linearly-polarized and circularly-polarized lights, respectively. The allowed response is indicated by \cm. The low of `CM' shows candidate materials where `(ML)`/'(BL)' means monolayer/bilayer. As for the point groups except for the cubic systems, some of the symbols \cm are closed in the parenthesis to indicate that the corresponding responses are not allowed in two-dimensional systems. Classification of the point groups with $\bm{H} = C_{2},C_{2v}$ is further divided according to the relation between the incident direction of light ($\bk$) and the primary axis of the point group.}
		\label{App_Table_magnetic_point_group}
		\centering
		\begin{tabular}{llccl}\hline
			\text{MPG} 			&$\mathcal{H}$	 	& $\updownarrow$ & $\circlearrowleft$ &\text{CM}  \\ \hline \hline
			\multicolumn{5}{l}{\text{(\T{}-symmetric groups)}}\\
			$11'$				&$C_1$   	& \cm               &\cm                    &  \\
			$21'$	&$C_2  \parallel \bk$ 	& \cmd               &\cmd                    &  \\
							&$C_2  \perp \bk$	& \cm               &\cm                    &  \\
			$m1'$				&$C_s$   	&\cm                &\cm                    &\text{WTe$_2$(BL)}  \\
			$2221'$			&$D_2$   	&\cmd                &\cmd                    &  \\ 
			$mm21'$			&$C_{2v}\parallel \bk$   	&\cmd                &\cmd                    &\multirow{2}{*}{\text{TaIrTe$_4$, MoTe$_2$\,($T_d$), WP$_2$}}  \\ 
						&$C_{2v} \perp \bk$  	&\cm                &\cm                    &  \\ 
			$41'$				&$C_4$   	&\cmd                &\cmd                    &  \\ 
			$\bar{4}1'$		&$S_4$   	&\cmd                &\cmd                    &  \\ 
			$4221'$			&$D_4$   	&\cmd                &\cmd                    &\text{(TaSe$_4$)$_2$I} \\ 
			$4mm1'$			&$C_{4v}$   	&\cmd                &\cmd                    &\text{BiTeI, TaP, TaAs, NbP}  \\ 
			$\bar{4}2m1'$		&$D_{2d}$   	&\cmd                &\cmd                    &  \\ 
			$31'$				&$C_3$   	&\cm                &\cmd                    &  \\ 
			$3m1'$				&$C_{3v}$   	&\cm                &\cmd                    &\text{Bi$_2$}\text{Se$_3$}\,(001),\text{LiOsO$_3$} \\ 
			$321'$			&$D_3$   	&\cm                &\cmd                    &\text{Bi}  \\ 
			$61'$				&$C_6$   	&\cmd                &\cmd                    &  \\ 
			$\bar{6}1'$		&$C_{3h}$   	&\cmd                &                    &  \\ 
			$6221'$				&$D_6$   	&\cmd                &\cmd                    &  \\ 
			$6mm1'$				&$C_{6v}$   	&\cmd                &\cmd                    &  \\ 
			$\bar{6}m21'$				&$D_{3h}$   	&\cm                &                    &\text{MoS$_2$(ML)}  \\ 
			$231'$				&$T$   	&\cm                &\cm                    &\text{RhSi}  \\ 
			$43m1'$				&$T_d$   	&\cm                &                    &\text{Ce$_3$Bi$_4$(Pt,Pd)$_3$}  \\  
			$4321'$				&$O$   	&                &\cm                    &\text{Li$_2$BPt$_3$}  \\ \hline
			\multicolumn{5}{l}{\text{(\PT{}-symmetric groups)}}\\
			$\bar{1}'$		&$C_1$    	& \cm               &\cm                    &\text{CaMn$_2$Bi$_2$}  \\
			$2'/m$	&$C_s$    	&\cm                &\cm                    &\text{SrMn$_2$As$_2$}  \\
			$2/m'$	&$C_2 \parallel \bk$   	&\cmd                &\cmd                    &\multirow{2}{*}{\text{Na$_2$RuO$_4$}}  \\
								&$C_2 \perp \bk$   	&\cm                &\cm                    &  \\
			$m'm'm'$	&$D_2$    	&\cmd                &\cmd                    &\text{LiMnPO$_4$}  \\
			$mmm'$	&$C_{2v} \parallel \bk $  	&\cmd                &\cmd                    &\multirow{2}{*}{\text{CuMnAs,~Mn$_2$Au}}  \\
								&$C_{2v} \perp \bk $   	&\cm                &\cm                    &  \\
			$4/m'$	&$C_4$    	&\cmd                &\cmd                    &  \\
			$4'/m'$	&$S_4$    	&\cmd                &\cmd                    &  \\
			$4/m'm'm'$	&$D_4$    	&\cmd                &\cmd                    &\text{Fe$_2$TeO$_6$}  \\
			$4/m'mm$	&$C_{4v}$    	&\cmd                &\cmd                    &  \\
			$4'/m'm'm$	&$D_{2d}$    	&\cmd                &\cmd                    &\text{BaMn$_2$As$_2$,~EuMnBi$_2$}  \\
			$\bar{3}'$	&$C_3$    	&\cm                &\cmd                    &  \\
			$\bar{3}'m$	&$C_{3v}$    	&\cm                &\cmd                    &\text{MnPS$_3$(ML)}  \\
			$\bar{3}'m'$	&$D_3$    	&\cm                &\cmd                    &\text{Cr$_2$O$_3$,~MnBi$_2$Te$_4$(BL)}  \\
			$6/m'$			&$C_6$    	&\cmd                &\cmd                    &  \\
			$6'/m$			&$C_{3h}$    	&\cmd                &                    &  \\
			$6/m'm'm'$	&$D_6$    	&\cmd                &\cmd                    &  \\
			$6/m'mm$	&$C_{6v}$    	&\cmd                &\cmd                    &  \\
			$6'/mmm'$	&$D_{3h}$    	&\cm                &                    &  \\
			$m'\bar{3}'$	&$T$    	&\cm                &\cm                    &\text{Cu$_3$TeO$_6$}  \\
			$m'\bar{3}'m$	&$T_d$    	&\cm                &                    &  \\
			$m'\bar{3}'m'$	&$O$    	&                &\cm                    & \\ \hline
		\end{tabular}

		\end{table}

\bibliography{photocurrent}

\end{document}